\documentclass[%
 reprint,
 amsmath,amssymb,
 aps,
]{revtex4-2}

\usepackage{graphicx}
\usepackage{dcolumn}
\usepackage{bm}
\usepackage{booktabs} 
\usepackage{multirow} 
\usepackage[normalem]{ulem}
\usepackage{siunitx}

\usepackage{xcolor}
\usepackage{subcaption}
\usepackage[font=small]{caption}
\usepackage{comment}
\usepackage{mathrsfs} 
\usepackage{tabularx} 

\newcommand{\bfX}{\mathbf{X}}
\newcommand{\bfY}{\mathbf{Y}}

\usepackage{subcaption} 

\newcommand{\cco}[1]{{\small \color{red} \tt [CO]}}

\begin{document}

\preprint{APS/123-QED}

\title{Flexible Uncertainty Calibration for Machine-Learned Interatomic Potentials}

\author{Cheuk Hin Ho}
\affiliation{Department of Mathematics, University of British Columbia, Vancouver, V6T1Z2, Canada}

\author{Christoph Ortner}
\affiliation{Department of Mathematics, University of British Columbia, Vancouver, V6T1Z2, Canada}

\author{Yangshuai Wang}%
\email{yswang@nus.edu.sg}
\affiliation{Department of Mathematics, National University of Singapore, 10 Lower Kent Ridge Road, 119076, Singapore}%

\date{\today}

\begin{abstract}
Reliable uncertainty quantification (UQ) is essential for developing machine-learned interatomic potentials (MLIPs) in predictive atomistic simulations. Conformal prediction (CP) is a statistical framework that constructs prediction intervals with guaranteed coverage under minimal assumptions, making it an attractive tool for UQ. However, existing CP techniques, while offering formal coverage guarantees, often lack accuracy, scalability, and adaptability to the complexity of atomic environments. In this work, we present a flexible uncertainty calibration framework for MLIPs, inspired by CP but reformulated as a parameterized optimization problem. This formulation enables the direct learning of environment-dependent quantile functions, producing sharper and more adaptive predictive intervals at negligible computational cost. Using the foundation model MACE-MP-0 as a representative case, we demonstrate the framework across diverse benchmarks, including ionic crystals, catalytic surfaces, and molecular systems. Our results show order-of-magnitude improvements in uncertainty–error correlation, enhanced data efficiency in active learning, and strong generalization performance, together with reliable transfer of calibrated uncertainties across distinct exchange–correlation functionals. This work establishes a principled and data-efficient approach to uncertainty calibration in MLIPs, providing a practical route toward more trustworthy and transferable atomistic simulations.
\end{abstract}

\maketitle


\section{Introduction}
\label{sec:intro}

Machine-learned interatomic potentials (MLIPs) have emerged as a powerful alternative to traditional empirical force fields and first-principles calculations, allowing large-scale and high-throughput atomistic simulations with near DFT accuracy at a fraction of the cost~\cite{behler2007generalized, bartok2010gaussian, witt2023acepotentials, batatia2022mace, shapeev2016moment, schutt2018schnet, batzner20223, cheng2024cartesian, liu2025fine, merchant2023scaling, zhang2024dpa}. Despite their success in diverse applications such as materials discovery and molecular dynamics, ensuring the reliability of MLIP predictions, particularly under out-of-distribution conditions, remains a critical challenge. This challenge has motivated a growing interest in uncertainty quantification (UQ), which provides a basic framework to assess predictive reliability and guide the reliable deployment of MLIP-based simulations~\cite{frenkel2023understanding, angelikopoulos2012bayesian, messerly2017uncertainty, kellner2024uncertainty, imbalzano2021uncertainty}.


A variety of strategies have been introduced to characterize and mitigate predictive errors. For a comprehensive overview, we refer to recent reviews~\cite{grasselli2025uncertainty, dai2025uncertainty}. Among existing methods, ensemble-based approaches are widely used, estimating predictive uncertainty from variance of quantities of interest (QoI) between models trained with different data splits~\cite{peterson2017addressing, musil2019fast}, potential forms~\cite{behler2014representing, xiao2018uncertainty}, or initializations~\cite{novikov2019improving, zhang2019active}. Distance-based methods evaluate the dissimilarity between new configurations and the training set using measures such as D-optimality~\cite{novikov2020mlip}, atomic fingerprints~\cite{botu2017machine}, or latent space distances~\cite{janet2019quantitative}. Probabilistic frameworks, including dropout-based Bayesian inference~\cite{xie2021advanced} and a posteriori adaptivity approaches~\cite{chen2022qm, wang2021posteriori}, also provide uncertainty estimates. However, these methods often lack proper calibration and are highly sensitive to model architecture and training choices~\cite{tan2023single}.
Despite their utility, many existing UQ approaches often fall short of delivering reliable accuracy. This is partly due to their lack of explicit modeling of prediction errors, which leads to weak alignment between estimated uncertainties and actual deviations. As a result, their effectiveness is limited in downstream tasks such as active learning~\cite{zaverkin2024uncertainty, hodapp2020operando}, transferability screening~\cite{dosinger2023efficient}, and stability assessment in molecular dynamics trajectories~\cite{van2023hyperactive, mismetti2024automated}. In addition, ensemble- and distance-based methods typically suffer from poor scalability, while Bayesian or dropout-based methods are often sensitive to architectural choices and prone to miscalibration.
These limitations highlight the need for principled, well-calibrated uncertainty estimates. Conformal prediction (CP) offers a promising solution: it is a model-agnostic framework that provides finite-sample guarantees for predictive intervals~\cite{angelopoulos2021gentle, shafer2008tutorial, yu2025conformal}. Although CP has recently been explored in atomistic modeling~\cite{hu2022robust, best2024uncertainty}, existing implementations suffer from two key drawbacks: (i) a decoupled training–calibration pipeline that limits accuracy and efficiency, and (ii) restricted expressiveness in adapting prediction intervals to complex local atomic environments.

To address limitations of conventional CP in atomistic modeling, we introduce a flexible calibration framework in which an analogue of the conformal quantile is learned as a smooth function of the local atomic environment rather than fixed as a global scalar. This enables predictive intervals to vary across space and adapt to complex structural features, improving both coverage and sharpness. Formally, the scalar quantile in the regular CP objective is replaced by a parameterized function trained with the pinball loss over the atoms in the calibration set. For robustness when nominal uncertainties are small, we further align the predicted interval directly to the reference force error using a physically motivated weight that emphasizes large deviations. The resulting objective is interpretable, numerically stable, and compatible with all local MLIP architectures.

The design is conceptually related to class-based CP~\cite{gibbs2025conformal} but enables fully data-driven and continuous adaptation within modern MLIP pipelines. Applied to the MACE-MP-0 foundation model, the framework yields substantial improvements over standard CP in uncertainty calibration on various benchmarks, including ionic crystals, catalytic surfaces, and molecular datasets. It provides adaptive, site-resolved confidence measures that closely track true errors and improves the selection of high-error configurations for active fine-tuning and molecular dynamics. The method introduces negligible computational overhead relative to the evaluation of the baseline model, remains data efficient with only a modest number of calibration configurations, and reliably transfers across distinct exchange-correlation functionals. Together, these properties establish a practical and principled approach to uncertainty calibration for MLIPs, supporting reliable large-scale atomistic simulations.

This paper is organized as follows. Section~\ref{sec:methods} introduces the regular CP framework, outlines its limitations for machine-learned interatomic potentials, and presents the class-based CP and the flexible calibration approach. Section~\ref{sec:results} reports numerical experiments across diverse datasets, demonstrating improved calibration accuracy, data efficiency, and generalization. Section~\ref{sec:summary} concludes with a summary of the main findings and discusses directions for future work. Additional details and supplementary numerical results are provided in Appendices~A–D.

\section{Methods}
\label{sec:methods}
In this section, we present new methods for calibrating heuristic uncertainty estimates in MLIP predictions. We begin by introducing regular conformal prediction to establish notation and background. We then review the generalization of the framework to class-based conformal prediction, designed to mitigate covariate shifts in testing distributions. We then propose a new uncertainty calibration approach, inspired by but not strictly within the framework of CP, and demonstrate its application to MLIPs in the context of calibrating forces uncertainty.

\subsection{Conformal Prediction}
\label{sec:sub:cp}
Conformal prediction~\cite{angelopoulos2021gentle} is a distribution-free and model-agnostic framework that calibrates uncertainty with guaranteed coverage under minimal assumptions. 

To begin with, let $(\Omega, \mathcal{F}, \mathbb{P})$ be a probability space. We consider a data-generating process governed by a (unknown) joint distribution $P$ on $\mathcal{X} \times \mathcal{Y}$, where $\mathcal{X}$ and $\mathcal{Y}$ denote the input and output spaces, respectively. We assume that $\mathcal{Y}$ is equipped with a norm $\|\cdot\|$ (possibly the Euclidean norm). Let $\mathcal{D}_{\rm train} = {(\mathbf{X}_{i}, \mathbf{Y}_{i})}_{i=1}^{N_{\rm train}}$ be a training dataset consisting of $N_{\rm train}$ i.i.d. samples drawn from $P$.

Given $\mathcal{D}_{\rm train}$, we train a predictive model $\tilde{f} : \mathcal{X} \to \mathcal{Y}$ with an associated heuristic uncertainty estimate $\sigma: \mathcal{X} \to \mathbb{R}{>0}$. Practically, the uncertainty estimate $\sigma(\mathbf{X})$ may be derived from ensemble methods, Monte Carlo dropout, Bayesian posterior variances, or other approximate epistemic uncertainty quantification schemes~\cite{lakshminarayanan2017simple, angelikopoulos2012bayesian, tan2023single, musil2019fast}. 

In many cases, $\sigma(\bfX)$ is inaccurate and requires calibration on a (finite) calibration set. Such a calibration set is analogously assumed to be comprised of i.i.d. samples from $P$, with an additional mild {\it exchangeability} assumption with test inputs~\cite{shaked1977concept}. Exchangeability means that the set of random variables $\mathcal{D}_{\rm cali} \cup \{\bfX_{\rm new}\}$ is permutation invariant for any testing configuration $\bfX_{\rm new}$.

To proceed with calibrating $\sigma(\bfX)$, we define a \emph{score function} $s : \mathcal{X} \times \mathcal{Y} \to \mathbb{R}_{\ge 0}$, 
\begin{equation}\label{eqn:general_score_func}
    s(\mathbf{X}, \mathbf{Y}) := \frac{\big\| \tilde{f}({\bf X}) - \mathbf{Y} \big\|}{\sigma(\mathbf{X})},
\end{equation}
which measures the discrepancy between the model prediction and the observed ground truth, normalized by the heuristic uncertainty estimate.

Using Eq.~\eqref{eqn:general_score_func}, calibration scores $\big\{ s_i := s(\mathbf{X}_i, \mathbf{Y}_i) \big\}_{i=1}^{N_{\text{cal}}}$ are computed for all samples in $\mathcal{D}_{\text{cal}}$. Given a fixed $\alpha \in (0, 1)$, setting 
\begin{equation}
\label{eqn:quantile}
    \hat{q} := \text{quantile}\left(\{s_i\}_{i=1}^{N_{\text{cal}}}, \frac{\lceil (N_{\text{cal}} + 1)(1 - \alpha) \rceil}{N_{\text{cal}}} \right) 
\end{equation}
ensures that
\begin{align}
\label{eqn:marginal_coverage_main}
1 - \alpha 
&\leq \mathbb{P}\big( \| \mathbf{Y}_{\rm new} - \tilde{f}(\bfX_{\rm new}) \| 
\leq \hat{q} \sigma(\bfX_{\rm new}) \big) \notag \\
&\leq 1 - \alpha + \frac{1}{N_{\rm cal}+1}.
\end{align}
Such a formulation, based on the leave-one-out augmented dataset~\cite{auddy2024approximate}, ensures the marginal coverage guarantee \eqref{eqn:marginal_coverage_main}. The scaled uncertainty $\hat{q} \sigma(\bfX_{\rm new})$ can be regarded as the calibrated uncertainty. See Appendix~\ref{sec:apd:standard_cp} for further details and Appendix~\ref{sec:apd:bayes} for its Bayesian interpretation.

A key property that serves as the main motivation of our proposed framework in Section~\ref{sec:sub:learnable_cp} is that regular conformal prediction can be framed as an optimization problem. As shown in~\cite{bai2022efficient}, the conformal quantile $\hat{q}$ in \eqref{eqn:quantile} can be equivalently obtained by solving
\begin{equation}\label{eqn:minimization_usual_cp}
\hat{q} = \mathop{\arg\min}_{q \in \mathbb{R}} \sum_{i=1}^{N_{\rm cal}} \ell_{\alpha}(q, s_i),
\end{equation}
where $\ell_{\alpha}$ is the pinball loss function defined by
\begin{equation}\label{eqn:pinball}
    \ell_\alpha(q, S) := 
\begin{cases}
(1 - \alpha)(S - q) & \text{if } S \geq q, \\
\alpha(q- S) & \text{if } S < q.
\end{cases}
\end{equation}
The hyperparameter $\alpha$ in \eqref{eqn:pinball} again governs the trade-off between underestimation and overestimation in the calibrated uncertainty, equivalent to the effect of $\alpha$ to the choice of quantile in \eqref{eqn:quantile}.

\subsection{Class-based Conformal Prediction}
\label{sec:sub:class_cp}
The conformal quantile $\hat{q}$, obtained as the solution to Eq.~\eqref{eqn:minimization_usual_cp}, serves as a global scaling factor for calibrating heuristic uncertainty estimates. Since $\hat{q}$ is shared across all inputs, regardless of the ``local'' quality of the heuristic uncertainty $\sigma(\mathbf{X})$, this fundamentally limit's the methods calibration accuracy.
Moreover, regular conformal prediction is highly sensitive to covariate shift, further degrading its performance in distributionally shifted settings \cite{gibbs2025conformal}. These limitations highlight the need for \emph{input-dependent rescaling}, where each test point $\mathbf{X}$ is assigned its own adjustment factor to better capture the heterogeneity of uncertainty across the input space.

\begin{figure*}[t]
    \centering
\includegraphics[width=0.95\textwidth]{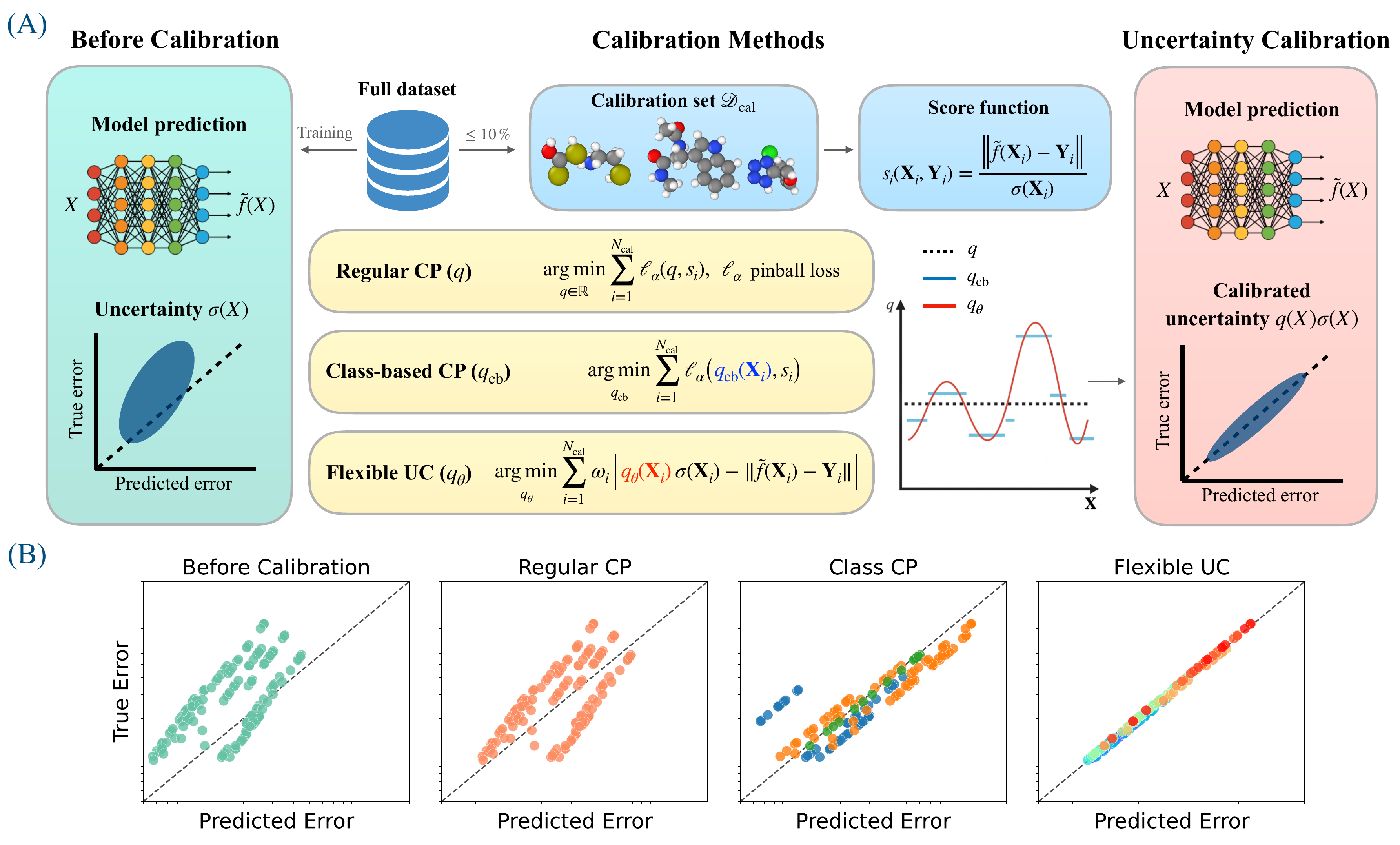}
    \caption{Conceptual illustration of the uncertainty calibration framework. (A) Workflow and calibration schemes. (B) Representative comparison of calibration outcomes. Before calibration and regular CP show systematic misalignment between predicted and true errors. 
Class CP improves alignment by grouping atomic environments into discrete classes, shown in different colors. 
Flexible UC further refines this by learning a smooth, environment-dependent mapping.}
    \label{fig:learnablecp-illustration}
\end{figure*}


A natural extension of regular conformal prediction is to allow the rescaling parameter to vary across different input classes $\xi \subseteq \mathcal{X}$ with $\mathcal{X} = \dot{\bigcup}_{\xi} \xi$ so that the calibration is becomes more fine-grained. One could first train a classification model that induces such a finite partition $\xi \in \Xi$ of the input space, and perform calibration on each class respectively. More precisely, calibration scores $\big\{ s^{\xi}_i := s(\mathbf{X}_i, \mathbf{Y}_i) \mid \bfX_i \in \xi, \xi \in \Xi \big\}_{i=1}^{N_{\xi, \text{cal}}}$ are computed for all samples in $\mathcal{D}_{\text{cal}}$. Now, fix $\alpha \in (0, 1)$, setting 
\begin{equation}
\label{eqn:quantile_class}
    \hat{q}_{\xi} := \text{quantile}\left(\{s^{\xi}_i\}_{i=1}^{N_{\text{cal}}}, \frac{\lceil (N^{\xi}_{\text{cal}} + 1)(1 - \alpha) \rceil}{N^{\xi}_{\text{cal}}} \right) 
\end{equation}
ensures that, as shown in~\cite{gibbs2025conformal, jung2022batch},
\begin{align}
\label{eqn:marginal_coverage_main_class}
1 - \alpha 
&\leq \mathbb{P}\big( \| \mathbf{Y}_{\rm new} - \tilde{f}(\bfX_{\rm new}) \| 
\leq \hat{q}_{\xi} \sigma(\bfX_{\rm new}) \big) \notag \\
&\leq 1 - \alpha + \frac{|\Xi|}{(N_{\rm cal}^\xi+1) \cdot \mathbb{P}(\bfX_{\rm new}\in\xi)}.
\end{align}
for any testing configuration $\bfX_{\rm new} \in \xi$. 

This can be thought of as rescaling uncertainties with a step function (w.r.t. a finite partition of input space) that is constant within each input class, see Appendix~\ref{sec:apd:class_cp} for details. Analogously to \eqref{eqn:minimization_usual_cp}, we can write this as an optimization problem
\begin{equation}
\label{eqn:group_min}
\hat{q}_{\rm cb} := \mathop{\arg\min}_{q_{\rm cb} \in \mathscr{F}_{\rm cb}} \sum_{i = 1}^{N_{\rm cal}} \ell_{\alpha}\big(q_{\rm cb}(\mathbf{X}_i), s_i\big),
\end{equation}
where the subscript ${\rm cb}$ is used to distinguish $\hat{q}_{\rm cb}$ from the quantile obtained in Eq.~\eqref{eqn:minimization_usual_cp}. In here, $\mathscr{F}_{\rm cb}$ denotes the class of step functions over a finite collection of classes $\Xi$, defined as
$$
\mathscr{F}_{\rm cb} := \left\{ x \mapsto \sum_{\xi \in \Xi} q_\xi \, \mathbf{1}\{x \in \Xi\} : q_\xi \in \mathbb{R},\, \forall \xi \in \Xi \right\},
$$
and $\mathbf{1}\{\cdot\}$ denotes the indicator function. This is equivalent to performing regular conformal prediction separately on each class $\xi$, thereby yielding a piecewise-constant function $\hat{q}_{\rm cb}(\mathbf{X})$ with $q_\xi = \hat{q}_{\xi}$ that assigns a distinct quantile value to each class. 

As shown in~\cite[Corollary 1]{gibbs2025conformal} (and briefly discussed in Appendix~\ref{sec:apd:cp_theory}), conformal prediction with class-dependent quantiles yields valid \emph{class-conditional coverage}, meaning the coverage guarantee holds separately within each class $\xi \in \Xi$.


\subsection{Flexible Uncertainty Calibration}
\label{sec:sub:learnable_cp}
As we shall see throughout the numerical experiments in Section~\ref{sec:results}, class-based CP provides relatively modest improvement to uncertainty calibration over regular CP. While the step-function formulation provides a useful extension and yields class-conditional coverage, it is natural to admit even more flexible calibration functions. 
Specifically, we will allow the quantile to vary smoothly as a function of the input, thereby capturing finer-grained variations in the uncertainty structure across the input space. A schematic illustration of the proposed uncertainty calibration framework and its workflow is provided in Figure~\ref{fig:learnablecp-illustration}. 

This could be achieved by replacing the step function class in \eqref{eqn:group_min} with an appropriate class of smooth functions $\mathscr{F}$, leading to the following optimization problem:
\begin{equation}
\label{eqn:learnable_ori_min}
\hat{q}_\theta := \mathop{\arg\min}_{q_{\theta} \in \mathscr{F}} \sum_{i = 1}^{N_{\rm cal}} \ell_{\alpha}\big(q_{\theta}(\mathbf{X}_i), s_i\big).
\end{equation}
For example, $\mathscr{F}$ may be chosen as the Barron space~\cite{weinan2019barron}, which encompasses functions that can be efficiently approximated by shallow neural networks. In this case, the quantile function $q_\theta$ is parameterized with learnable parameters $\theta$, allowing for input-dependent calibration that adapts continuously across the input domain. 

Such an extension, is a generalization of class partitioning of input space in Section~\ref{sec:sub:class_cp} to infinite dimensional. However, as shown in~\cite{foygel2021limits, vovk2012conditional}, exact conditional coverage over an infinite-dimensional input space is infeasible, since the error bound in~\eqref{eqn:marginal_coverage_main} scales with the number of partition classes. To mitigate this, \cite{gibbs2025conformal} introduces a surrogate calibration objective, yielding a relaxed but theoretically justified conditional guarantee. See Appendix~\ref{sec:apd:flexible} for more explanations. Apart from the above issue, we make the observation that the hyperparameter $\alpha$ is rather restrictive that can only be chosen to bias under/over-estimation. Although the coverage property entails from $\alpha$ is theoretically attractive, practically one might prefer general improvement of quality of uncertainty instead of an exact coverage.

With this in mind, we propose a weighted objective functional that generalizes \eqref{eqn:learnable_ori_min} and aligns the predicted uncertainty with the observed prediction error. Specifically, we define  
\begin{align}
\label{eqn:weighted_minimize_q}
    \hat{q}_\theta 
    &= \mathop{\arg\min}_{q_\theta \in \mathscr{F}} 
       \sum_{i = 1}^{N_{\rm cal}} 
       \widetilde{w}(\mathbf{X}_i, \mathbf{Y}_i)\,
       \big| q_\theta(\mathbf{X}_i) - s(\mathbf{X}_i, \mathbf{Y}_i) \big|,
\end{align}
where $\widetilde{w}(\mathbf{X}_i, \mathbf{Y}_i)$ is a weight function. Note that when $\widetilde{w}$ is defined as  
\begin{equation}
    \widetilde{w}(\mathbf{X}_i, \mathbf{Y}_i) =
    \begin{cases}
        1 - \alpha, & \text{if } q_\theta(\mathbf{X}_i) - s(\mathbf{X}_i, \mathbf{Y}_i) \geq 0, \\
        \alpha, & \text{otherwise},
    \end{cases}
\end{equation}
the minimization problem \eqref{eqn:weighted_minimize_q} reduces to \eqref{eqn:learnable_ori_min}. In practice, we use an equivalent formulation of \eqref{eqn:weighted_minimize_q}
\begin{equation}
\label{eqn:weighted_minimize_q_err_form}
    \hat{q}_\theta = \mathop{\arg\min}_{q_\theta \in \mathscr{F}} 
       \sum_{i = 1}^{N_{\rm cal}} 
       w(\mathbf{X}_i, \mathbf{Y}_i)\,
       \big| q_\theta \,\sigma(\mathbf{X}_i) - \| \tilde{f}(\mathbf{X}_i) - \mathbf{Y}_i \| \big|,
\end{equation}
where $\widetilde{w}(\mathbf{X}_i, \mathbf{Y}_i)=w(\mathbf{X}_i, \mathbf{Y}_i) / \sigma(\mathbf{X}_i)$ so that the weights $w(\mathbf{X}_i, \mathbf{Y}_i)$ can be chosen with the same unit as the error instead of the score, which is more intuitive. The weight is typically chosen to emphasize samples with larger observed errors.

This formulation prioritizes aligning predicted uncertainties with observed errors rather than enforcing theoretical coverage guarantees, thereby improving calibration quality in both overestimation and underestimation, as further demonstrated in Section~\ref{sec:results}. The overall method is illustrated in Fig.~\ref{fig:learnablecp-illustration}, and will be used consistently throughout our numerical experiments.

\subsection{Application to Forces Uncertainty}
\label{sec:sub:force_uncertainty}

The framework introduced in the previous section is highly general and agnostic to the specific nature of the prediction task. In this section, we specialize the formulation to the context of machine-learned interatomic potentials (MLIPs), with a particular focus on the calibration of force uncertainties that are central to applications such as molecular dynamics and active learning. Importantly, the same methodology can be readily extended to other atom-centered quantities (e.g. charges), however, calibrating global quantities (e.g., total energy) is more challenging due to the increased variance stemming from size-extensive effects and the typically limited size of available calibration datasets.

To begin with, consider a MLIP model trained to approximate the potential energy \( E \in \mathbb{R} \) of a given atomic configuration $\{{\bm r}_i, Z_i\}_{i = 1}^{N_{\rm at}}$. If $\widetilde{E}\big(\{{\bm r}_i\}_{i = 1}^{N_{\rm at}}\big)$ is the predicted energy, then $\widetilde{{\bf F}}_i := - \nabla_{{\bm r}_i} \widetilde{E} \in \mathbb{R}^3$ is the predicted force. For a {\em local} MLIP model, we can write 
\begin{equation} \label{eq:force_localenv}
    \widetilde{\mathbf{F}}_i := \tilde{f}(\mathbf{X}_i),
\end{equation}
where ${\mathbf{X}}_i$ denotes the local atomic environment of atom \( i \). For non-conservative force fields \eqref{eq:force_localenv} is applied directly without reference to the MLIP energy. For future reference, we use $\mathcal{X}$ to denote the class of all atomic environments in a system of interest, so that the predicted atomic forces $\tilde{f} : \mathcal{X} \to \mathbb{R}^3$. 
We assume that each prediction \( \widetilde{\mathbf{F}}_i \) is accompanied by an associated uncertainty estimate \( \sigma_F(\mathbf{X}_i) \in \mathbb{R}_{>0} \), derived from ensemble variance, dropout variance, posterior approximation (e.g., Laplace or variational inference), or other heuristic epistemic uncertainty estimators~\cite{lakshminarayanan2017simple, angelikopoulos2012bayesian, tan2023single, musil2019fast, perez2025uncertainty}. Table~\ref{tab:mlips_uq} summarizes representative MLIP architectures and their commonly used UQ approaches, which in this work provide the baseline uncertainty estimates $\sigma_F(\mathbf{X}_i)$.

To calibrate these per-atom force uncertainties, we define a site-specific conformal score function based on the pair \( (\mathbf{X}_i, \mathbf{F}_i) \), where \( \mathbf{F}_i \in \mathbb{R}^3 \) is the ground-truth DFT force on atom \( i \):
\begin{equation}
\label{eqn:score_F}
s_F(\mathbf{X}_i, \mathbf{F}_i) := \frac{ \big\| \widetilde{\mathbf{F}}_i - \mathbf{F}_i \big\| }{ \sigma_F(\mathbf{X}_i) }.
\end{equation}
This formulation follows directly from the general conformal score in Eq.~\eqref{eqn:general_score_func}, applied at the atomic level by replacing the generic output \(\mathbf{Y}\) with the per-atom force vector \(\mathbf{F}_i\). The resulting score \( s_F \in \mathbb{R}_{\ge 0} \) quantifies the normalized discrepancy between the predicted and true forces, scaled by the model’s heuristic uncertainty.

In the calibration framework, each atomic environment \( \mathbf{X}_i \) is mapped to a descriptor vector, which serves as the input to the quantile model $q_{\theta}$. For traditional descriptors such as atom-centered symmetry functions (ACSF)~\cite{behler2007generalized}, smooth overlap of atomic positions (SOAP)~\cite{bartok2013representing}, or atomic cluster expansion (ACE)~\cite{drautz2020atomic, liu2022towards}, these feature vectors are fixed functions of local geometry. In contrast, equivariant neural networks (ENNs) like MACE learn symmetry-adapted representations jointly with the potential, providing richer descriptors. In our case, we use a simple feed-forward neural network that maps trained MACE descriptors to quantile estimates. This demonstrates that the proposed calibration approach naturally accommodates both fixed and learned descriptors, and is therefore broadly applicable across MLIP architectures.



\begin{table*}
\centering
\setlength{\tabcolsep}{4pt} 
\caption{Representative machine-learned interatomic potentials (MLIPs) and their commonly used uncertainty quantification (UQ) methods with descriptions and references.}
\label{tab:mlips_uq}
\begin{tabular}{l c c c c}
\toprule
\textbf{MLIPs} & \textbf{Representative UQ} & \textbf{Descriptions} & \textbf{Foundation Model} & \textbf{References} \\
\midrule
ACE & Bayesian & Bayesian linear regression & No & \cite{van2023hyperactive} \\
DP & Ensemble & Model ensemble (different random seeds) & No & \cite{zhang2019active} \\
EquiformerV2 & Feature distance & Distance in latent feature space & Yes & \cite{sakai2024active} \\
GAP & Bayesian & Gaussian process regression & No & \cite{musil2019fast} \\
HDNNP & Dropout & Monte Carlo dropout & No & \cite{wen2020uncertainty} \\
MACE & Ensemble & Model ensemble (different random seeds) & No & \cite{perego2024data} \\
MACE-MP-0 & LLPR & Local linear prediction rigidities & Yes & \cite{bigi2024prediction} \\ 
MTP & D-optimality & D-optimality criterion in active learning & No & \cite{podryabinkin2023mlip} \\
NEP & Ensemble & Model ensemble (different random seeds) & No & \cite{fan2022gpumd} \\
NequIP & Bayesian & Gaussian mixture posterior & No & \cite{zhu2023fast} \\
SevenNet & Ensemble & Multi-head committee models & Yes & \cite{beck2025multi} \\
SNAP & POPS & Pointwise optimal parameter sets & No & \cite{swinburne2025} \\
\bottomrule
\end{tabular}
\end{table*}


\section{Results}
\label{sec:results}

To demonstrate the effectiveness of the proposed uncertainty calibration framework, we conduct a series of numerical experiments based on the state-of-the-art MACE architecture~\cite{batatia2022mace}, which enables us to undertake a broad range of tests within a single framework. MACE extends the Atomic Cluster Expansion (ACE)~\cite{drautz2020atomic} by incorporating higher-order equivariant message passing through tensor products (see Appendix~\ref{sec:apd:mace} for details). We employ the recently released foundation model MACE-MP-0b3~\cite{batatia2023foundation}, trained on the MPtraj dataset, together with LLPR-based uncertainty estimates~\cite{bigi2024prediction} as the baseline (see Appendix~\ref{sec:apd:appendix} for background). 

Our evaluation proceeds in four stages, each addressing a different aspect of calibration. 
(1) We begin by examining baseline performance on both task-specific and general-purpose datasets, and by assessing the computational efficiency of the approach (Section~\ref{sec:sub:various_datasets}). (2) We then move to a more challenging setting, testing how well the method generalizes to unseen atomic environments in catalytic reaction pathways (Section~\ref{sec:sub:gen}). (3) Next, we demonstrate how calibrated uncertainties can guide fine-tuning in molecular dynamics simulations, highlighting their value for practical workflows (Section~\ref{sec:sub:md}). (4) Finally, we test whether calibrated uncertainties can be transferred across different exchange–correlation functionals, an important aspect of multi-fidelity and multi-functional training (Section~\ref{sec:sub:transfer}).

In each case study, we compare three calibration strategies. Regular CP applies a single global quantile, obtained by solving Eq.~\eqref{eqn:quantile} with $\alpha = 0.5$. Class-based CP partitions local atomic environments into discrete groups. We typically set $N_{\rm class} = 20$ if there is no further explicit mentioned and we construct the partition using a Gaussian mixture model~\cite{rasmussen1999infinite}. For datasets with moderately diverse environments (cf.~Section~\ref{sec:sub:transfer}), the number of classes is empirically increased until the score distributions are well separated, beyond which further refinement has little effect. Flexible UC, our proposed approach, employs a feedforward neural network that takes MACE descriptors as input and outputs a multiplicative scaling factor applied to the baseline estimate $\sigma$. Full implementation details are given in Appendix~\ref{sec:apd:sec:apd:hyperparameters}.

\subsection{Uncertainty Calibration with Atomistic Foundation Model}
\label{sec:sub:various_datasets}

We first evaluate and calibrate uncertainties predicted by the pre-trained atomistic foundation model MACE-MP-0b3 using the LLPR baseline. No additional training or fine-tuning is performed; our focus here is solely on post hoc calibration of the existing model predictions.

As a starting point, we begin with the LiCl dataset from~\cite{sivaraman2021automated}, a prototypical ionic compound with simple composition and well-separated local atomic environments. Its moderate size and distinct structural motifs make it an ideal test case for evaluating the initial performance of uncertainty quantification methods. Beyond this case-specific benchmark, we also include evaluations on larger public datasets such as MPtraj~\cite{jain2013commentary} and MATPES~\cite{kaplan2025foundational} to demonstrate the broader applicability of our method.

\begin{figure}
    \centering
    \includegraphics[width=1.0\linewidth]{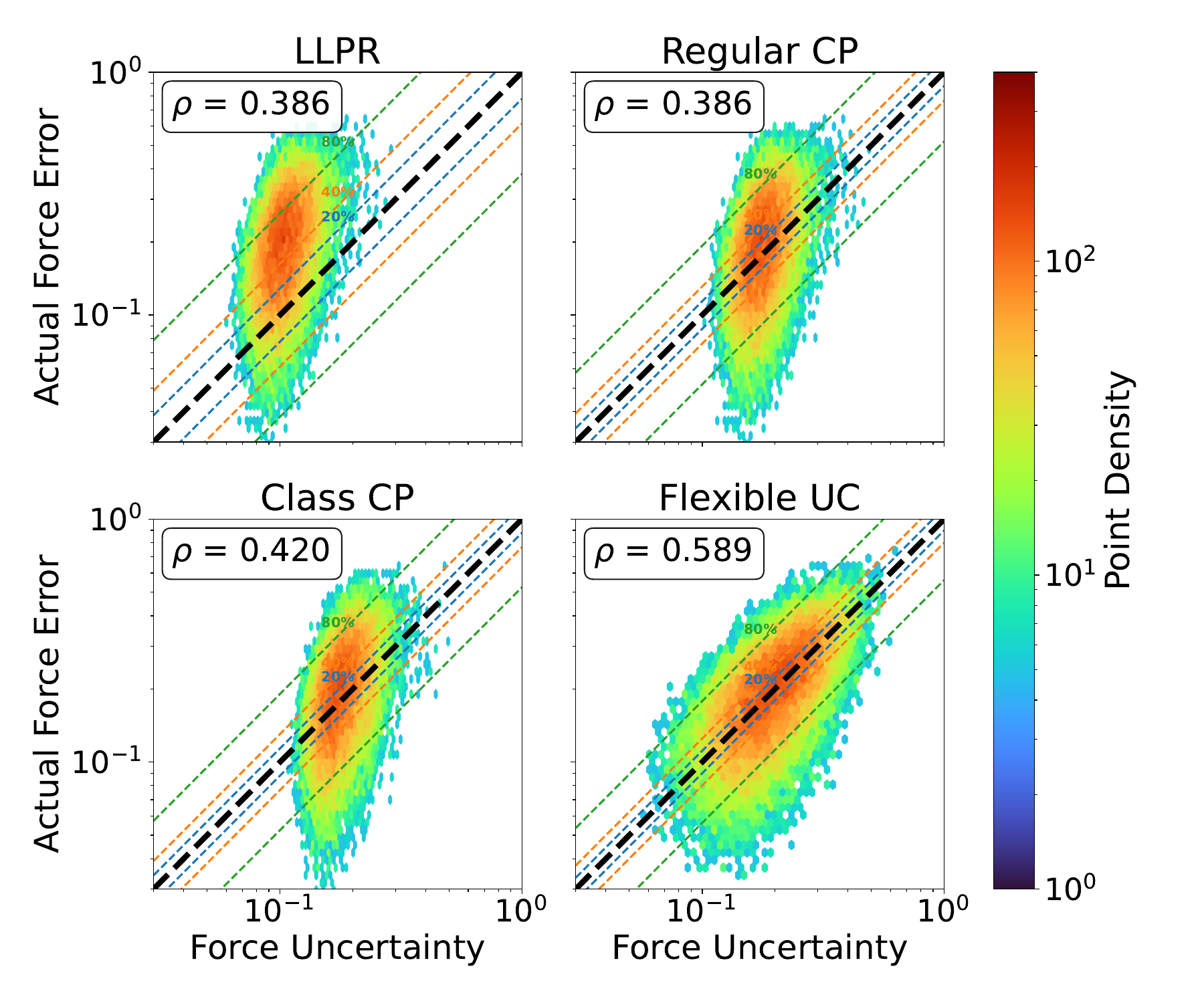}
    \caption{Comparison of uncertainties from LLPR, regular CP, class-based CP, and flexible UC on LiCl dataset. $\rho$ denotes the spearman rank correlation coefficient.
    }
    \label{fig:LiCl_benchmark}
\end{figure}

\paragraph{Calibration.}
$10 \%$ of configurations are drawn randomly (uniform) from the dataset to form the calibration set.
Figure~\ref{fig:LiCl_benchmark} presents a comparison of predicted force uncertainties versus actual force errors across four uncertainty estimation/calibration schemes: LLPR (without CP), regular CP, class-based CP, and our proposed flexible uncertainty calibration. To quantify the alignment between predicted uncertainty and empirical error, we report the Spearman rank correlation coefficient $\rho$ (see Appendix~\ref{sec:apd:Spearman_rank_coefficient} for definition), which is widely used to assess monotonic relationships in uncertainty quantification~\cite{benesty2009pearson}.

The results indicate that regular CP offers limited calibration performance, reflected by the identical Spearman coefficients ($\rho = 0.386$) for the LLPR baseline and CP. The associated error–uncertainty distributions only show a difference in scale. The class-based CP yields a modest improvement by incorporating four empirically defined atomic classes, leading to a slight quantitative improvement to $\rho = 0.420$. In contrast, our proposed flexible calibration framework qualitatively improves the monotonic alignment between predicted uncertainties and actual errors, and achieves a markedly higher Spearman coefficient of $\rho = 0.589$.

These results reveal a key limitation of regular CP in atomistic systems~\cite{best2024uncertainty}: while global rescaling adjusts uncertainty magnitudes, it fails to improve their alignment with actual errors. In contrast, our flexible, data-driven calibration better captures local error patterns. Results on MPtraj and MATPES is supplemented to support the broader applicability of our method (see Appendix~\ref{sec:apd:mptraj_matpes}) for the results.

\paragraph{Efficiency.}
An essential requirement for a practical uncertainty calibration method is that it should not introduce significant computational overhead relative to baseline model evaluation (i.e., model prediction and LLPR uncertainty estimation in this work). To assess efficiency, we benchmarked the evaluation time of the calibrated quantile $\hat{q}_\theta$ when applied alongside the MACE–LLPR model on the LiCl dataset. The results, summarized in Table~\ref{tab:evaluation_time}, show that $\hat{q}_\theta$ contributes only 0.02\% of the total evaluation time. In other words, the additional cost of calibration is negligible compared with baseline LLPR inference. This demonstrates that the proposed method delivers reliable uncertainty estimates at virtually no extra computational expense, ensuring its suitability for large-scale MD simulations and active learning workflows.

\begin{table}[ht!]
\centering
\setlength{\tabcolsep}{12pt} 
\caption{Evaluation time (in seconds) of different calibration methods on the LiCl benchmark. The timings are obtained by evaluating correpsonding architecture on the whole training set. Both class-based CP and the proposed flexible UC introduce negligible overhead compared with baseline MACE–LLPR evaluation or a standard MACE evaluation. This also demonstrates the need for a more efficient baseline uncertainty estimate.}
\label{tab:evaluation_time}
\begin{tabular}{l c}
\toprule
\textbf{Method} & \textbf{Evaluation Time (s)} \\
\midrule
MACE              & 0.0182 ($\pm$ 0.0339) \\
MACE-LLPR              & 0.9766 ($\pm$ 0.0516) \\
Regular/Class CP    & 0.0001 ($\pm$ 0.0001) \\
Flexible UC  & 0.0002 ($\pm$ 0.0001) \\
\bottomrule
\end{tabular}
\end{table}

\subsection{Generalization to Unseen Atomic Environments: Catalytic Reaction Pathways}
\label{sec:sub:gen}

To assess the generalization capability of the flexible uncertainty calibration method and to demonstrate that the choice of calibration set $\mathcal{D}_{\rm cal}$ does not need to closely match the target distribution, we evaluate its performance on a catalytic reaction pathway dataset introduced in~\cite{schaaf2023accurate}. This dataset, which includes both non-doped and Pt-doped catalytic surfaces, is chosen to test the method's robustness against domain shifts in local atomic environments. Specifically, the model is calibrated on non-doped surfaces and used to estimate uncertainties on Pt-doped configurations. We hypothesize that our method can generalize across such shifts due to the shared chemical descriptors (e.g., atomic species)~\cite{lopanitsyna2023modeling} embedded in both the model representation and the quantile predictor $\hat{q}_{\theta}(\mathbf{X}_i)$.

\begin{figure*}
    \centering
    \includegraphics[width=0.95\linewidth]{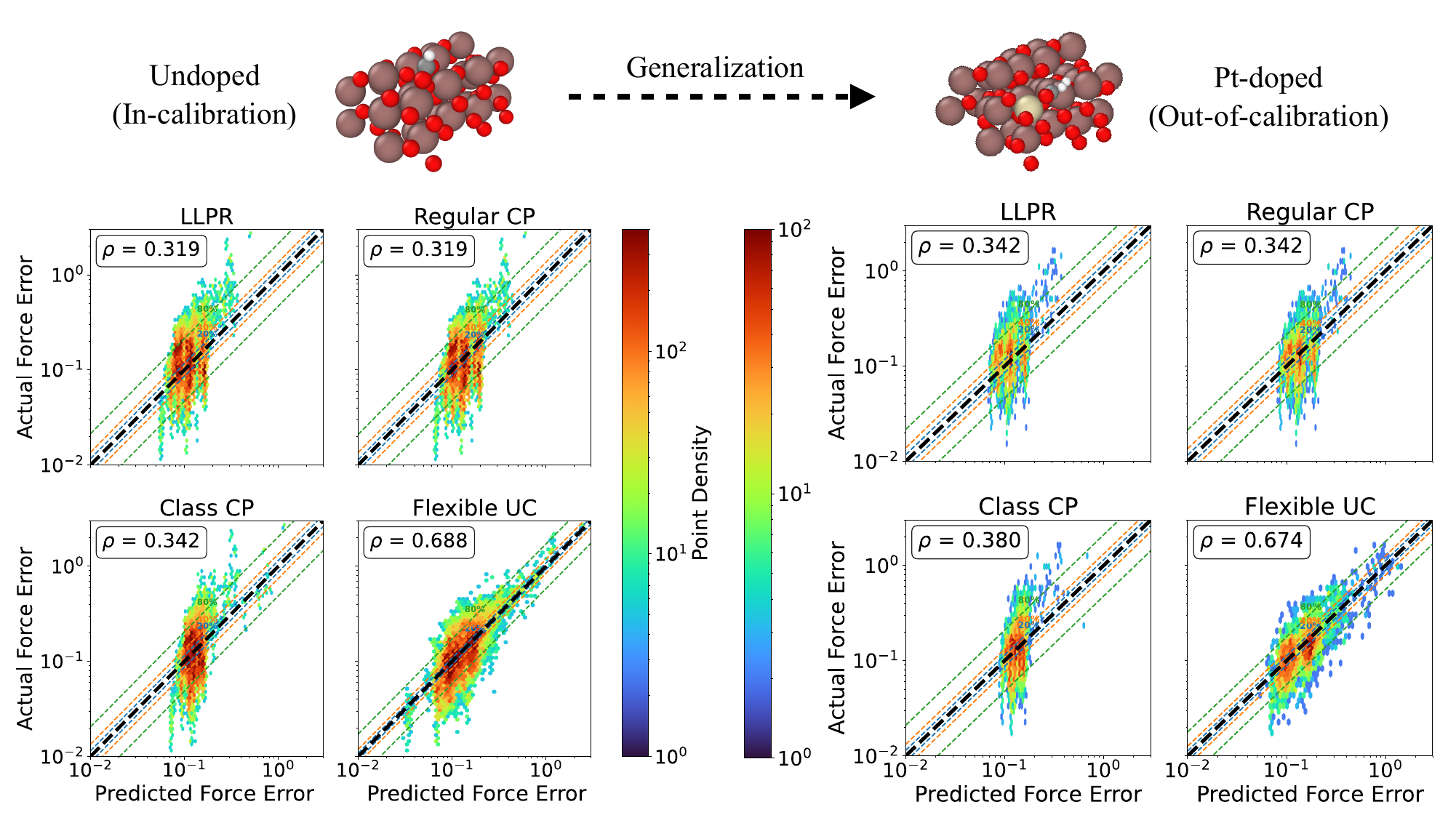}
    \caption{Uncertainty estimates from LLPR, regular CP, class CP, and flexible UC on catalytic surface data. {\bf Left:} undoped configurations (in-calibration). {\bf Right:} Pt-doped configurations (out-of-calibration). Regular and class-based CP show similar behavior in both cases, whereas flexible UC substantially improves generalization to unseen Pt-doped environments.}
    \label{fig:catalyst_composite_1}
\end{figure*}

We begin by evaluating the calibration performance on non-doped catalytic surfaces with only $5 \%$ of the available data. We use five classes for class CP. As shown in Figure~\ref{fig:catalyst_composite_1}, both regular CP and class-based CP yield limited improvements: regular CP leaves the error–uncertainty correlation largely unchanged, while class-based CP leads to only a modest increase in the Spearman rank coefficient. In contrast, the flexible uncertainty calibration method achieves much better alignment between predicted uncertainties and actual errors, reflected by a Pearson correlation coefficient of $0.688$. These results are consistent with our observations on the LiCl dataset (cf. Section~\ref{sec:sub:various_datasets}).

To assess generalization to previously unseen atomic environments, we evaluate the calibrated uncertainties on Pt-doped surfaces. Notably, the model was never exposed to these Pt-containing configurations during calibration. Despite this, the flexible method maintains strong performance, with the Pearson coefficient increasing from $0.347$ (before calibration) to $0.677$ (after calibration), and a visibly tighter error–uncertainty scatter. These results suggest that our approach is not heavily dependent on the specific choice of calibration set $\mathcal{D}_{\rm cal}$, and can generalize robustly across varying chemical compositions.

\begin{figure*}
    \includegraphics[width=0.95\linewidth]{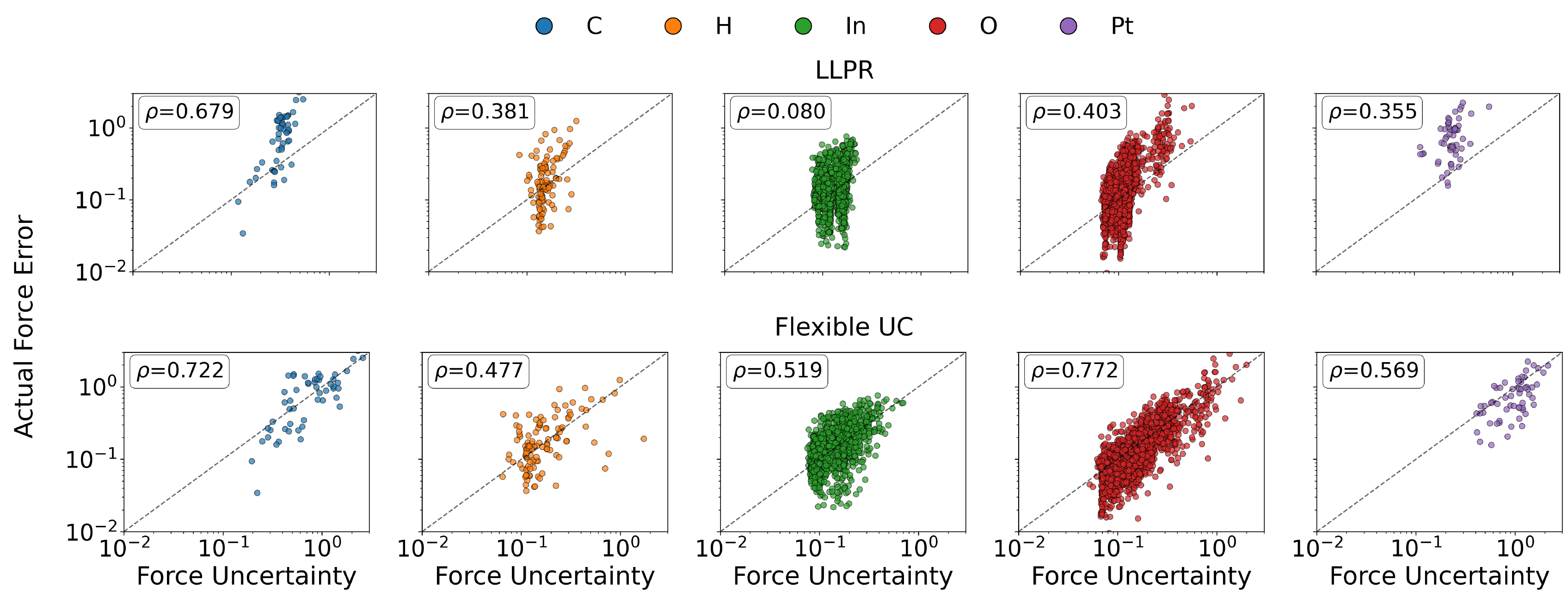}
    \caption{Species-resolved calibration of force uncertainties on Pt-doped surfaces.
Predicted force uncertainties versus actual errors are shown for different elements. {\bf Top:} regular conformal prediction. {\bf Bottom:} flexible calibration. Although Pt atoms are excluded from the calibration set, the proposed method yields improved correlation ($\rho$) across all species.}
\label{fig:catalyst_composite}
\end{figure*}

To further examine the generalization performance, Figure~\ref{fig:catalyst_composite} presents species-resolved scatter plots of predicted force errors versus calibrated uncertainties. The top row corresponds to LLPR (before calibration), and the bottom row to the flexible uncertainty calibration method. Notably, Pt atoms are absent from the calibration set. As shown, all species exhibit improved alignment between error and uncertainty after calibration. In particular, the flexible method generalizes effectively to unseen atomic environments, yielding improved correlation for Pt atoms. This demonstrates the robustness of the method to extrapolation beyond the calibration distribution.

\begin{figure}
    \centering
    \includegraphics[width=0.9\linewidth]{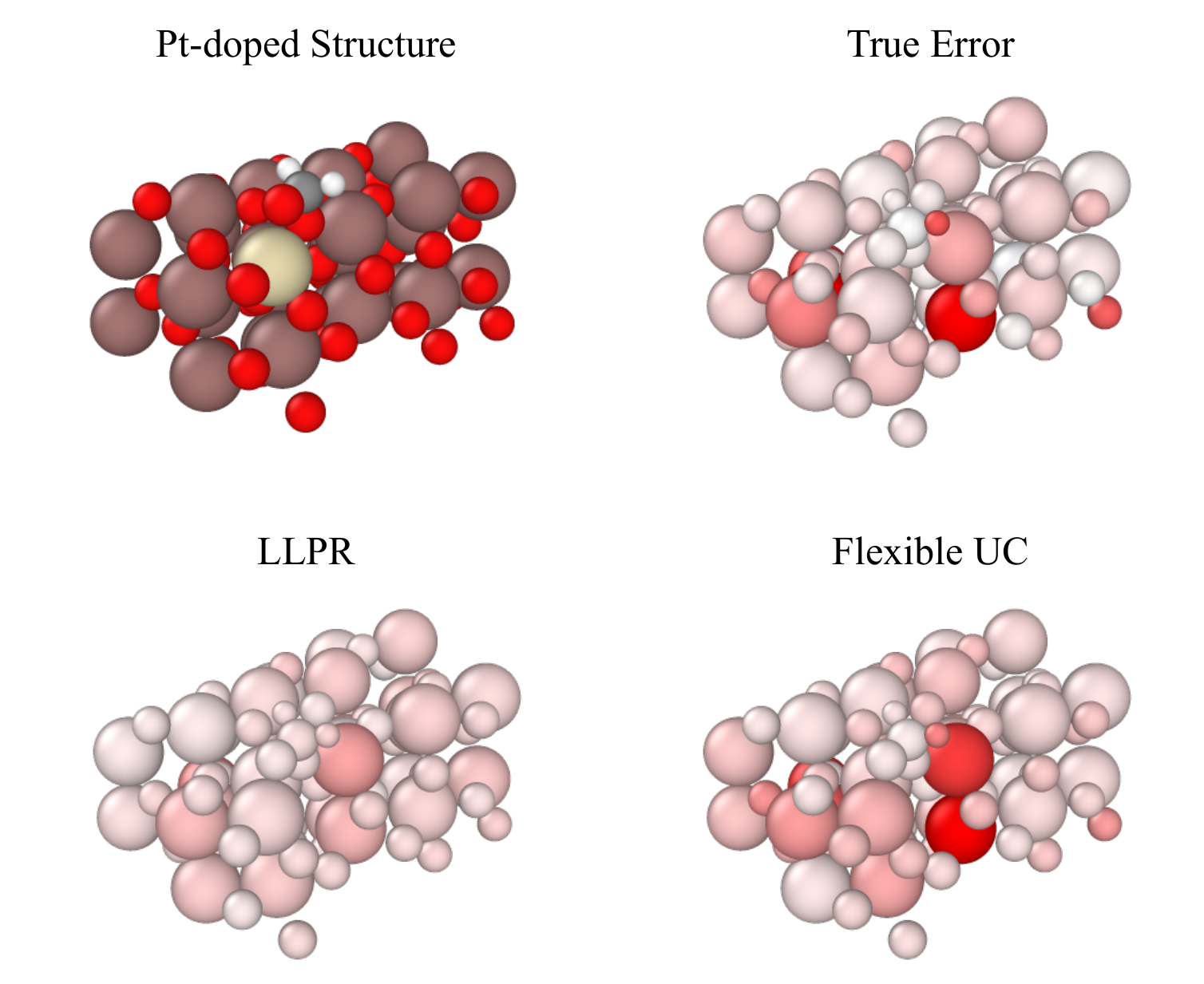}
    \caption{Uncertainty estimates from LLPR and flexible UC, together with true force errors, for a representative Pt-doped surface configuration (atomic structure shown on the left). Errors and uncertainties are assigned on a per-atom basis, with darker colors indicating atoms exhibiting larger values.}
\label{fig:catalyst_generalization_atomic}
\end{figure}

To further highlight the advantages of flexible calibration, Figure~\ref{fig:catalyst_generalization_atomic} compares per-atom predicted uncertainties with reference force errors for a representative Pt-doped configuration. Flexible UC yields sharper and more accurate localization of high-error atoms, closely matching the true error distribution. In contrast, LLPR uncertainties are diffuse and show weak correlation with the actual errors. This improvement is particularly important for large-scale simulations, where direct DFT validation is infeasible for detecting erroneous sites. The poor performance of LLPR, in some cases even worse than random selection, suggests a systematic bias in its uncertainty estimates toward misleading configurations. Using LLPR without calibration may therefore result in catastrophic or ineffective active learning. By providing reliable alignment between errors and uncertainties, flexible UC establishes a practical foundation for robust fine-tuning and enables more effective active learning strategies, as illustrated in the next section.

\subsection{Uncertainty-Guided Fine-Tuning in Molecular Dynamics}
\label{sec:sub:md}
To demonstrate the practical impact of uncertainty calibration, we investigate its role in active fine-tuning within molecular dynamics (MD) simulations. MD trajectories are a natural testbed for active learning, since they continuously generate new atomic configurations whose predictive errors are a priori unknown~\cite{van2023hyperactive}. A central challenge in this setting is to reliably detect high-error configurations, which are the most informative candidates for model refinement. Without calibrated uncertainty, active learning risks either missing critical erroneous configurations or selecting misleading ones (as demonstrated in Table~\ref{tab:benchmark_accuracy}). While this limitation is well recognized in the broader active learning literature~\cite{van2023hyperactive, podryabinkin2023mlip, zhang2019active}, it has not, to our knowledge, been systematically examined in the context of fine-tuning MLIP foundation models.

To evaluate this capability, we initiate an NVT trajectory from a randomly selected configuration in the catalyst dataset described in the previous section. We sample uniformly from the first 75\% of the trajectory to form a calibration set, train a flexible uncertainty calibration model on this subset, and then assess performance on the remaining 25\% of the trajectory. This setup mimics a realistic active learning workflow, where calibration must be established on a limited portion of the data while robust uncertainty estimates are required for unseen configurations throughout the simulation.

We performed single-point DFT evaluations every 4~ps along the testing trajectory to ensure the configurations are sufficiently decorrelated. Consecutive evaluations were grouped into segments, which we define as windows of a given size. Within each window, the configuration with the largest absolute force error relative to the DFT reference was identified as the most erroneous. The ability of each method to correctly detect these configurations is reported in Table~\ref{tab:benchmark_accuracy}. The results show that flexible UC consistently outperforms LLPR across all window sizes, reliably identifying high-error configurations that LLPR often fails to capture. Notably, even for larger windows where the increased number of configurations introduces additional uncertainty, flexible UC still achieves up to 50\% accuracy, compared to 0\% for LLPR. 

These results highlight the practical significance of flexible uncertainty calibration: in an uncertainty guided fine-tuning, or active learning loops from iterative exploration with simulations, the ability to pinpoint erroneous configurations is crucial for targeted data acquisition and efficient model refinement.

\begin{table}[ht!]
\centering
\setlength{\tabcolsep}{10pt} 
\caption{Identification accuracy of the most erroneous configurations 
before (LLPR) and after calibration (Flexible UC) across different window sizes.}
\label{tab:benchmark_accuracy}
\begin{tabular}{l c c}
\toprule
\textbf{Window Size} & \textbf{LLPR} & \textbf{Flexible UC} \\
\midrule
12 ps & 30.0 \% & 70.0 \% \\
16 ps & 12.5 \% & 62.5 \% \\
20 ps & 0.0 \%  & 50.0 \% \\
\bottomrule
\end{tabular}
\end{table}

\begin{figure*}
    \centering
    \includegraphics[width=0.75\linewidth]{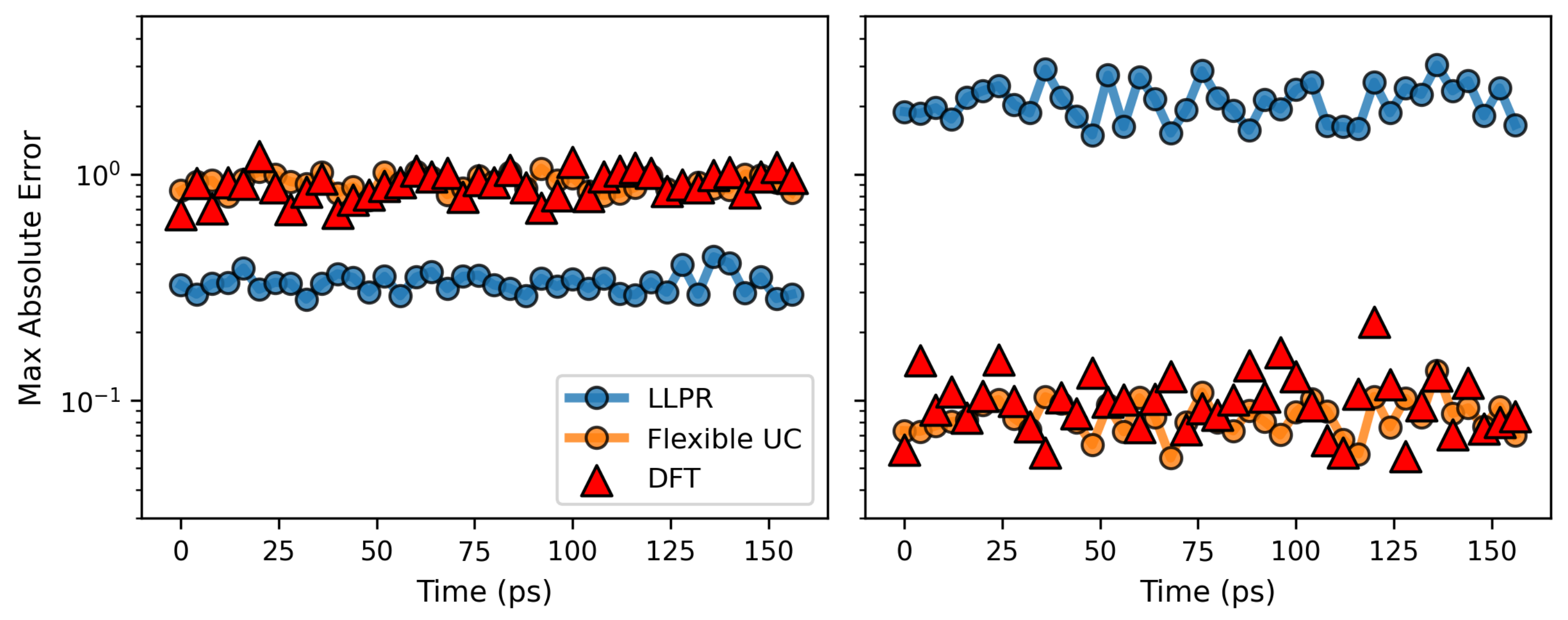}
\caption{Temporal evolution of predictive uncertainty and actual errors during an NVT simulation. {\bf Left:} Representative frames before fine-tuning, with and without calibration. {\bf Right:} Representative frames after fine-tuning, with and without calibration. Calibration improves error alignment, supporting active fine-tuning.}
    \label{fig:uncertainty_nvt}
\end{figure*}

To further assess the impact of calibration on predictive uncertainties, we carried out additional NVT simulations initialized from the same structure. Configurations were sampled along the trajectory, followed by a single round of fine-tuning and subsequent uncertainty calibration using the proposed flexible UC framework. Figure~\ref{fig:uncertainty_nvt} summarizes representative frames before and after fine-tuning. Three key observations emerge: (i) prediction errors decrease substantially after fine-tuning; (ii) the original LLPR uncertainty, which previously underestimated the true error, becomes more conservative after fine-tuning; and (iii) once calibrated, the uncertainty aligns much more closely with the actual error, again confirming the effectiveness of the flexible UC method. 

To the best of our knowledge these findings demonstrate, for the first time, that large-scale atomistic foundation models can be fine-tuned within an active learning workflow while maintaining reliable error detection. Although we illustrate only a single fine-tuning step, flexible UC establishes a solid basis for uncertainty-guided active learning. A complete fine-tuning scheme will be reported in future work, as the present study focuses on introducing and validating the flexible UC framework.

\subsection{Uncertainty Transfer Across Exchange–Correlation Functionals}
\label{sec:sub:transfer}
A further advantage of the proposed framework lies in its ability to transfer calibrated uncertainty estimates across distinct data distributions, in particular across exchange–correlation (XC) functionals. This capability is essential when constructing large-scale training sets that combine data generated from different functionals. To illustrate this point, we consider a model pre-trained on PBE and examine its transfer to alternative functionals such as PBEsol \cite{terentjev2018dispersion} and $\omega$B97M-V/def2-TZVPD \cite{levine2025open}. For a given atomic index $i$, the relevant discrepancy can be expressed as
\begin{equation}
\label{eqn:transfer_xc_forces_error}
\big|\widetilde{{\bf F}}_i^{\mathrm{PBE}} - {\bf F}_i^{\mathrm{XC}}\big|
= \big|\widetilde{{\bf F}}_i^{\mathrm{PBE}} - {\bf F}_i^{\mathrm{PBE}} - \varepsilon_i^{\mathrm{corr}}\big|,
\end{equation}
where $\widetilde{{\bf F}}_i^{\mathrm{PBE}}$ denotes the force predicted by the PBE-pretrained MLIP, ${\bf F}_i^{\mathrm{PBE}}$ the corresponding DFT reference, and $\varepsilon_i^{\mathrm{corr}}$ the functional correction that accounts for systematic differences between XC approximations. Configurations of interest therefore correspond either to regions where the PBE model error $\big|\widetilde{{\bf F}}_i^{\mathrm{PBE}}-{\bf F}_i^{\mathrm{PBE}}\big|$ is large, or where the cross-functional shift $|\varepsilon_i^{\mathrm{corr}}|$ is significant. Under the mild assumption that cancellation between these two terms is negligible, maximizing \eqref{eqn:transfer_xc_forces_error} is equivalent to both or either one of PBE model error and cross-functional shift. Which our flexible calibration framework provides a principled and cost-efficient means of estimating~\eqref{eqn:transfer_xc_forces_error} without exhaustive labeling. 

This transfer setting is particularly important because uncertainty calibration is not only required to correct model confidence within a single functional, but also to maintain consistency across heterogeneous datasets that underpin emerging foundation models. By exploiting calibrated error estimates from the source functional and learning only a lightweight correction term from a small number of target-functional calculations, one can significantly reduce the labeling cost while retaining reliable uncertainty estimates. In practice, this enables systematic fine-tuning across functionals and supports the integration of diverse data sources into a unified MLIP framework, thereby extending the applicability of atomistic foundation models to broader chemical and physical domains. To the best of our knowledge, this constitutes the first systematic demonstration of uncertainty calibration for cross-functional transfer.

To demonstrate this transferability, we consider the MACE-MP-0b3 model with LLPR-based uncertainties. By construction, LLPR captures prediction errors relative to the PBE functional (consistent with the MPtraj training data), but it does not directly reflect discrepancies introduced by other exchange–correlation functionals. In this context, our framework enables PBE-based uncertainty estimates to be transferred to the error measure in~\eqref{eqn:transfer_xc_forces_error}, thereby extending their applicability to different functionals. We evaluate this capability on two challenging benchmarks: the HEA25 dataset~\cite{owen2024complexity}, computed with the PBEsol functional~\cite{terentjev2018dispersion}, and a subset of the Open Molecule 2025 dataset~\cite{levine2025open}, generated at the $\omega$B97M-V/def2-TZVPD level of theory, which departs substantially from PBE.

\paragraph{HEA25.}

\begin{figure}
    \centering
    \includegraphics[width=1.0\linewidth]{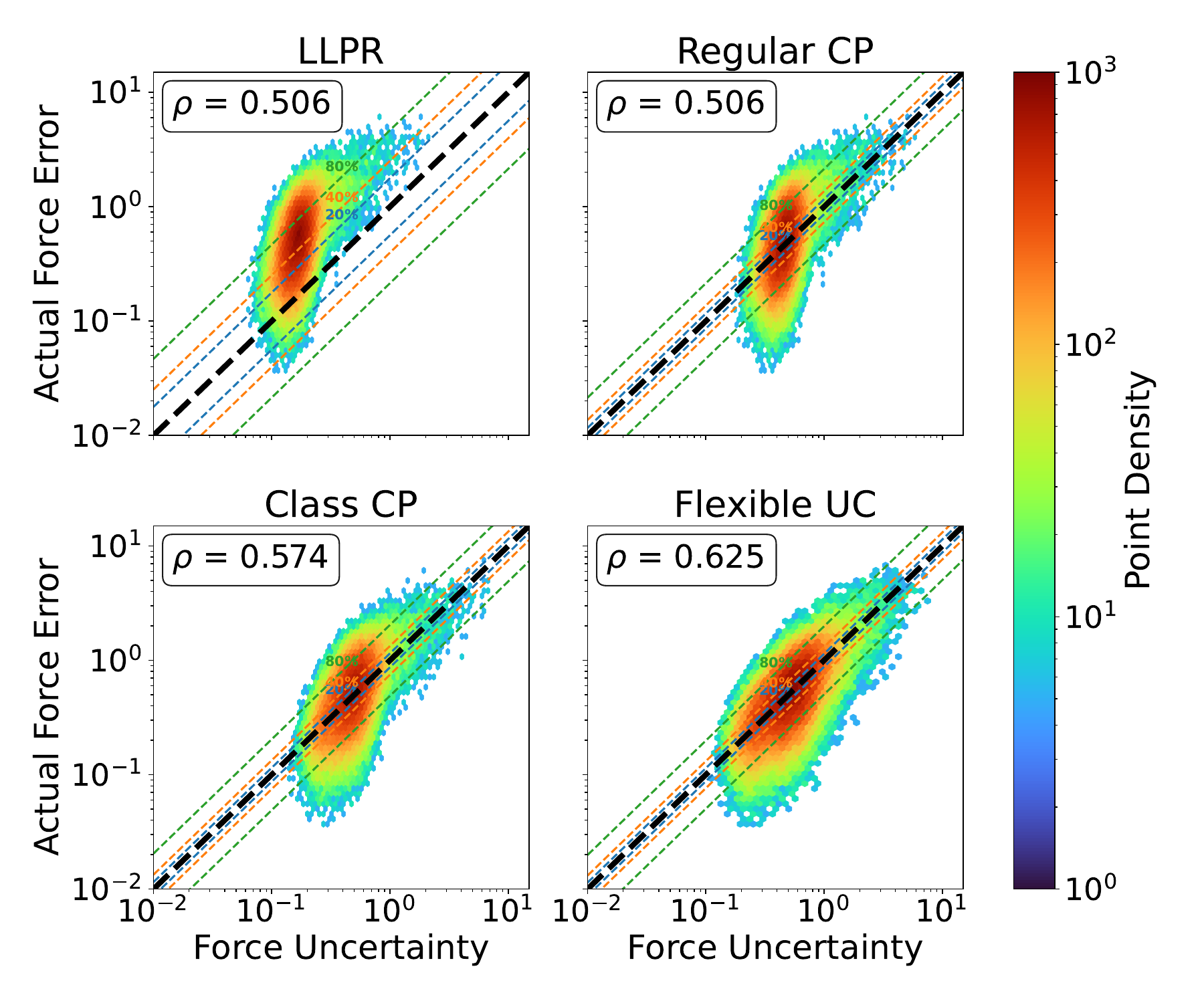}
    \caption{Calibration performance on the HEA25 dataset computed with the PBEsol functional, comparing uncertainty estimates from LLPR, regular CP, class-based CP, and flexible UC. Flexible UC provides the most significant improvement in correlating predicted uncertainties with actual errors.}
    \label{fig:hea25_benchmark}
\end{figure}

Figure~\ref{fig:hea25_benchmark} reports calibration results on the HEA25 dataset, which targets chemically complex, high-entropy alloys and was computed with the PBEsol functional. Since PBEsol is relatively close to the PBE functional used for pre-training, both LLPR and regular CP already provide moderately reasonable uncertainty estimates. In particular, LLPR is able to capture high-error configurations (e.g., errors on the order of 1 eV/\AA) with fair accuracy. Class-based CP provides a slight improvement by leveraging empirical partitioning of the input space, but the correlation gain remains limited. In contrast, the flexible calibration approach yields a marked improvement in aligning predicted uncertainties with actual errors, as reflected by the higher Spearman rank correlation coefficient ($\rho = 0.625$). This demonstrates that even for systems relatively close to the training distribution, flexible UC substantially enhances the reliability and robustness of uncertainty estimates.

\paragraph{Open Molecule 2025.}

\begin{figure}[ht]
    \centering
    \includegraphics[width=1.0\linewidth]{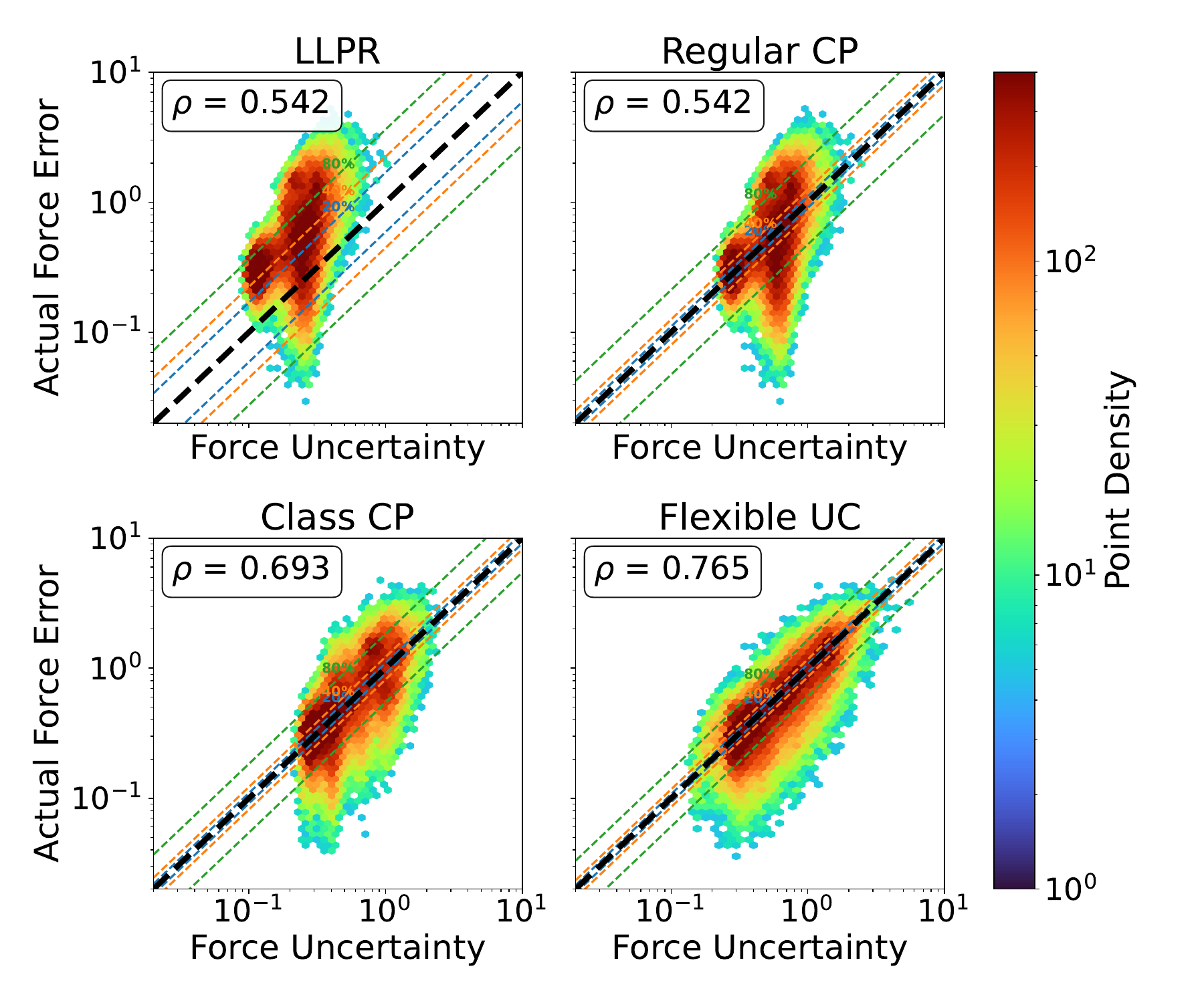}
    \caption{Predicted force uncertainties versus actual force errors on the Open Molecule 2025 dataset. Flexible UC achieves the strongest correlation between predicted uncertainties and observed errors under substantial distribution shifts.}
    \label{fig:omol_benchmark}
\end{figure}

\begin{figure}[ht]
    \centering
    \includegraphics[width=0.925\linewidth]{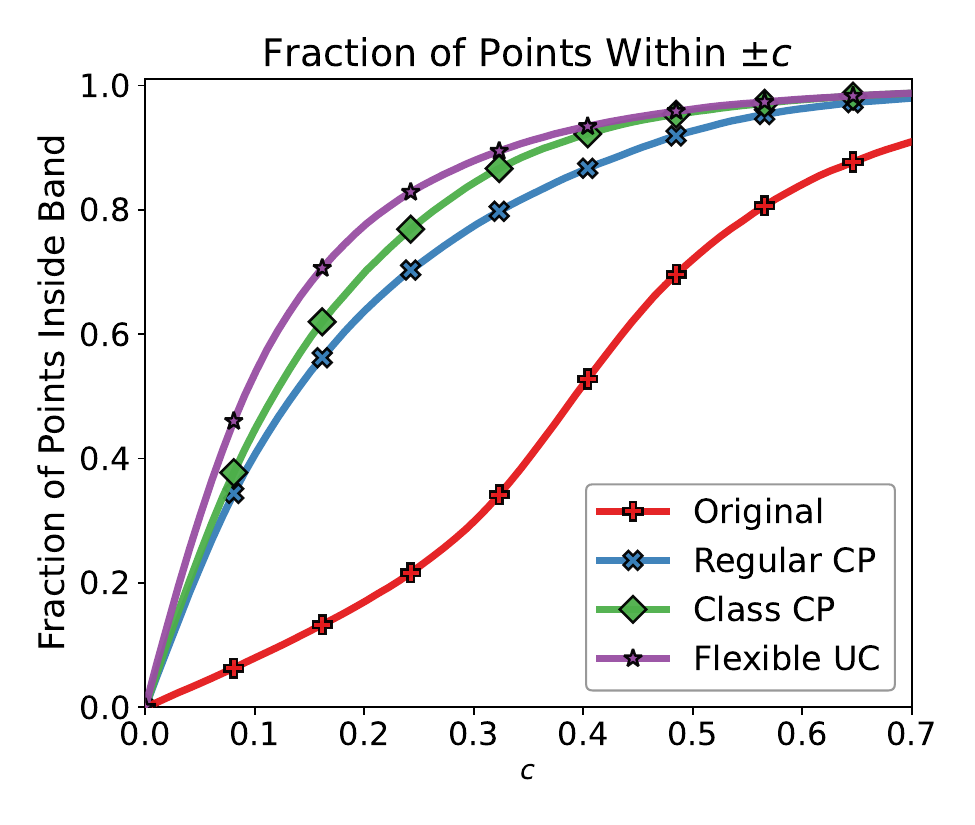}
    \caption{Calibration coverage on the Open Molecule 2025 dataset. The curves show the 
    of atomic forces lying within uncertainty intervals of varying width. Flexible UC provides the most reliable calibration compared to baseline methods.}
    \label{fig:omol_coverage}
\end{figure}

To further evaluate transferability under substantial distribution shifts, we benchmark the proposed method on a subset of the Open Molecule 2025 dataset~\cite{levine2025open}, which differs significantly from MPtraj in both chemical composition and the XC functional. From the validation split, we randomly select 2,000 charge-neutral molecular configurations, corresponding to roughly 90,000 atomic sites in total. The results are summarized in Figure~\ref{fig:omol_benchmark}. In this case, LLPR-based uncertainties fail to capture the correlation between predicted and true errors, as most predictions collapse into a narrow uncertainty range. Regular CP provides no improvement beyond LLPR, while class-based CP achieves moderate gains, likely reflecting the greater separability of local atomic environments in molecular systems compared to extended solids. The proposed flexible UC framework again delivers the best correlation on this challenging benchmark. Furthermore, coverage analysis (Figure~\ref{fig:omol_coverage}) confirms that flexible UC consistently provides better-calibrated intervals, with a larger fraction of points falling within prescribed confidence bands. 

These results demonstrate the robustness of flexible UC in handling strong distributional shifts, highlighting its potential as a general calibration strategy for atomistic foundation models.

\begin{figure}[htbp]
    \centering
    \includegraphics[width=0.9\linewidth]{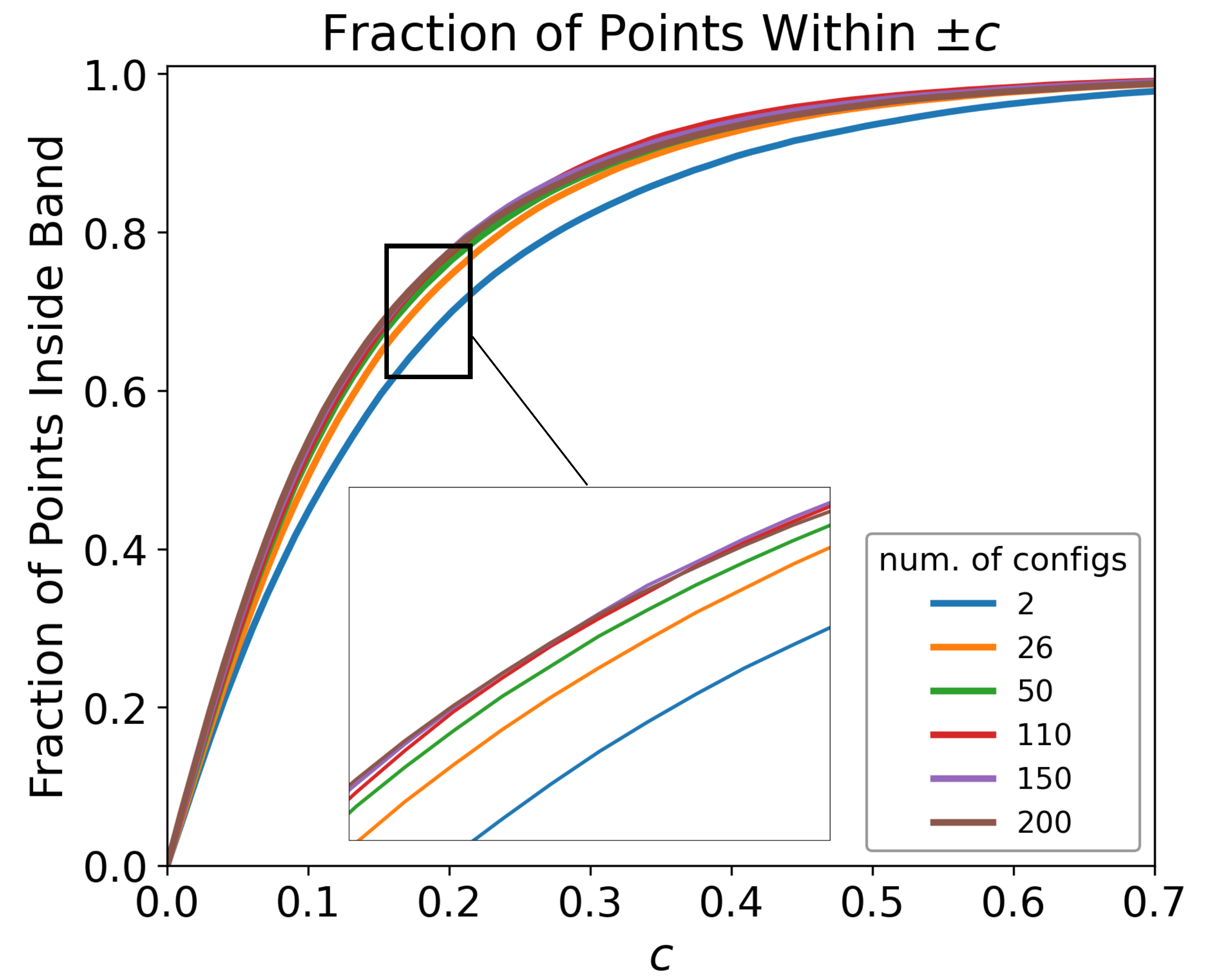}
    \caption{Convergence of the coverage curve on the Open Molecule 2025 dataset as a function of calibration set size. Reliable calibration is achieved with only a small number of configurations.}
    \label{fig:convergence_nmol}
\end{figure}

\begin{figure}[htbp]
    \centering
    \includegraphics[width=0.9\linewidth]{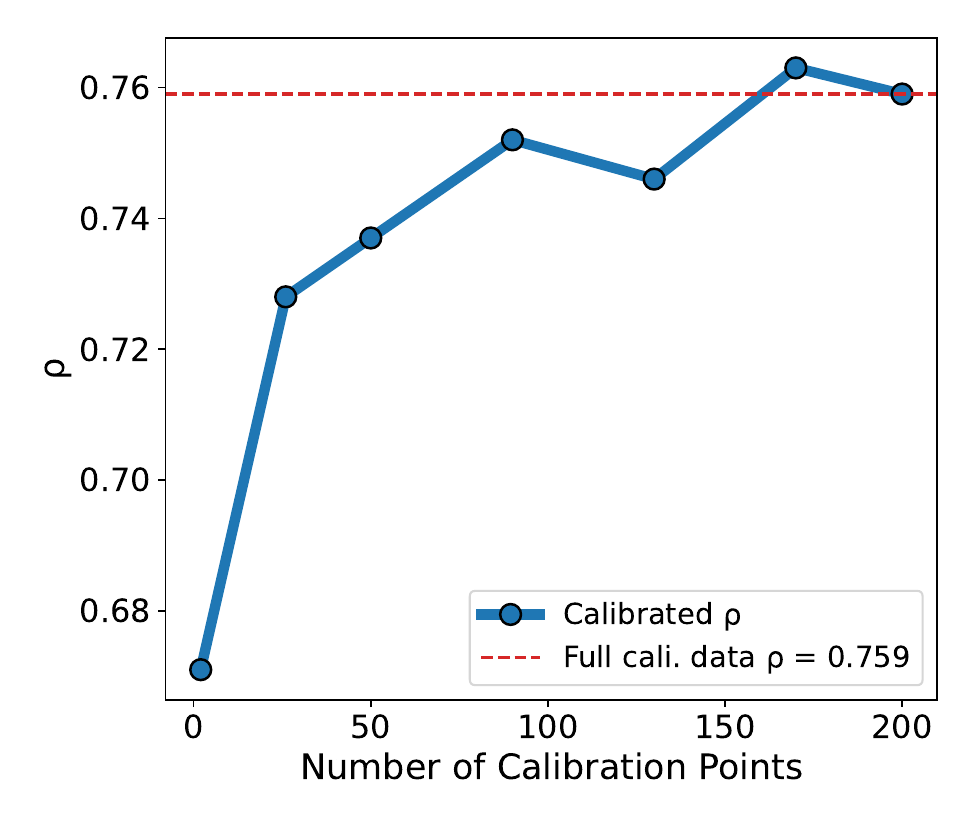}
    \caption{Convergence of the Pearson correlation coefficient on the Open Molecule 2025 dataset. Performance stabilizes after approximately 50 calibration configurations.}
    \label{fig:convergence_omol}
\end{figure}

\paragraph{Calibration efficiency.} 
In our final test, we assess the efficiency of the flexible UC scheme by varying the number of calibration configurations. 
As shown in Figures~\ref{fig:convergence_nmol} and \ref{fig:convergence_omol}, both the coverage curve and the Pearson correlation coefficient improve rapidly with additional calibration data and plateau after only about 25 calibration configurations. This behavior reflects the fact that the effective calibration size scales with the total number of atoms rather than the number of configurations alone, thereby amplifying the information contained in each configuration. These results demonstrate that the proposed method is highly data efficient, making it practical even in scenarios where reference calculations are expensive and only limited calibration data are available.

\section{Summary and Outlook}
\label{sec:summary}

We have presented a flexible uncertainty calibration framework for MLIPs, which reformulates conformal prediction into a learnable, environment-dependent quantile model. By embedding the quantile function into the predictive pipeline and optimizing it end-to-end, the method produces adaptive and well-calibrated uncertainty estimates that closely align with true errors. Applied to the MACE-MP-0 foundation model, the framework yields significant improvements in calibration error while being extremely data efficient during training and incurring negligible computational overhead during evaluation. Extensive benchmarks across ionic crystals, catalytic surfaces, and molecular datasets demonstrate the method’s robustness, and practical tests show its value in active fine-tuning, molecular dynamics, and cross-xc-functional transfer.

Looking ahead, several promising directions arise. First, extending the framework to jointly calibrate multiple physical quantities (energy, forces, stresses) could provide unified statistical guarantees across coupled observables. Second, leveraging site-resolved calibrated uncertainties for large-scale or long-time molecular dynamics simulations would enable robust configuration screening and reliable uncertainty-guided active learning. Third, applying the method to data-scarce regimes such as phase transitions, chemical reactions, or battery interfaces could establish reliable uncertainty quantification in scenarios where extrapolation risks are most severe. Taken together, this work establishes flexible uncertainty calibration as a practical and principled approach to improving the reliability of MLIPs, and paves the way toward uncertainty-aware foundation models for atomistic simulation. 

\begin{acknowledgments}
We thank Dr. Sanggyu Chong (École Polytechnique Fédérale de Lausanne) for sharing the implementation details of LLPR, which supported this work. 
\end{acknowledgments}


\appendix

\section{Conformal Prediction Theory}
\label{sec:apd:cp_theory}

This appendix presents additional theoretical foundations and technical details related to the conformal prediction framework, which are not fully discussed in the main text. 

\subsection{Regular Conformal Prediction}
\label{sec:apd:standard_cp}

We define a coverage function $\mathcal{C} : \mathcal{X} \rightrightarrows \mathcal{Y}$ as a \emph{set-valued mapping}, meaning that for each input $\mathbf{X} \in \mathcal{X}$, the output $\mathcal{C}(\mathbf{X}) \subseteq \mathcal{Y}$ is a measurable subset of $\mathcal{Y}$ representing the predicted confidence region. Given a user-specified confidence level $\alpha \in (0,1)$, the function $\mathcal{C}$ is constructed such that, for any test point $(\mathbf{X}_{\rm new}, \mathbf{Y}_{\rm new}) \sim P$, the following marginal coverage condition holds:
\begin{equation}\label{eqn:marginal_coverage}
    \mathbb{P}\big( \mathbf{Y}_{\rm new} \in \mathcal{C}(\mathbf{X}_{\rm new}) \big) \geq 1 - \alpha.
\end{equation}

Under finite-sample exchangeability, this condition can be more precisely bounded via the conformal calibration framework~\cite{romano2019conformalized} as
\begin{equation*}
1 - \alpha \leq \mathbb{P}\big( \mathbf{Y}_{\rm new} \in \mathcal{C}(\mathbf{X}_{\rm new}) \big)
\leq 1 - \alpha + \frac{1}{N_{\rm cal}+1},
\end{equation*}
ensuring that the coverage holds marginally over the randomness in $\mathbf{X}_{\rm new}$ and $\mathcal{D}_{\rm cal}$. The upper bound $1 - \alpha + 1/(N_{\rm cal}+1)$ arises from the discrete nature of empirical quantile estimation under finite calibration size, and reflects the worst-case marginal coverage achievable under the exchangeability assumption.

Using Eq.~\eqref{eqn:general_score_func}, we compute calibration scores $\{ s_i := s(\mathbf{X}_i, \mathbf{Y}_i) \}_{i=1}^{N_{\text{cal}}}$ for all entries in $\mathcal{D}_{\text{cal}}$. To enforce marginal coverage at confidence level $1 - \alpha$, we compute the $(1 - \alpha)$ empirical quantile $\hat{q} \in \mathbb{R}_{\ge 0}$ defined by
\begin{equation*}
    \hat{q} := \text{quantile}\left(\{s_i\}_{i=1}^{N_{\text{cal}}}, \frac{\lceil (N_{\text{cal}} + 1)(1 - \alpha) \rceil}{N_{\text{cal}}} \right).
\end{equation*}
This quantile formulation ensures that the conformal region satisfies the marginal coverage guarantee stated in Eq.~\eqref{eqn:marginal_coverage} under the assumption of exchangeability.

The conformal prediction region at a new input $\mathbf{X}_{\text{new}}$ is then given by:
\begin{equation}
\label{eqn:conformal_region}
    \mathcal{C}(\mathbf{X}_{\text{new}}) := \left\{ \mathbf{Y} \in \mathcal{Y} \;\middle|\; s(\mathbf{X}_{\text{new}}, \mathbf{Y}) \leq \hat{q} \right\}.
\end{equation}
In other words, $\mathcal{C}(\mathbf{X}_{\text{new}})$ contains all plausible outputs $\mathbf{Y}$ that are within the $(1 - \alpha)$ tolerance range defined by the quantile of calibration residuals.

\subsection{Class-based Conformal Prediction}
\label{sec:apd:class_cp}

To ensure valid conditional coverage within a class-based conformal prediction framework, we must rigorously define the associated coverage function, as motivated in Section~\ref{sec:methods} (cf.~Eq.~\eqref{eqn:conformal_region}).

Following the analytical framework in~\cite{gibbs2025conformal}, and starting from Eq.~\eqref{eqn:marginal_coverage_main}, we recall that the new conformity score $s_{\rm new}(\bfY) := s(\bfX_{\rm new}, \bfY)$ is not compared to the $(1-\alpha)$ empirical quantile of the calibration scores $\{s_i\}_{i=1}^{N_{\rm cal}}$, but instead to the adjusted quantile at level 
\[
\big( \left\lceil (N_{\rm cal} + 1) \cdot (1 - \alpha) \right\rceil / N_{\rm cal} \big).
\]
This adjusted threshold arises from augmenting the calibration dataset with the unknown new score $s_{\rm new}$.

To incorporate this perspective into quantile regression for conformity scores, we propose fitting a model $\hat{q}_{\rm cb}$ on an augmented dataset that includes a guess $s$ for the unobserved $s_{\rm new}$. Formally, we define $\hat{q}_s$ as the minimizer of the empirical quantile loss:
\begin{align}
\hat{q}_{\rm cb}^{s} := \mathop{\arg\min}_{q_{\rm cb} \in \mathscr{F}_{\rm cb}} \, &\frac{1}{N_{\rm cal} + 1} \sum_{i=1}^{N_{\rm cal}} \ell_{\alpha}\big(q_{\rm cb}(\bfX_i), s_i\big) \nonumber \\
&+ \frac{1}{N_{\rm cal} + 1} \ell_{\alpha}\big(q_{\rm cb}(\bfX_{\rm new}), s\big),
\end{align}
where $\mathscr{F}_{\rm cb}$ denotes the class of quantile regression functions (detailed below), and $\ell_\alpha(\cdot,\cdot)$ is the standard asymmetric pinball loss associated with quantile level $\alpha$.

The prediction set associated with input $\bfX_{\rm new}$ is then defined as
\begin{equation}
\label{eqn:apd:class_cp_coverage}
\mathcal{C}(\bfX_{\rm new}) := \left\{ \bfY \in \mathcal{Y} : s(\bfX_{\rm new}, \bfY) \leq \hat{q}_{\rm cb}^{s_{\rm new}}(\bfX_{\rm new}) \right\}.
\end{equation}

The function class $\mathscr{F}_{\rm cb}$ is chosen to be the space of classwise constant (i.e., step) functions defined over a finite partition $\Xi$ of the covariate space:
\[
\mathscr{F}_{\rm cb} := \left\{ x \mapsto \sum_{\xi \in \Xi} q_\xi \cdot \mathbf{1}\{x \in \xi\} : q_\xi \in \mathbb{R},\, \forall \xi \in \Xi \right\},
\]
where $\mathbf{1}\{\cdot\}$ denotes the indicator function.

Under this construction, it is shown in~\cite[Corollary 1]{gibbs2025conformal} that the prediction set $\mathcal{C}(\bfX_{\rm new})$ satisfies the following conditional coverage guarantee for any class $\xi \in \Xi$:
\[
\mathbb{P}\left( \bfY_{\rm new} \in \mathcal{C}(\bfX_{\rm new}) \,\middle|\, \bfX_{\rm new} \in \xi \right) \geq 1 - \alpha.
\]
Moreover, if the conditional distribution of the conformity score $s \mid \bfX$ is continuous, then the coverage error admits the following upper bound:
\begin{align}
\label{eqn:apd:class_cp_upper_bound}
\mathbb{P}&\left( \bfY_{\rm new} \in \mathcal{C}(\bfX_{\rm new}) \,\middle|\, \bfX_{\rm new} \in \xi \right) \nonumber \\
&\leq 1 - \alpha + \frac{|\Xi|}{(N_{\rm cal} + 1) \cdot \mathbb{P}(\bfX_{\rm new} \in \xi)}.
\end{align}
This bound explicitly characterizes the effect of class granularity and calibration set size on the worst-case conditional coverage deviation.

\subsection{Flexible Uncertainty Calibration}
\label{sec:apd:flexible}

For flexible uncertainty calibration, the goal is to construct valid prediction sets without restricting to a finite-dimensional function class. However, prior work~\cite{foygel2021limits, vovk2012conditional} shows that exact conditional coverage is unattainable in general infinite-dimensional settings. We therefore seek relaxed guarantees under suitable assumptions, analogous to the relaxation introduced earlier.

To this end, note that direct optimization over an unrestricted function class is intractable. To address this issue and ensure well-posedness of the quantile regression problem, we introduce regularization. Specifically, for a chosen regularization functional $\mathcal{R}(\cdot)$, we define:
\begin{align}
\hat{q}^s_{\theta} := \mathop{\arg\min}_{q_{\theta} \in \mathscr{F}} &\left( \frac{1}{N_{\rm cal}+1} \sum_{i=1}^{N_{\rm cal}} \ell_\alpha\big(q_{\theta}(\bfX_i), s_i\big) \right. \nonumber \\
&\left. + \frac{1}{N_{\rm cal}+1} \ell_\alpha\big(q_{\theta}(\bfX_{\rm new}), s\big) + \mathcal{R}(q_{\theta}) \right).
\end{align}
Here, the regularization term $\mathcal{R}(q_\theta)$ promotes desirable properties such as smoothness or sparsity, and serves to control the effective complexity of the function class $\mathscr{F}$.

Having obtained $\hat{q}_\theta^s$, we define the prediction set as:
\[
\mathcal{C}(\bfX_{\rm new}) := \big\{ \bfY \in \mathcal{Y} : s(\bfX_{\rm new}, \bfY) \leq \hat{q}_\theta^{s_{\rm new}}(\bfX_{\rm new}) \big\}.
\]
Under mild regularity conditions on $\mathcal{R}(\cdot)$, a relaxed conditional coverage guarantee can be established, as formalized in~\cite[Theorem 3]{gibbs2025conformal}. In particular, the deviation from the target coverage level $1 - \alpha$ is shown to depend on the smoothness and capacity control induced by the regularizer.

We further note that flexible uncertainty calibration can naturally incorporate weighted loss formulations (cf.~Eq.~\eqref{eqn:weighted_minimize_q}), which may be viewed as inducing an implicit regularization on the quantile estimator. This extension suggests the possibility of relaxed coverage guarantees, although a formal statistical analysis is left for future work.

\subsection{Bayesian Interpretation}
\label{sec:apd:bayes}

The connection between the proposed calibration scheme and the Bayesian framework can be interpreted through the lens of uncertainty quantification in predictive distributions. In particular, we may view the heuristic uncertainty $\sigma(\mathbf{X})$ used in conformal calibration as an input-dependent scaling of the predictive variance under a Bayesian posterior.

Let $\Theta \in \mathbb{R}^p$ denote the neural network weights, and let $\mathcal{D}_{\rm train}$ denote the training dataset. A common approximation in Bayesian neural networks is to assume a Gaussian posterior over parameters via Laplace approximation around the maximum a posteriori (MAP) estimate $\Theta_0$:
\begin{equation}
p(\Theta \mid \mathcal{D}_{\rm train}) \approx \mathcal{N}(\Theta_0, \boldsymbol{\Sigma}),
\end{equation}
where $\boldsymbol{\Sigma}$ is the inverse Hessian of the negative log-posterior evaluated at $\Theta_0$, i.e.,
\[
\boldsymbol{\Sigma}^{-1} \approx \nabla^2_{\Theta} \left[ -\log p(\mathcal{D}_{\rm train}~|~\Theta) - \log p(\Theta) \right]_{\Theta = \Theta_0}.
\]

Under this posterior approximation, the predictive distribution at a new input $\mathbf{X}^*$ is given by:
\begin{equation}
p(\mathbf{Y} \mid \mathbf{X}^*, \mathcal{D}_{\rm train}) \approx \mathcal{N}\left( \hat{f}_{\Theta_0}(\mathbf{X}^*),\; J(\mathbf{X}^*) \boldsymbol{\Sigma} J(\mathbf{X}^*)^\top \right),
\end{equation}
where $J(\mathbf{X}^*) = \nabla_\Theta \hat{f}_\Theta(\mathbf{X}^*) \in \mathbb{R}^{d_y \times p}$ is the Jacobian of the model output with respect to the parameters, evaluated at $\Theta_0$.

In our framework, we do not explicitly compute $\boldsymbol{\Sigma}$ or $J(\mathbf{X}^*)$; instead, we apply conformal calibration using a scalar uncertainty proxy $\sigma(\mathbf{X}^*)$. This can be interpreted as approximating the predictive variance by a scalar-scaled version of the global posterior covariance:
\begin{equation}
p(\mathbf{Y} \mid \mathbf{X}^*, \mathcal{D}_{\rm train}) \approx \mathcal{N}\left( \hat{f}_{\Theta_0}(\mathbf{X}^*),\; \alpha^2(\mathbf{X}^*) \boldsymbol{\Sigma} \right),
\end{equation}
where $\alpha(\mathbf{X}^*)$ is a data-dependent scaling factor learned via conformal calibration. The role of $\alpha(\mathbf{X}^*)$ is to adjust the (possibly misspecified) model-derived uncertainty to achieve correct frequentist coverage.

Conformal prediction can be interpreted as a nonparametric correction to a misspecified Bayesian posterior, where $\alpha(\mathbf{X}^*)$ accounts for local variability not captured by the global covariance $\boldsymbol{\Sigma}$. This guarantees valid uncertainty quantification regardless of the quality of the base estimate $\sigma(\mathbf{X})$, whether obtained from dropout, ensembles, or other approximations.

\section{MACE Architecture}
\label{sec:apd:mace}

MACE~\cite{batatia2022mace} is an E(3)-equivariant message passing neural network (EMPNN) model that uses higher-body-order messages. The expressiveness of the model is improved by using efficient multi-body messages instead of two-body messages. Multi-body messages reduce the number of network layers required to achieve the same expressiveness under two-body messages, resulting in a fast and highly parallelizable model. The MACE architecture follows the general framework of MPNNs and includes three parts: message construction, update, and readout.

\textbf{Message passing:} In message construction, MACE combines equivariant message passing with efficient many-body messages. The edges are embedded using a learnable radial basis $R^{(t)}_{kl_1l_2l_3}$ and a set of spherical harmonics $Y^{m_1}_{l_1}$, and the self-interaction is performed on the features $h^{(t)}_{j,\tilde{k}l_2m_2}$ with learnable weights $W^{(t)}_{\tilde{k}kl_2}$. $\bm{A}^{(t)}_{i,kl_3m_3}$ is the two-body feature obtained by pooling neighbor atoms:
\begin{align}
A^{(t)}_{i,kl_3m_3} = \sum_{l_1m_1,l_2m_2} &C^{l_3m_3}_{l_1m_1,l_2m_2} \sum_{j \in \mathcal{N}(i)} R^{(t)}_{kl_1l_2l_3}(r_{ji}) \cdot \nonumber \\
& \cdot Y^{m_1}_{l_1}(\hat{r}_{ji}) \sum_{\tilde{k}} W^{(t)}_{\tilde{k}kl_2} h^{(t)}_{j,\tilde{k}l_2m_2}, 
\end{align}
where $C^{l_3m_3}_{l_1m_1,l_2m_2}$ are the standard Clebsch--Gordan coefficients~\cite{de2018octet}. The key operation of MACE is to construct a multi-body feature $\bm{B}^{(t)}_{i,\eta_\nu kLM}$ through the tensor product of the two-body features $A^{(t)}_{i,kl_3m_3}$:
\begin{equation}
B^{(t)}_{i,\eta_\nu kLM} = \sum_{lm} C^{LM}_{\eta_\nu,lm} \prod_{\xi=1}^{\nu} \sum_{\tilde{k}} w^{(t)}_{\tilde{k}k l_\xi} A^{(t)}_{i,\tilde{k} l_\xi m_\xi}, 
\end{equation}
where the coupling coefficients $C^{LM}_{\eta_\nu}$ correspond to the generalized Clebsch--Gordan coefficients and $w^{(t)}_{\tilde{k}k l_\xi}$ is the learnable weight for the self-interaction of $A^{(t)}_{i,kl_3m_3}$. Finally, the message $m^{(t)}_i$ can be written as a linear expansion:
\begin{equation}
m^{(t)}_{i,kLM} = \sum_{\nu} \sum_{\eta_\nu} W^{(t)}_{z_i; kL,\eta_\nu} B^{(t)}_{i,\eta_\nu kLM}.
\end{equation}

\textbf{Update:} The update is a linear function of the message and the residual connection~\cite{he2016deep}:
\begin{align}
h^{(t+1)}_{i,kLM} &= U^k_t(\sigma^{(t)}_i, m^{(t)}_i) \nonumber \\
&= \sum_{\tilde{k}} W^{(t)}_{kL,\tilde{k}} m^{(t)}_{i,\tilde{k}LM} + \sum_{\tilde{k}} W^{(t)}_{z_i,\tilde{k}} h^{(t)}_{i,\tilde{k}LM}. 
\end{align}

\textbf{Readout:} The readout is a mapping from the invariant part of the node features to a hierarchical decomposition of site energies:
\begin{equation}
E_i = E^{(0)}_i + E^{(1)}_i + \cdots + E^{(T)}_i, 
\end{equation}
where
\begin{equation}
E^{(t)}_i = \mathcal{R}_t(h^{(t)}_i) = 
\begin{cases}
\sum_{\tilde{k}} W^{(t)}_{\text{readout}, \tilde{k}} h^{(t)}_{i,\tilde{k}00}, & t < T, \\
\text{MLP}^{(t)}_{\text{readout}} \left( \{ h^{(t)}_{i,k00} \}_k \right), & t = T.
\end{cases} \tag{19}
\end{equation}
The readout only depends on the invariant features $h^{(t)}_{i,k00}$, making the site energy contribution $E^{(t)}_i$ invariant.

\section{LLPR Uncertainty Quantification}
\label{sec:apd:appendix}

This section introduces the LLPR uncertainty quantification method~\cite{bigi2024prediction}, which defines prediction rigidities via a constrained optimization formulation. We consider a standard regression task with training data  
\(\mathcal{D} = \{ (\mathbf{x}_i, y_i) \}_{i=1}^{N_{\text{train}}}\), where \(\mathbf{x}_i \in \mathbb{R}^d\) are input features and \(y_i \in \mathbb{R}\) are scalar targets. The model prediction is denoted by \(\tilde{y}(\mathbf{x}_i, \mathbf{w})\), and the empirical loss is given by  
\[
\mathcal{L}(\mathbf{w}) = \sum_{i=1}^{N_{\text{train}}} \ell\big( \tilde{y}(\mathbf{x}_i, \mathbf{w}), y_i \big).
\]

The concept of prediction rigidity is based on evaluating how sensitive a model prediction \(\tilde{y}(\mathbf{x}_i, \mathbf{w})\) is to perturbations away from its optimal value \(\tilde{y}(\mathbf{x}_\star, \mathbf{w}_\circ)\), where \(\mathbf{w}_\circ\) denotes the optimized model parameters~\cite{chong2023robustness}. To quantify this sensitivity for a specific test point \(\mathbf{x}_\star\), a modified loss function \(\mathcal{L}_c\) is introduced by adding a Lagrange multiplier term that enforces a constraint on the predicted value \(\tilde{y}(\mathbf{x}_\star, \mathbf{w})\), forcing it to take an arbitrary value \(\epsilon_\star\):
\begin{equation}
\mathcal{L}_c(\mathbf{w}, \lambda, \epsilon_\star) = \mathcal{L}(\mathbf{w}) + \lambda \big( \epsilon_\star - \tilde{y}(\mathbf{x}_\star, \mathbf{w}) \big).
\end{equation}

After solving the constrained optimization problem  
\(\partial \mathcal{L}_c / \partial \mathbf{w} = 0\) and \(\partial \mathcal{L}_c / \partial \lambda = 0\),  
the prediction rigidity of \(\tilde{y}(\mathbf{x}_\star, \mathbf{w}_\circ)\) is defined as
\[
R_\star = \left. \frac{\partial^2 \mathcal{L}_c(\epsilon_\star)}{\partial \epsilon_\star^2} \right|_{\epsilon_\star = \tilde{y}(\mathbf{x}_\star, \mathbf{w}_\circ)}.
\]
This quantity reflects the sensitivity of the minimized loss to perturbations in the predicted value \(\tilde{y}(\mathbf{x}_\star, \mathbf{w}_\circ)\). A larger \(R_\star\) implies higher model confidence in the prediction. To compute this analytically, we apply a second-order approximation of the loss around the optimum \(\mathbf{w}_\circ\):
\[
\mathcal{L}(\mathbf{w}) \approx \mathcal{L}(\mathbf{w}_\circ) + \frac{1}{2} (\mathbf{w} - \mathbf{w}_\circ)^\top \mathbf{H}_\circ (\mathbf{w} - \mathbf{w}_\circ),
\]
where the linear term vanishes due to the first-order optimality condition. Linearizing the prediction near \(\mathbf{w}_\circ\), the constrained loss becomes~\cite[Section 3.2]{bigi2024prediction}
\[
\mathcal{L}_c(\epsilon_\star) \approx \mathcal{L}(\mathbf{w}_\circ) + \frac{1}{2} \frac{ \left( \epsilon_\star - \tilde{y}(\mathbf{x}_\star, \mathbf{w}_\circ) \right)^2 }{ \left. \frac{\partial \tilde{y}_\star}{\partial \mathbf{w}} \right|_{\mathbf{w}_\circ}^\top \mathbf{H}_\circ^{-1} \left. \frac{\partial \tilde{y}_\star}{\partial \mathbf{w}} \right|_{\mathbf{w}_\circ} }.
\]
The rigidity \(R_\star\) then has the closed-form:
\[
R_\star = \left( \left. \frac{\partial \tilde{y}_\star}{\partial \mathbf{w}} \right|_{\mathbf{w}_\circ}^\top \mathbf{H}_\circ^{-1} \left. \frac{\partial \tilde{y}_\star}{\partial \mathbf{w}} \right|_{\mathbf{w}_\circ} \right)^{-1}.
\]

The Hessian \(\mathbf{H}_\circ\) can be precomputed from the training set. While computing the exact Hessian is expensive due to second-order derivatives, a practical and widely used approximation is the generalized Gauss–Newton form:
\begin{equation}
\label{eq:hessian_approx}
\begin{aligned}
\mathbf{H}_\circ &= \frac{\partial^2 \mathcal{L}}{\partial \mathbf{w} \partial \mathbf{w}^\top}
= \frac{\partial}{\partial \mathbf{w}} \sum_i \frac{\partial \ell_i}{\partial \tilde{y}_i} \frac{\partial \tilde{y}_i}{\partial \mathbf{w}^\top} \\
&= \sum_i \left( \frac{\partial \ell_i}{\partial \tilde{y}_i} \frac{\partial^2 \tilde{y}_i}{\partial \mathbf{w} \partial \mathbf{w}^\top}
+ \frac{\partial \tilde{y}_i}{\partial \mathbf{w}} \frac{\partial^2 \ell_i}{\partial \tilde{y}_i^2} \frac{\partial \tilde{y}_i}{\partial \mathbf{w}^\top} \right) \\
&\approx \sum_i \frac{\partial \tilde{y}_i}{\partial \mathbf{w}} \frac{\partial^2 \ell_i}{\partial \tilde{y}_i^2} \frac{\partial \tilde{y}_i}{\partial \mathbf{w}^\top},
\end{aligned}
\end{equation}
where all derivatives are evaluated at \(\mathbf{w}_\circ\). The approximation neglects the first term under the assumption that the model is well-trained (\(\partial \ell_i / \partial \tilde{y}_i \approx 0\)). This form is commonly used in optimization and UQ~\cite{schraudolph2002fast, holzmuller2023framework, levenberg1944method, marquardt1963algorithm}.

\section{Supplementary Information}
\label{sec:ablation}

\subsection{Spearman Rank Correlation Coefficient}
\label{sec:apd:Spearman_rank_coefficient}

The Spearman rank correlation coefficient ($\rho$) is a nonparametric statistic that measures the strength and direction of a monotonic relationship between two variables. Unlike the Pearson coefficient, which evaluates linear dependence, Spearman correlation is based solely on ranked values. In the context of uncertainty quantification, $\rho$ provides a direct assessment of whether larger predicted uncertainties are consistently associated with larger observed errors, thereby serving as a standard metric for evaluating the quality of uncertainty ranking.  

Given two sets of observations, predicted uncertainties $\{u_i\}_{i=1}^N$ and observed errors $\{e_i\}_{i=1}^N$, we assign ranks $R(u_i)$ and $R(e_i)$ to each value. The coefficient $\rho$ is then defined as the Pearson correlation between these ranks:  
\begin{equation}
\rho = \frac{\sum_{i=1}^N \left(R(u_i) - \overline{R(u)}\right)\left(R(e_i) - \overline{R(e)}\right)}{\sqrt{\sum_{i=1}^N \left(R(u_i) - \overline{R(u)}\right)^2} \, \sqrt{\sum_{i=1}^N \left(R(e_i) - \overline{R(e)}\right)^2}}.
\label{eq:Pearson}
\end{equation}

In the absence of ties, $\rho$ admits closed form  
\begin{equation}
\rho = 1 - \frac{6 \sum_{i=1}^N d_i^2}{N(N^2-1)},
\end{equation}
where $d_i = R(u_i) - R(e_i)$ is the rank difference for each observation.  

A value of $\rho = 1$ corresponds to perfect agreement between uncertainty ranking and error ranking, $\rho = -1$ indicates perfect disagreement, and $\rho = 0$ signifies no monotonic association. In practice, higher $\rho$ values indicate better alignment between predicted uncertainties and true errors, and are thus desirable for evaluating the effectiveness of calibration schemes.

\subsection{Flexible UC on Other Public Datasets}
\label{sec:apd:mptraj_matpes}

\begin{figure}
\centering
\includegraphics[width=1.0\linewidth]{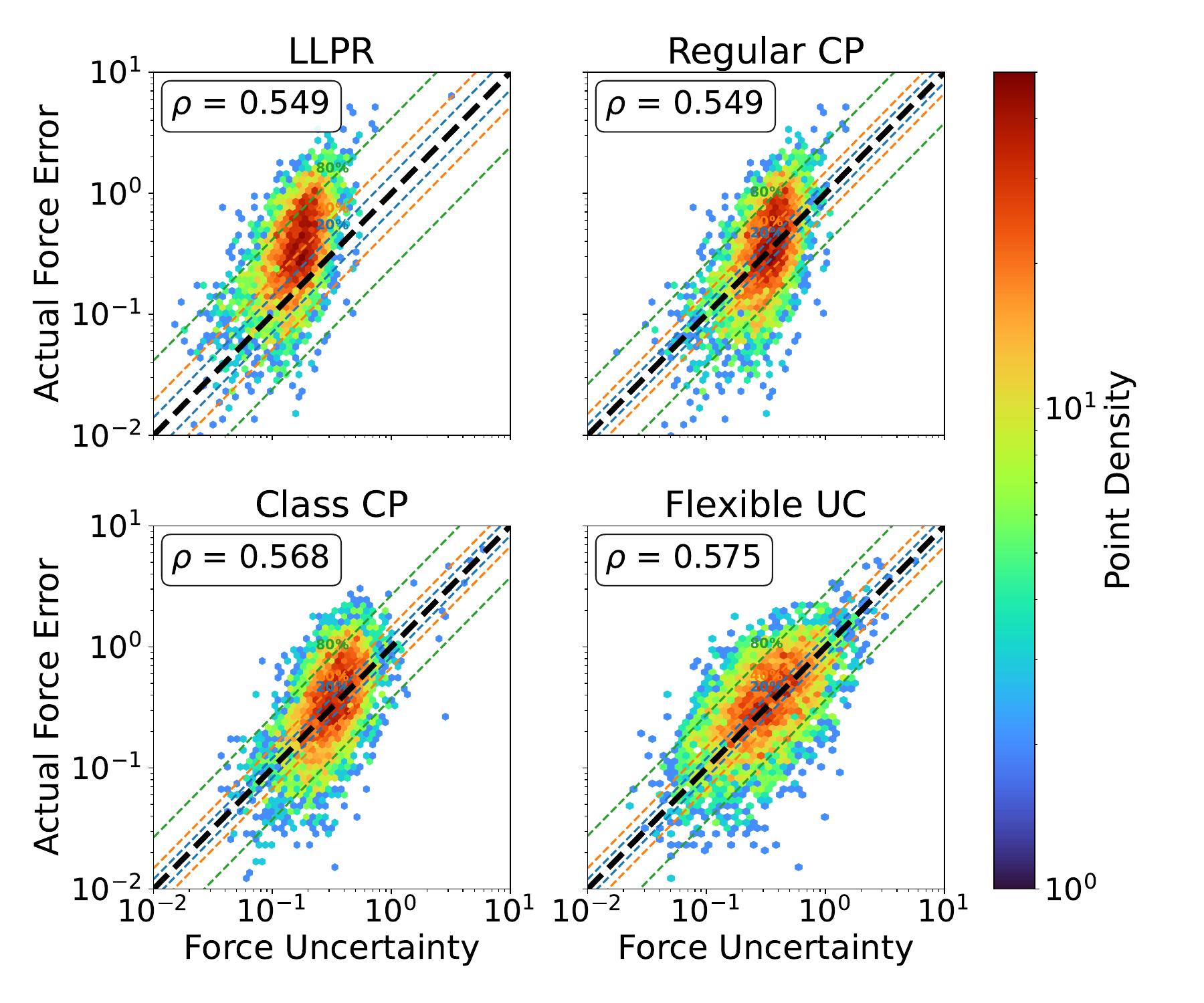}
\caption{Uncertainty estimates from LLPR, regular CP, class-based CP, and flexible UC on the MATPES dataset.}
\label{fig:matpes_cp}
\end{figure}

\begin{figure}
\centering
\includegraphics[width=1.0\linewidth]{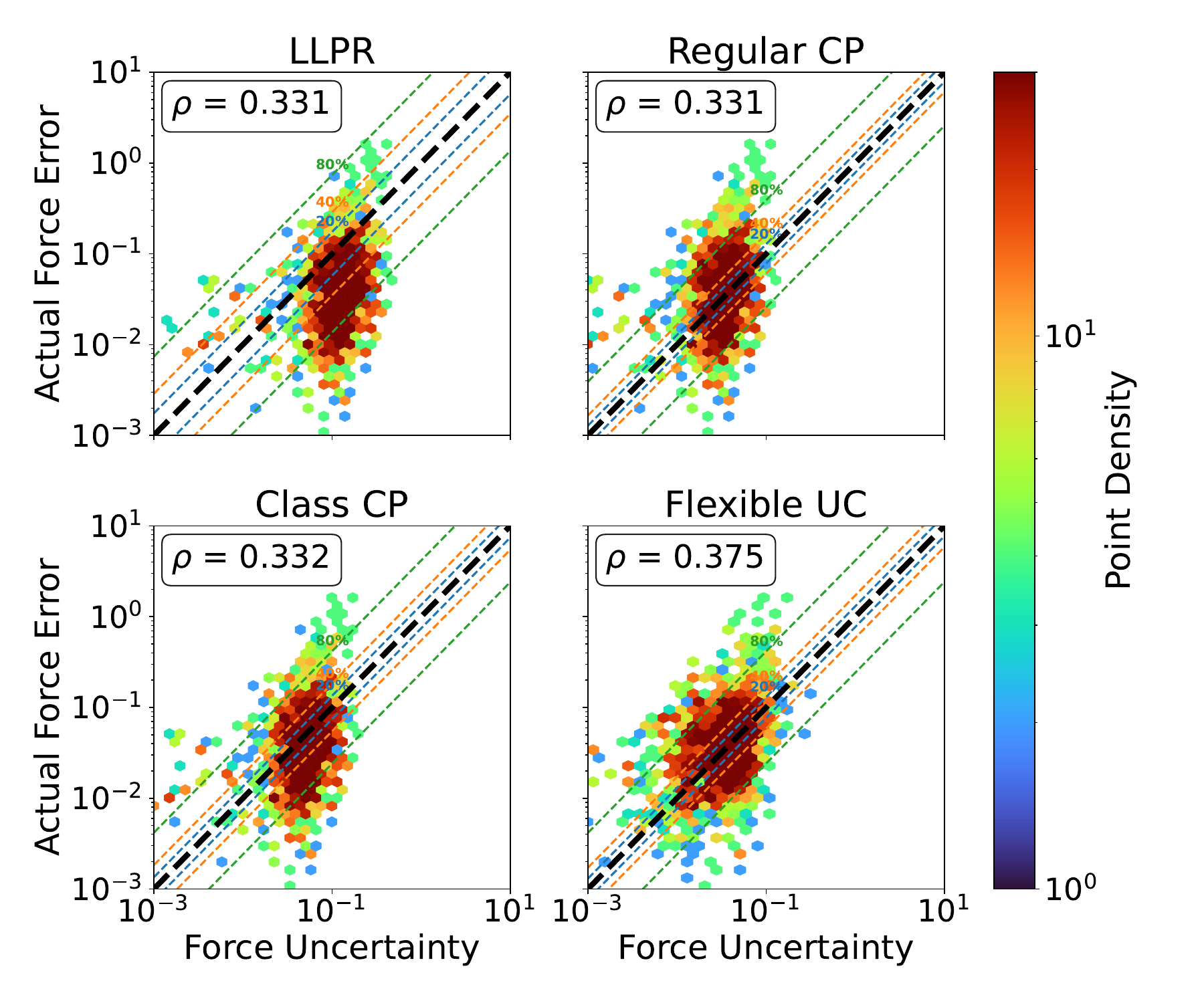}
\caption{Uncertainty estimates from LLPR, regular CP, class-based CP, and flexible UC on the MPtraj dataset.}
\label{fig:mp_cp}
\end{figure}

We further evaluate the proposed calibration scheme on two widely used public datasets, MPtraj and MATPES, with results shown in Figures~\ref{fig:matpes_cp} and \ref{fig:mp_cp}. These datasets are representative benchmarks for molecular and materials modeling and are often used for training and validation of foundation MLIPs. In this context, calibration provides a useful test of whether the proposed framework delivers consistent benefits beyond the custom benchmarks considered in the main text.

In both MPtraj and MATPES, the improvements achieved by flexible uncertainty calibration are less pronounced compared with the LiCl and catalytic surface benchmarks discussed in Section~\ref{sec:sub:various_datasets}. This outcome is expected, since the distributional gap between calibration and testing sets within MPtraj, or between MPtraj and MATPES, is relatively small. As a result, both qualitative improvements in the error–uncertainty correlation and quantitative gains in rank-based metrics remain modest. Put differently, when training and testing environments are already closely matched, the baseline LLPR and regular CP strategies provide reasonably reliable estimates, leaving less room for additional gains from more flexible calibration.

Nevertheless, the experiments on MPtraj and MATPES provide two important insights. First, they confirm that flexible calibration does not deteriorate performance even in low-shift regimes, thereby maintaining consistency with standard approaches. Second, they highlight that the advantages of flexible UC are most significant when substantial distributional shifts are present, such as in ionic systems or heterogeneous catalytic environments. In practice, this suggests that flexible UC may be especially valuable in active learning and transfer settings, where models are repeatedly exposed to configurations far from the calibration distribution.

Taken together, these results demonstrate that flexible UC provides a general and robust framework: it delivers substantial improvements when extrapolation is required, while ensuring comparable performance to existing methods when calibration and testing domains are similar.

\subsection{Illustration of the Weighted Loss Function}
\label{sec:apd:loss_func_weight}

To penalize high-error configurations more strongly, we introduce a weighted loss in which each data point $(\mathbf{X}_i, \mathbf{Y}_i)$ is assigned a weight $w$ depending on its prediction error. Formally,
\begin{equation}
w := w_{c_0}(\epsilon) = {\rm sigmoid}\!\big(c_0 (\epsilon - 0.05)\big) + 0.3 ,
\end{equation}
where $\epsilon := \|\tilde{f}(\mathbf{X}_i) - \mathbf{Y}_i\|$ denotes the prediction error and $c_0$ is a tunable hyperparameter. This design smoothly increases $w$ as the error grows, thereby emphasizing configurations with larger deviations.  

\begin{figure}[ht!]
    \centering
    \includegraphics[width=0.9\linewidth]{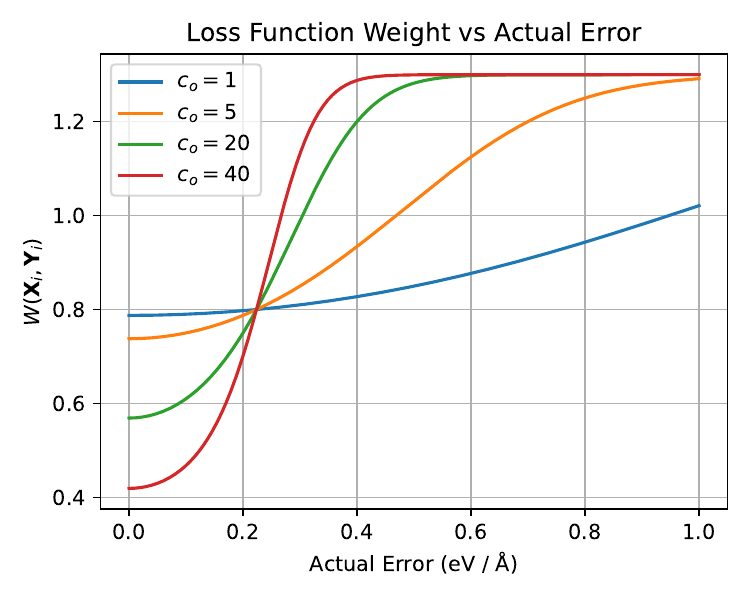}
    \caption{Weighted loss function $w_{c_0}(\epsilon)$ for different choices of $c_0$.}
    \label{fig:weighted_loss_plot}
\end{figure}

Figure~\ref{fig:weighted_loss_plot} shows the dependence of $w_{c_0}(\epsilon)$ on $\epsilon$ across different $c_0$ values. In all experiments we set $c_0 = 40$, and we found that moderate changes of this parameter do not significantly affect the results. 

\subsection{Descriptor-based Quantile Model and Implementation Details}
\label{sec:apd:sec:apd:hyperparameters}

We briefly outline the implementation details of flexible UC and class-based CP. 
For flexible UC, we adopt a deliberately simple design for the quantile model. 
A feedforward neural network learns a multiplicative adjustment $q(\mathbf{X})$ 
to the baseline uncertainty estimate $\sigma$, based on MACE descriptors 
$h(\mathbf{X}) \in \mathbb{R}^d$. The model computes
\begin{equation}
q(\mathbf{X}) = \sigma \cdot \text{Softplus}(f_\theta(h(\mathbf{X}))),
\end{equation}
where
\begin{equation}
f_\theta(h(\mathbf{X})) = W_3 \, \phi\!\left(W_2 \, \phi(W_1 h(\mathbf{X}) + b_1) + b_2\right) + b_3,
\end{equation}
with $\phi(\cdot)$ denoting the ReLU activation. 
The hidden dimension of each dense layer is set to 64, except for the final scalar 
output prior to the Softplus, which ensures positivity of the predicted scaling factor. 
All quantile models are trained using the Adam optimizer with learning rate $10^{-3}$, 
implemented in \texttt{PyTorch}. 

For class-based CP, we set the number of classes to $N_{\rm class} = 20$ and partition 
local atomic environments using the \texttt{BayesianGaussianMixture} implementation 
in \texttt{scikit-learn}. In datasets with highly diverse environments (e.g., MATPES, 
MPtraj, OMOL), the number of classes is empirically increased until a clear separation 
of score distributions is achieved; further refinement beyond this point yields 
negligible changes.


\bibliography{apssamp}

\begin{thebibliography}{82}%
\makeatletter
\providecommand \@ifxundefined [1]{%
 \@ifx{#1\undefined}
}%
\providecommand \@ifnum [1]{%
 \ifnum #1\expandafter \@firstoftwo
 \else \expandafter \@secondoftwo
 \fi
}%
\providecommand \@ifx [1]{%
 \ifx #1\expandafter \@firstoftwo
 \else \expandafter \@secondoftwo
 \fi
}%
\providecommand \natexlab [1]{#1}%
\providecommand \enquote  [1]{``#1''}%
\providecommand \bibnamefont  [1]{#1}%
\providecommand \bibfnamefont [1]{#1}%
\providecommand \citenamefont [1]{#1}%
\providecommand \href@noop [0]{\@secondoftwo}%
\providecommand \href [0]{\begingroup \@sanitize@url \@href}%
\providecommand \@href[1]{\@@startlink{#1}\@@href}%
\providecommand \@@href[1]{\endgroup#1\@@endlink}%
\providecommand \@sanitize@url [0]{\catcode `\\12\catcode `\$12\catcode `\&12\catcode `\#12\catcode `\^12\catcode `\_12\catcode `\%12\relax}%
\providecommand \@@startlink[1]{}%
\providecommand \@@endlink[0]{}%
\providecommand \url  [0]{\begingroup\@sanitize@url \@url }%
\providecommand \@url [1]{\endgroup\@href {#1}{\urlprefix }}%
\providecommand \urlprefix  [0]{URL }%
\providecommand \Eprint [0]{\href }%
\providecommand \doibase [0]{https://doi.org/}%
\providecommand \selectlanguage [0]{\@gobble}%
\providecommand \bibinfo  [0]{\@secondoftwo}%
\providecommand \bibfield  [0]{\@secondoftwo}%
\providecommand \translation [1]{[#1]}%
\providecommand \BibitemOpen [0]{}%
\providecommand \bibitemStop [0]{}%
\providecommand \bibitemNoStop [0]{.\EOS\space}%
\providecommand \EOS [0]{\spacefactor3000\relax}%
\providecommand \BibitemShut  [1]{\csname bibitem#1\endcsname}%
\let\auto@bib@innerbib\@empty
\bibitem [{\citenamefont {Behler}\ and\ \citenamefont {Parrinello}(2007)}]{behler2007generalized}%
  \BibitemOpen
  \bibfield  {author} {\bibinfo {author} {\bibfnamefont {J.}~\bibnamefont {Behler}}\ and\ \bibinfo {author} {\bibfnamefont {M.}~\bibnamefont {Parrinello}},\ }\bibfield  {title} {\bibinfo {title} {Generalized neural-network representation of high-dimensional potential-energy surfaces},\ }\href@noop {} {\bibfield  {journal} {\bibinfo  {journal} {Physical review letters}\ }\textbf {\bibinfo {volume} {98}},\ \bibinfo {pages} {146401} (\bibinfo {year} {2007})}\BibitemShut {NoStop}%
\bibitem [{\citenamefont {Bart{\'o}k}\ \emph {et~al.}(2010)\citenamefont {Bart{\'o}k}, \citenamefont {Payne}, \citenamefont {Kondor},\ and\ \citenamefont {Cs{\'a}nyi}}]{bartok2010gaussian}%
  \BibitemOpen
  \bibfield  {author} {\bibinfo {author} {\bibfnamefont {A.~P.}\ \bibnamefont {Bart{\'o}k}}, \bibinfo {author} {\bibfnamefont {M.~C.}\ \bibnamefont {Payne}}, \bibinfo {author} {\bibfnamefont {R.}~\bibnamefont {Kondor}},\ and\ \bibinfo {author} {\bibfnamefont {G.}~\bibnamefont {Cs{\'a}nyi}},\ }\bibfield  {title} {\bibinfo {title} {Gaussian approximation potentials: The accuracy of quantum mechanics, without the electrons},\ }\href@noop {} {\bibfield  {journal} {\bibinfo  {journal} {Physical review letters}\ }\textbf {\bibinfo {volume} {104}},\ \bibinfo {pages} {136403} (\bibinfo {year} {2010})}\BibitemShut {NoStop}%
\bibitem [{\citenamefont {Witt}\ \emph {et~al.}(2023)\citenamefont {Witt}, \citenamefont {van~der Oord}, \citenamefont {Gel{\v{z}}inyt{\.e}}, \citenamefont {J{\"a}rvinen}, \citenamefont {Ross}, \citenamefont {Darby}, \citenamefont {Ho}, \citenamefont {Baldwin}, \citenamefont {Sachs}, \citenamefont {Kermode} \emph {et~al.}}]{witt2023acepotentials}%
  \BibitemOpen
  \bibfield  {author} {\bibinfo {author} {\bibfnamefont {W.~C.}\ \bibnamefont {Witt}}, \bibinfo {author} {\bibfnamefont {C.}~\bibnamefont {van~der Oord}}, \bibinfo {author} {\bibfnamefont {E.}~\bibnamefont {Gel{\v{z}}inyt{\.e}}}, \bibinfo {author} {\bibfnamefont {T.}~\bibnamefont {J{\"a}rvinen}}, \bibinfo {author} {\bibfnamefont {A.}~\bibnamefont {Ross}}, \bibinfo {author} {\bibfnamefont {J.~P.}\ \bibnamefont {Darby}}, \bibinfo {author} {\bibfnamefont {C.~H.}\ \bibnamefont {Ho}}, \bibinfo {author} {\bibfnamefont {W.~J.}\ \bibnamefont {Baldwin}}, \bibinfo {author} {\bibfnamefont {M.}~\bibnamefont {Sachs}}, \bibinfo {author} {\bibfnamefont {J.}~\bibnamefont {Kermode}}, \emph {et~al.},\ }\bibfield  {title} {\bibinfo {title} {Acepotentials. jl: A julia implementation of the atomic cluster expansion},\ }\href@noop {} {\bibfield  {journal} {\bibinfo  {journal} {The Journal of Chemical Physics}\ }\textbf {\bibinfo {volume} {159}} (\bibinfo {year} {2023})}\BibitemShut {NoStop}%
\bibitem [{\citenamefont {Batatia}\ \emph {et~al.}(2022)\citenamefont {Batatia}, \citenamefont {Kovacs}, \citenamefont {Simm}, \citenamefont {Ortner},\ and\ \citenamefont {Cs{\'a}nyi}}]{batatia2022mace}%
  \BibitemOpen
  \bibfield  {author} {\bibinfo {author} {\bibfnamefont {I.}~\bibnamefont {Batatia}}, \bibinfo {author} {\bibfnamefont {D.~P.}\ \bibnamefont {Kovacs}}, \bibinfo {author} {\bibfnamefont {G.}~\bibnamefont {Simm}}, \bibinfo {author} {\bibfnamefont {C.}~\bibnamefont {Ortner}},\ and\ \bibinfo {author} {\bibfnamefont {G.}~\bibnamefont {Cs{\'a}nyi}},\ }\bibfield  {title} {\bibinfo {title} {Mace: Higher order equivariant message passing neural networks for fast and accurate force fields},\ }\href@noop {} {\bibfield  {journal} {\bibinfo  {journal} {Advances in Neural Information Processing Systems}\ }\textbf {\bibinfo {volume} {35}},\ \bibinfo {pages} {11423} (\bibinfo {year} {2022})}\BibitemShut {NoStop}%
\bibitem [{\citenamefont {Shapeev}(2016)}]{shapeev2016moment}%
  \BibitemOpen
  \bibfield  {author} {\bibinfo {author} {\bibfnamefont {A.~V.}\ \bibnamefont {Shapeev}},\ }\bibfield  {title} {\bibinfo {title} {Moment tensor potentials: A class of systematically improvable interatomic potentials},\ }\href@noop {} {\bibfield  {journal} {\bibinfo  {journal} {Multiscale Modeling \& Simulation}\ }\textbf {\bibinfo {volume} {14}},\ \bibinfo {pages} {1153} (\bibinfo {year} {2016})}\BibitemShut {NoStop}%
\bibitem [{\citenamefont {Sch{\"u}tt}\ \emph {et~al.}(2018)\citenamefont {Sch{\"u}tt}, \citenamefont {Sauceda}, \citenamefont {Kindermans}, \citenamefont {Tkatchenko},\ and\ \citenamefont {M{\"u}ller}}]{schutt2018schnet}%
  \BibitemOpen
  \bibfield  {author} {\bibinfo {author} {\bibfnamefont {K.~T.}\ \bibnamefont {Sch{\"u}tt}}, \bibinfo {author} {\bibfnamefont {H.~E.}\ \bibnamefont {Sauceda}}, \bibinfo {author} {\bibfnamefont {P.-J.}\ \bibnamefont {Kindermans}}, \bibinfo {author} {\bibfnamefont {A.}~\bibnamefont {Tkatchenko}},\ and\ \bibinfo {author} {\bibfnamefont {K.-R.}\ \bibnamefont {M{\"u}ller}},\ }\bibfield  {title} {\bibinfo {title} {Schnet--a deep learning architecture for molecules and materials},\ }\href@noop {} {\bibfield  {journal} {\bibinfo  {journal} {The Journal of Chemical Physics}\ }\textbf {\bibinfo {volume} {148}} (\bibinfo {year} {2018})}\BibitemShut {NoStop}%
\bibitem [{\citenamefont {Batzner}\ \emph {et~al.}(2022)\citenamefont {Batzner}, \citenamefont {Musaelian}, \citenamefont {Sun}, \citenamefont {Geiger}, \citenamefont {Mailoa}, \citenamefont {Kornbluth}, \citenamefont {Molinari}, \citenamefont {Smidt},\ and\ \citenamefont {Kozinsky}}]{batzner20223}%
  \BibitemOpen
  \bibfield  {author} {\bibinfo {author} {\bibfnamefont {S.}~\bibnamefont {Batzner}}, \bibinfo {author} {\bibfnamefont {A.}~\bibnamefont {Musaelian}}, \bibinfo {author} {\bibfnamefont {L.}~\bibnamefont {Sun}}, \bibinfo {author} {\bibfnamefont {M.}~\bibnamefont {Geiger}}, \bibinfo {author} {\bibfnamefont {J.~P.}\ \bibnamefont {Mailoa}}, \bibinfo {author} {\bibfnamefont {M.}~\bibnamefont {Kornbluth}}, \bibinfo {author} {\bibfnamefont {N.}~\bibnamefont {Molinari}}, \bibinfo {author} {\bibfnamefont {T.~E.}\ \bibnamefont {Smidt}},\ and\ \bibinfo {author} {\bibfnamefont {B.}~\bibnamefont {Kozinsky}},\ }\bibfield  {title} {\bibinfo {title} {E (3)-equivariant graph neural networks for data-efficient and accurate interatomic potentials},\ }\href@noop {} {\bibfield  {journal} {\bibinfo  {journal} {Nature Communnications}\ }\textbf {\bibinfo {volume} {13}},\ \bibinfo {pages} {2453} (\bibinfo {year} {2022})}\BibitemShut {NoStop}%
\bibitem [{\citenamefont {Cheng}(2024)}]{cheng2024cartesian}%
  \BibitemOpen
  \bibfield  {author} {\bibinfo {author} {\bibfnamefont {B.}~\bibnamefont {Cheng}},\ }\bibfield  {title} {\bibinfo {title} {Cartesian atomic cluster expansion for machine learning interatomic potentials},\ }\href@noop {} {\bibfield  {journal} {\bibinfo  {journal} {npj Comput. Mater.}\ }\textbf {\bibinfo {volume} {10}},\ \bibinfo {pages} {157} (\bibinfo {year} {2024})}\BibitemShut {NoStop}%
\bibitem [{\citenamefont {Liu}\ \emph {et~al.}(2025)\citenamefont {Liu}, \citenamefont {Zeng}, \citenamefont {Luo}, \citenamefont {Wang}, \citenamefont {Zhao},\ and\ \citenamefont {Xu}}]{liu2025fine}%
  \BibitemOpen
  \bibfield  {author} {\bibinfo {author} {\bibfnamefont {X.}~\bibnamefont {Liu}}, \bibinfo {author} {\bibfnamefont {K.}~\bibnamefont {Zeng}}, \bibinfo {author} {\bibfnamefont {Z.}~\bibnamefont {Luo}}, \bibinfo {author} {\bibfnamefont {Y.}~\bibnamefont {Wang}}, \bibinfo {author} {\bibfnamefont {T.}~\bibnamefont {Zhao}},\ and\ \bibinfo {author} {\bibfnamefont {Z.}~\bibnamefont {Xu}},\ }\bibfield  {title} {\bibinfo {title} {Fine-tuning universal machine-learned interatomic potentials: A tutorial on methods and applications},\ }\href@noop {} {\bibfield  {journal} {\bibinfo  {journal} {arXiv preprint arXiv:2506.21935}\ } (\bibinfo {year} {2025})}\BibitemShut {NoStop}%
\bibitem [{\citenamefont {Merchant}\ \emph {et~al.}(2023)\citenamefont {Merchant}, \citenamefont {Batzner}, \citenamefont {Schoenholz}, \citenamefont {Aykol}, \citenamefont {Cheon},\ and\ \citenamefont {Cubuk}}]{merchant2023scaling}%
  \BibitemOpen
  \bibfield  {author} {\bibinfo {author} {\bibfnamefont {A.}~\bibnamefont {Merchant}}, \bibinfo {author} {\bibfnamefont {S.}~\bibnamefont {Batzner}}, \bibinfo {author} {\bibfnamefont {S.~S.}\ \bibnamefont {Schoenholz}}, \bibinfo {author} {\bibfnamefont {M.}~\bibnamefont {Aykol}}, \bibinfo {author} {\bibfnamefont {G.}~\bibnamefont {Cheon}},\ and\ \bibinfo {author} {\bibfnamefont {E.~D.}\ \bibnamefont {Cubuk}},\ }\bibfield  {title} {\bibinfo {title} {Scaling deep learning for materials discovery},\ }\href@noop {} {\bibfield  {journal} {\bibinfo  {journal} {Nature}\ }\textbf {\bibinfo {volume} {624}},\ \bibinfo {pages} {80} (\bibinfo {year} {2023})}\BibitemShut {NoStop}%
\bibitem [{\citenamefont {Zhang}\ \emph {et~al.}(2024)\citenamefont {Zhang}, \citenamefont {Liu}, \citenamefont {Zhang}, \citenamefont {Zhang}, \citenamefont {Cai}, \citenamefont {Bi}, \citenamefont {Du}, \citenamefont {Qin}, \citenamefont {Peng}, \citenamefont {Huang} \emph {et~al.}}]{zhang2024dpa}%
  \BibitemOpen
  \bibfield  {author} {\bibinfo {author} {\bibfnamefont {D.}~\bibnamefont {Zhang}}, \bibinfo {author} {\bibfnamefont {X.}~\bibnamefont {Liu}}, \bibinfo {author} {\bibfnamefont {X.}~\bibnamefont {Zhang}}, \bibinfo {author} {\bibfnamefont {C.}~\bibnamefont {Zhang}}, \bibinfo {author} {\bibfnamefont {C.}~\bibnamefont {Cai}}, \bibinfo {author} {\bibfnamefont {H.}~\bibnamefont {Bi}}, \bibinfo {author} {\bibfnamefont {Y.}~\bibnamefont {Du}}, \bibinfo {author} {\bibfnamefont {X.}~\bibnamefont {Qin}}, \bibinfo {author} {\bibfnamefont {A.}~\bibnamefont {Peng}}, \bibinfo {author} {\bibfnamefont {J.}~\bibnamefont {Huang}}, \emph {et~al.},\ }\bibfield  {title} {\bibinfo {title} {Dpa-2: a large atomic model as a multi-task learner},\ }\href@noop {} {\bibfield  {journal} {\bibinfo  {journal} {npj Computational Materials}\ }\textbf {\bibinfo {volume} {10}},\ \bibinfo {pages} {293} (\bibinfo {year} {2024})}\BibitemShut {NoStop}%
\bibitem [{\citenamefont {Frenkel}\ and\ \citenamefont {Smit}(2023)}]{frenkel2023understanding}%
  \BibitemOpen
  \bibfield  {author} {\bibinfo {author} {\bibfnamefont {D.}~\bibnamefont {Frenkel}}\ and\ \bibinfo {author} {\bibfnamefont {B.}~\bibnamefont {Smit}},\ }\href@noop {} {\emph {\bibinfo {title} {Understanding molecular simulation: from algorithms to applications}}}\ (\bibinfo  {publisher} {Elsevier},\ \bibinfo {year} {2023})\BibitemShut {NoStop}%
\bibitem [{\citenamefont {Angelikopoulos}\ \emph {et~al.}(2012)\citenamefont {Angelikopoulos}, \citenamefont {Papadimitriou},\ and\ \citenamefont {Koumoutsakos}}]{angelikopoulos2012bayesian}%
  \BibitemOpen
  \bibfield  {author} {\bibinfo {author} {\bibfnamefont {P.}~\bibnamefont {Angelikopoulos}}, \bibinfo {author} {\bibfnamefont {C.}~\bibnamefont {Papadimitriou}},\ and\ \bibinfo {author} {\bibfnamefont {P.}~\bibnamefont {Koumoutsakos}},\ }\bibfield  {title} {\bibinfo {title} {Bayesian uncertainty quantification and propagation in molecular dynamics simulations: a high performance computing framework},\ }\href@noop {} {\bibfield  {journal} {\bibinfo  {journal} {The Journal of chemical physics}\ }\textbf {\bibinfo {volume} {137}} (\bibinfo {year} {2012})}\BibitemShut {NoStop}%
\bibitem [{\citenamefont {Messerly}\ \emph {et~al.}(2017)\citenamefont {Messerly}, \citenamefont {Knotts},\ and\ \citenamefont {Wilding}}]{messerly2017uncertainty}%
  \BibitemOpen
  \bibfield  {author} {\bibinfo {author} {\bibfnamefont {R.~A.}\ \bibnamefont {Messerly}}, \bibinfo {author} {\bibfnamefont {T.~A.}\ \bibnamefont {Knotts}},\ and\ \bibinfo {author} {\bibfnamefont {W.~V.}\ \bibnamefont {Wilding}},\ }\bibfield  {title} {\bibinfo {title} {Uncertainty quantification and propagation of errors of the lennard-jones 12-6 parameters for n-alkanes},\ }\href@noop {} {\bibfield  {journal} {\bibinfo  {journal} {The Journal of chemical physics}\ }\textbf {\bibinfo {volume} {146}} (\bibinfo {year} {2017})}\BibitemShut {NoStop}%
\bibitem [{\citenamefont {Kellner}\ and\ \citenamefont {Ceriotti}(2024)}]{kellner2024uncertainty}%
  \BibitemOpen
  \bibfield  {author} {\bibinfo {author} {\bibfnamefont {M.}~\bibnamefont {Kellner}}\ and\ \bibinfo {author} {\bibfnamefont {M.}~\bibnamefont {Ceriotti}},\ }\bibfield  {title} {\bibinfo {title} {Uncertainty quantification by direct propagation of shallow ensembles},\ }\href@noop {} {\bibfield  {journal} {\bibinfo  {journal} {Machine Learning: Science and Technology}\ }\textbf {\bibinfo {volume} {5}},\ \bibinfo {pages} {035006} (\bibinfo {year} {2024})}\BibitemShut {NoStop}%
\bibitem [{\citenamefont {Imbalzano}\ \emph {et~al.}(2021)\citenamefont {Imbalzano}, \citenamefont {Zhuang}, \citenamefont {Kapil}, \citenamefont {Rossi}, \citenamefont {Engel}, \citenamefont {Grasselli},\ and\ \citenamefont {Ceriotti}}]{imbalzano2021uncertainty}%
  \BibitemOpen
  \bibfield  {author} {\bibinfo {author} {\bibfnamefont {G.}~\bibnamefont {Imbalzano}}, \bibinfo {author} {\bibfnamefont {Y.}~\bibnamefont {Zhuang}}, \bibinfo {author} {\bibfnamefont {V.}~\bibnamefont {Kapil}}, \bibinfo {author} {\bibfnamefont {K.}~\bibnamefont {Rossi}}, \bibinfo {author} {\bibfnamefont {E.~A.}\ \bibnamefont {Engel}}, \bibinfo {author} {\bibfnamefont {F.}~\bibnamefont {Grasselli}},\ and\ \bibinfo {author} {\bibfnamefont {M.}~\bibnamefont {Ceriotti}},\ }\bibfield  {title} {\bibinfo {title} {Uncertainty estimation for molecular dynamics and sampling},\ }\href@noop {} {\bibfield  {journal} {\bibinfo  {journal} {The Journal of chemical physics}\ }\textbf {\bibinfo {volume} {154}} (\bibinfo {year} {2021})}\BibitemShut {NoStop}%
\bibitem [{\citenamefont {Grasselli}\ \emph {et~al.}(2025)\citenamefont {Grasselli}, \citenamefont {Chong}, \citenamefont {Kapil}, \citenamefont {Bonfanti},\ and\ \citenamefont {Rossi}}]{grasselli2025uncertainty}%
  \BibitemOpen
  \bibfield  {author} {\bibinfo {author} {\bibfnamefont {F.}~\bibnamefont {Grasselli}}, \bibinfo {author} {\bibfnamefont {S.}~\bibnamefont {Chong}}, \bibinfo {author} {\bibfnamefont {V.}~\bibnamefont {Kapil}}, \bibinfo {author} {\bibfnamefont {S.}~\bibnamefont {Bonfanti}},\ and\ \bibinfo {author} {\bibfnamefont {K.}~\bibnamefont {Rossi}},\ }\bibfield  {title} {\bibinfo {title} {Uncertainty in the era of machine learning for atomistic modeling},\ }\href@noop {} {\bibfield  {journal} {\bibinfo  {journal} {arXiv preprint arXiv:2503.09196}\ } (\bibinfo {year} {2025})}\BibitemShut {NoStop}%
\bibitem [{\citenamefont {Dai}\ \emph {et~al.}(2025)\citenamefont {Dai}, \citenamefont {Adhikari},\ and\ \citenamefont {Wen}}]{dai2025uncertainty}%
  \BibitemOpen
  \bibfield  {author} {\bibinfo {author} {\bibfnamefont {J.}~\bibnamefont {Dai}}, \bibinfo {author} {\bibfnamefont {S.}~\bibnamefont {Adhikari}},\ and\ \bibinfo {author} {\bibfnamefont {M.}~\bibnamefont {Wen}},\ }\bibfield  {title} {\bibinfo {title} {Uncertainty quantification and propagation in atomistic machine learning},\ }\href@noop {} {\bibfield  {journal} {\bibinfo  {journal} {Reviews in Chemical Engineering}\ }\textbf {\bibinfo {volume} {41}},\ \bibinfo {pages} {333} (\bibinfo {year} {2025})}\BibitemShut {NoStop}%
\bibitem [{\citenamefont {Peterson}\ \emph {et~al.}(2017)\citenamefont {Peterson}, \citenamefont {Christensen},\ and\ \citenamefont {Khorshidi}}]{peterson2017addressing}%
  \BibitemOpen
  \bibfield  {author} {\bibinfo {author} {\bibfnamefont {A.~A.}\ \bibnamefont {Peterson}}, \bibinfo {author} {\bibfnamefont {R.}~\bibnamefont {Christensen}},\ and\ \bibinfo {author} {\bibfnamefont {A.}~\bibnamefont {Khorshidi}},\ }\bibfield  {title} {\bibinfo {title} {Addressing uncertainty in atomistic machine learning},\ }\href@noop {} {\bibfield  {journal} {\bibinfo  {journal} {Physical Chemistry Chemical Physics}\ }\textbf {\bibinfo {volume} {19}},\ \bibinfo {pages} {10978} (\bibinfo {year} {2017})}\BibitemShut {NoStop}%
\bibitem [{\citenamefont {Musil}\ \emph {et~al.}(2019)\citenamefont {Musil}, \citenamefont {Willatt}, \citenamefont {Langovoy},\ and\ \citenamefont {Ceriotti}}]{musil2019fast}%
  \BibitemOpen
  \bibfield  {author} {\bibinfo {author} {\bibfnamefont {F.}~\bibnamefont {Musil}}, \bibinfo {author} {\bibfnamefont {M.~J.}\ \bibnamefont {Willatt}}, \bibinfo {author} {\bibfnamefont {M.~A.}\ \bibnamefont {Langovoy}},\ and\ \bibinfo {author} {\bibfnamefont {M.}~\bibnamefont {Ceriotti}},\ }\bibfield  {title} {\bibinfo {title} {Fast and accurate uncertainty estimation in chemical machine learning},\ }\href@noop {} {\bibfield  {journal} {\bibinfo  {journal} {Journal of chemical theory and computation}\ }\textbf {\bibinfo {volume} {15}},\ \bibinfo {pages} {906} (\bibinfo {year} {2019})}\BibitemShut {NoStop}%
\bibitem [{\citenamefont {Behler}(2014)}]{behler2014representing}%
  \BibitemOpen
  \bibfield  {author} {\bibinfo {author} {\bibfnamefont {J.}~\bibnamefont {Behler}},\ }\bibfield  {title} {\bibinfo {title} {Representing potential energy surfaces by high-dimensional neural network potentials},\ }\href@noop {} {\bibfield  {journal} {\bibinfo  {journal} {Journal of Physics: Condensed Matter}\ }\textbf {\bibinfo {volume} {26}},\ \bibinfo {pages} {183001} (\bibinfo {year} {2014})}\BibitemShut {NoStop}%
\bibitem [{\citenamefont {Xiao}\ \emph {et~al.}(2018)\citenamefont {Xiao}, \citenamefont {Li},\ and\ \citenamefont {Wang}}]{xiao2018uncertainty}%
  \BibitemOpen
  \bibfield  {author} {\bibinfo {author} {\bibfnamefont {W.}~\bibnamefont {Xiao}}, \bibinfo {author} {\bibfnamefont {Y.}~\bibnamefont {Li}},\ and\ \bibinfo {author} {\bibfnamefont {P.}~\bibnamefont {Wang}},\ }\bibfield  {title} {\bibinfo {title} {Uncertainty quantification of machine learning potentials for atomistic simulation},\ }in\ \href@noop {} {\emph {\bibinfo {booktitle} {AIAA Non-Deterministic Approaches Conference}}}\ (\bibinfo {year} {2018})\ pp.\ \bibinfo {pages} {2018--2166}\BibitemShut {NoStop}%
\bibitem [{\citenamefont {Novikov}\ and\ \citenamefont {Shapeev}(2019)}]{novikov2019improving}%
  \BibitemOpen
  \bibfield  {author} {\bibinfo {author} {\bibfnamefont {I.~S.}\ \bibnamefont {Novikov}}\ and\ \bibinfo {author} {\bibfnamefont {A.~V.}\ \bibnamefont {Shapeev}},\ }\bibfield  {title} {\bibinfo {title} {Improving accuracy of interatomic potentials: more physics or more data? a case study of silica},\ }\href@noop {} {\bibfield  {journal} {\bibinfo  {journal} {Materials Today Communications}\ }\textbf {\bibinfo {volume} {18}},\ \bibinfo {pages} {74} (\bibinfo {year} {2019})}\BibitemShut {NoStop}%
\bibitem [{\citenamefont {Zhang}\ \emph {et~al.}(2019)\citenamefont {Zhang}, \citenamefont {Lin}, \citenamefont {Wang}, \citenamefont {Car},\ and\ \citenamefont {E}}]{zhang2019active}%
  \BibitemOpen
  \bibfield  {author} {\bibinfo {author} {\bibfnamefont {L.}~\bibnamefont {Zhang}}, \bibinfo {author} {\bibfnamefont {D.-Y.}\ \bibnamefont {Lin}}, \bibinfo {author} {\bibfnamefont {H.}~\bibnamefont {Wang}}, \bibinfo {author} {\bibfnamefont {R.}~\bibnamefont {Car}},\ and\ \bibinfo {author} {\bibfnamefont {W.}~\bibnamefont {E}},\ }\bibfield  {title} {\bibinfo {title} {Active learning of uniformly accurate interatomic potentials for materials simulation},\ }\href@noop {} {\bibfield  {journal} {\bibinfo  {journal} {Physical Review Materials}\ }\textbf {\bibinfo {volume} {3}},\ \bibinfo {pages} {023804} (\bibinfo {year} {2019})}\BibitemShut {NoStop}%
\bibitem [{\citenamefont {Novikov}\ \emph {et~al.}(2020)\citenamefont {Novikov}, \citenamefont {Gubaev}, \citenamefont {Podryabinkin},\ and\ \citenamefont {Shapeev}}]{novikov2020mlip}%
  \BibitemOpen
  \bibfield  {author} {\bibinfo {author} {\bibfnamefont {I.~S.}\ \bibnamefont {Novikov}}, \bibinfo {author} {\bibfnamefont {K.}~\bibnamefont {Gubaev}}, \bibinfo {author} {\bibfnamefont {E.~V.}\ \bibnamefont {Podryabinkin}},\ and\ \bibinfo {author} {\bibfnamefont {A.~V.}\ \bibnamefont {Shapeev}},\ }\bibfield  {title} {\bibinfo {title} {The mlip package: moment tensor potentials with mpi and active learning},\ }\href@noop {} {\bibfield  {journal} {\bibinfo  {journal} {Machine Learning: Science and Technology}\ }\textbf {\bibinfo {volume} {2}},\ \bibinfo {pages} {025002} (\bibinfo {year} {2020})}\BibitemShut {NoStop}%
\bibitem [{\citenamefont {Botu}\ \emph {et~al.}(2017)\citenamefont {Botu}, \citenamefont {Batra}, \citenamefont {Chapman},\ and\ \citenamefont {Ramprasad}}]{botu2017machine}%
  \BibitemOpen
  \bibfield  {author} {\bibinfo {author} {\bibfnamefont {V.}~\bibnamefont {Botu}}, \bibinfo {author} {\bibfnamefont {R.}~\bibnamefont {Batra}}, \bibinfo {author} {\bibfnamefont {J.}~\bibnamefont {Chapman}},\ and\ \bibinfo {author} {\bibfnamefont {R.}~\bibnamefont {Ramprasad}},\ }\bibfield  {title} {\bibinfo {title} {Machine learning force fields: construction, validation, and outlook},\ }\href@noop {} {\bibfield  {journal} {\bibinfo  {journal} {The Journal of Physical Chemistry C}\ }\textbf {\bibinfo {volume} {121}},\ \bibinfo {pages} {511} (\bibinfo {year} {2017})}\BibitemShut {NoStop}%
\bibitem [{\citenamefont {Janet}\ \emph {et~al.}(2019)\citenamefont {Janet}, \citenamefont {Duan}, \citenamefont {Yang}, \citenamefont {Nandy},\ and\ \citenamefont {Kulik}}]{janet2019quantitative}%
  \BibitemOpen
  \bibfield  {author} {\bibinfo {author} {\bibfnamefont {J.~P.}\ \bibnamefont {Janet}}, \bibinfo {author} {\bibfnamefont {C.}~\bibnamefont {Duan}}, \bibinfo {author} {\bibfnamefont {T.}~\bibnamefont {Yang}}, \bibinfo {author} {\bibfnamefont {A.}~\bibnamefont {Nandy}},\ and\ \bibinfo {author} {\bibfnamefont {H.~J.}\ \bibnamefont {Kulik}},\ }\bibfield  {title} {\bibinfo {title} {A quantitative uncertainty metric controls error in neural network-driven chemical discovery},\ }\href@noop {} {\bibfield  {journal} {\bibinfo  {journal} {Chemical science}\ }\textbf {\bibinfo {volume} {10}},\ \bibinfo {pages} {7913} (\bibinfo {year} {2019})}\BibitemShut {NoStop}%
\bibitem [{\citenamefont {Xie}\ \emph {et~al.}(2021)\citenamefont {Xie}, \citenamefont {Ma}, \citenamefont {Lei}, \citenamefont {Zhang}, \citenamefont {Xue}, \citenamefont {Tan},\ and\ \citenamefont {Guo}}]{xie2021advanced}%
  \BibitemOpen
  \bibfield  {author} {\bibinfo {author} {\bibfnamefont {J.}~\bibnamefont {Xie}}, \bibinfo {author} {\bibfnamefont {Z.}~\bibnamefont {Ma}}, \bibinfo {author} {\bibfnamefont {J.}~\bibnamefont {Lei}}, \bibinfo {author} {\bibfnamefont {G.}~\bibnamefont {Zhang}}, \bibinfo {author} {\bibfnamefont {J.-H.}\ \bibnamefont {Xue}}, \bibinfo {author} {\bibfnamefont {Z.-H.}\ \bibnamefont {Tan}},\ and\ \bibinfo {author} {\bibfnamefont {J.}~\bibnamefont {Guo}},\ }\bibfield  {title} {\bibinfo {title} {Advanced dropout: A model-free methodology for bayesian dropout optimization},\ }\href@noop {} {\bibfield  {journal} {\bibinfo  {journal} {IEEE Transactions on Pattern Analysis and Machine Intelligence}\ }\textbf {\bibinfo {volume} {44}},\ \bibinfo {pages} {4605} (\bibinfo {year} {2021})}\BibitemShut {NoStop}%
\bibitem [{\citenamefont {Chen}\ \emph {et~al.}(2022)\citenamefont {Chen}, \citenamefont {Ortner},\ and\ \citenamefont {Wang}}]{chen2022qm}%
  \BibitemOpen
  \bibfield  {author} {\bibinfo {author} {\bibfnamefont {H.}~\bibnamefont {Chen}}, \bibinfo {author} {\bibfnamefont {C.}~\bibnamefont {Ortner}},\ and\ \bibinfo {author} {\bibfnamefont {Y.}~\bibnamefont {Wang}},\ }\bibfield  {title} {\bibinfo {title} {Qm/mm methods for crystalline defects. part 3: Machine-learned mm models},\ }\href@noop {} {\bibfield  {journal} {\bibinfo  {journal} {Multiscale Modeling \& Simulation}\ }\textbf {\bibinfo {volume} {20}},\ \bibinfo {pages} {1490} (\bibinfo {year} {2022})}\BibitemShut {NoStop}%
\bibitem [{\citenamefont {Wang}\ \emph {et~al.}(2021)\citenamefont {Wang}, \citenamefont {Chen}, \citenamefont {Liao}, \citenamefont {Ortner}, \citenamefont {Wang},\ and\ \citenamefont {Zhang}}]{wang2021posteriori}%
  \BibitemOpen
  \bibfield  {author} {\bibinfo {author} {\bibfnamefont {Y.}~\bibnamefont {Wang}}, \bibinfo {author} {\bibfnamefont {H.}~\bibnamefont {Chen}}, \bibinfo {author} {\bibfnamefont {M.}~\bibnamefont {Liao}}, \bibinfo {author} {\bibfnamefont {C.}~\bibnamefont {Ortner}}, \bibinfo {author} {\bibfnamefont {H.}~\bibnamefont {Wang}},\ and\ \bibinfo {author} {\bibfnamefont {L.}~\bibnamefont {Zhang}},\ }\bibfield  {title} {\bibinfo {title} {A posteriori error estimates for adaptive qm/mm coupling methods},\ }\href@noop {} {\bibfield  {journal} {\bibinfo  {journal} {SIAM Journal on Scientific Computing}\ }\textbf {\bibinfo {volume} {43}},\ \bibinfo {pages} {A2785} (\bibinfo {year} {2021})}\BibitemShut {NoStop}%
\bibitem [{\citenamefont {Tan}\ \emph {et~al.}(2023)\citenamefont {Tan}, \citenamefont {Urata}, \citenamefont {Goldman}, \citenamefont {Dietschreit},\ and\ \citenamefont {G{\'o}mez-Bombarelli}}]{tan2023single}%
  \BibitemOpen
  \bibfield  {author} {\bibinfo {author} {\bibfnamefont {A.~R.}\ \bibnamefont {Tan}}, \bibinfo {author} {\bibfnamefont {S.}~\bibnamefont {Urata}}, \bibinfo {author} {\bibfnamefont {S.}~\bibnamefont {Goldman}}, \bibinfo {author} {\bibfnamefont {J.~C.}\ \bibnamefont {Dietschreit}},\ and\ \bibinfo {author} {\bibfnamefont {R.}~\bibnamefont {G{\'o}mez-Bombarelli}},\ }\bibfield  {title} {\bibinfo {title} {Single-model uncertainty quantification in neural network potentials does not consistently outperform model ensembles},\ }\href@noop {} {\bibfield  {journal} {\bibinfo  {journal} {npj Computational Materials}\ }\textbf {\bibinfo {volume} {9}},\ \bibinfo {pages} {225} (\bibinfo {year} {2023})}\BibitemShut {NoStop}%
\bibitem [{\citenamefont {Zaverkin}\ \emph {et~al.}(2024)\citenamefont {Zaverkin}, \citenamefont {Holzm{\"u}ller}, \citenamefont {Christiansen}, \citenamefont {Errica}, \citenamefont {Alesiani}, \citenamefont {Takamoto}, \citenamefont {Niepert},\ and\ \citenamefont {K{\"a}stner}}]{zaverkin2024uncertainty}%
  \BibitemOpen
  \bibfield  {author} {\bibinfo {author} {\bibfnamefont {V.}~\bibnamefont {Zaverkin}}, \bibinfo {author} {\bibfnamefont {D.}~\bibnamefont {Holzm{\"u}ller}}, \bibinfo {author} {\bibfnamefont {H.}~\bibnamefont {Christiansen}}, \bibinfo {author} {\bibfnamefont {F.}~\bibnamefont {Errica}}, \bibinfo {author} {\bibfnamefont {F.}~\bibnamefont {Alesiani}}, \bibinfo {author} {\bibfnamefont {M.}~\bibnamefont {Takamoto}}, \bibinfo {author} {\bibfnamefont {M.}~\bibnamefont {Niepert}},\ and\ \bibinfo {author} {\bibfnamefont {J.}~\bibnamefont {K{\"a}stner}},\ }\bibfield  {title} {\bibinfo {title} {Uncertainty-biased molecular dynamics for learning uniformly accurate interatomic potentials},\ }\href@noop {} {\bibfield  {journal} {\bibinfo  {journal} {npj Computational Materials}\ }\textbf {\bibinfo {volume} {10}},\ \bibinfo {pages} {83} (\bibinfo {year} {2024})}\BibitemShut {NoStop}%
\bibitem [{\citenamefont {Hodapp}\ and\ \citenamefont {Shapeev}(2020)}]{hodapp2020operando}%
  \BibitemOpen
  \bibfield  {author} {\bibinfo {author} {\bibfnamefont {M.}~\bibnamefont {Hodapp}}\ and\ \bibinfo {author} {\bibfnamefont {A.}~\bibnamefont {Shapeev}},\ }\bibfield  {title} {\bibinfo {title} {In operando active learning of interatomic interaction during large-scale simulations},\ }\href@noop {} {\bibfield  {journal} {\bibinfo  {journal} {Machine Learning: Science and Technology}\ }\textbf {\bibinfo {volume} {1}},\ \bibinfo {pages} {045005} (\bibinfo {year} {2020})}\BibitemShut {NoStop}%
\bibitem [{\citenamefont {D{\"o}singer}\ \emph {et~al.}(2023)\citenamefont {D{\"o}singer}, \citenamefont {Hodapp}, \citenamefont {Peil}, \citenamefont {Reichmann}, \citenamefont {Razumovskiy}, \citenamefont {Scheiber},\ and\ \citenamefont {Romaner}}]{dosinger2023efficient}%
  \BibitemOpen
  \bibfield  {author} {\bibinfo {author} {\bibfnamefont {C.}~\bibnamefont {D{\"o}singer}}, \bibinfo {author} {\bibfnamefont {M.}~\bibnamefont {Hodapp}}, \bibinfo {author} {\bibfnamefont {O.}~\bibnamefont {Peil}}, \bibinfo {author} {\bibfnamefont {A.}~\bibnamefont {Reichmann}}, \bibinfo {author} {\bibfnamefont {V.}~\bibnamefont {Razumovskiy}}, \bibinfo {author} {\bibfnamefont {D.}~\bibnamefont {Scheiber}},\ and\ \bibinfo {author} {\bibfnamefont {L.}~\bibnamefont {Romaner}},\ }\bibfield  {title} {\bibinfo {title} {Efficient descriptors and active learning for grain boundary segregation},\ }\href@noop {} {\bibfield  {journal} {\bibinfo  {journal} {Physical Review Materials}\ }\textbf {\bibinfo {volume} {7}},\ \bibinfo {pages} {113606} (\bibinfo {year} {2023})}\BibitemShut {NoStop}%
\bibitem [{\citenamefont {van~der Oord}\ \emph {et~al.}(2023)\citenamefont {van~der Oord}, \citenamefont {Sachs}, \citenamefont {Kov{\'a}cs}, \citenamefont {Ortner},\ and\ \citenamefont {Cs{\'a}nyi}}]{van2023hyperactive}%
  \BibitemOpen
  \bibfield  {author} {\bibinfo {author} {\bibfnamefont {C.}~\bibnamefont {van~der Oord}}, \bibinfo {author} {\bibfnamefont {M.}~\bibnamefont {Sachs}}, \bibinfo {author} {\bibfnamefont {D.~P.}\ \bibnamefont {Kov{\'a}cs}}, \bibinfo {author} {\bibfnamefont {C.}~\bibnamefont {Ortner}},\ and\ \bibinfo {author} {\bibfnamefont {G.}~\bibnamefont {Cs{\'a}nyi}},\ }\bibfield  {title} {\bibinfo {title} {Hyperactive learning for data-driven interatomic potentials},\ }\href@noop {} {\bibfield  {journal} {\bibinfo  {journal} {npj Computational Materials}\ }\textbf {\bibinfo {volume} {9}},\ \bibinfo {pages} {168} (\bibinfo {year} {2023})}\BibitemShut {NoStop}%
\bibitem [{\citenamefont {Mismetti}\ and\ \citenamefont {Hodapp}(2024)}]{mismetti2024automated}%
  \BibitemOpen
  \bibfield  {author} {\bibinfo {author} {\bibfnamefont {L.}~\bibnamefont {Mismetti}}\ and\ \bibinfo {author} {\bibfnamefont {M.}~\bibnamefont {Hodapp}},\ }\bibfield  {title} {\bibinfo {title} {Automated atomistic simulations of dissociated dislocations with ab initio accuracy},\ }\href@noop {} {\bibfield  {journal} {\bibinfo  {journal} {Physical Review B}\ }\textbf {\bibinfo {volume} {109}},\ \bibinfo {pages} {094120} (\bibinfo {year} {2024})}\BibitemShut {NoStop}%
\bibitem [{\citenamefont {Angelopoulos}\ and\ \citenamefont {Bates}(2021)}]{angelopoulos2021gentle}%
  \BibitemOpen
  \bibfield  {author} {\bibinfo {author} {\bibfnamefont {A.~N.}\ \bibnamefont {Angelopoulos}}\ and\ \bibinfo {author} {\bibfnamefont {S.}~\bibnamefont {Bates}},\ }\bibfield  {title} {\bibinfo {title} {A gentle introduction to conformal prediction and distribution-free uncertainty quantification},\ }\href@noop {} {\bibfield  {journal} {\bibinfo  {journal} {arXiv preprint arXiv:2107.07511}\ } (\bibinfo {year} {2021})}\BibitemShut {NoStop}%
\bibitem [{\citenamefont {Shafer}\ and\ \citenamefont {Vovk}(2008)}]{shafer2008tutorial}%
  \BibitemOpen
  \bibfield  {author} {\bibinfo {author} {\bibfnamefont {G.}~\bibnamefont {Shafer}}\ and\ \bibinfo {author} {\bibfnamefont {V.}~\bibnamefont {Vovk}},\ }\bibfield  {title} {\bibinfo {title} {A tutorial on conformal prediction.},\ }\href@noop {} {\bibfield  {journal} {\bibinfo  {journal} {Journal of Machine Learning Research}\ }\textbf {\bibinfo {volume} {9}} (\bibinfo {year} {2008})}\BibitemShut {NoStop}%
\bibitem [{\citenamefont {Yu}\ \emph {et~al.}(2025)\citenamefont {Yu}, \citenamefont {Ho},\ and\ \citenamefont {Wang}}]{yu2025conformal}%
  \BibitemOpen
  \bibfield  {author} {\bibinfo {author} {\bibfnamefont {Y.}~\bibnamefont {Yu}}, \bibinfo {author} {\bibfnamefont {C.~H.}\ \bibnamefont {Ho}},\ and\ \bibinfo {author} {\bibfnamefont {Y.}~\bibnamefont {Wang}},\ }\bibfield  {title} {\bibinfo {title} {A conformal prediction framework for uncertainty quantification in physics-informed neural networks},\ }\href@noop {} {\bibfield  {journal} {\bibinfo  {journal} {arXiv preprint arXiv:2509.13717}\ } (\bibinfo {year} {2025})}\BibitemShut {NoStop}%
\bibitem [{\citenamefont {Hu}\ \emph {et~al.}(2022)\citenamefont {Hu}, \citenamefont {Musielewicz}, \citenamefont {Ulissi},\ and\ \citenamefont {Medford}}]{hu2022robust}%
  \BibitemOpen
  \bibfield  {author} {\bibinfo {author} {\bibfnamefont {Y.}~\bibnamefont {Hu}}, \bibinfo {author} {\bibfnamefont {J.}~\bibnamefont {Musielewicz}}, \bibinfo {author} {\bibfnamefont {Z.~W.}\ \bibnamefont {Ulissi}},\ and\ \bibinfo {author} {\bibfnamefont {A.~J.}\ \bibnamefont {Medford}},\ }\bibfield  {title} {\bibinfo {title} {Robust and scalable uncertainty estimation with conformal prediction for machine-learned interatomic potentials},\ }\href@noop {} {\bibfield  {journal} {\bibinfo  {journal} {Machine Learning: Science and Technology}\ }\textbf {\bibinfo {volume} {3}},\ \bibinfo {pages} {045028} (\bibinfo {year} {2022})}\BibitemShut {NoStop}%
\bibitem [{\citenamefont {Best}\ \emph {et~al.}(2024)\citenamefont {Best}, \citenamefont {Sullivan},\ and\ \citenamefont {Kermode}}]{best2024uncertainty}%
  \BibitemOpen
  \bibfield  {author} {\bibinfo {author} {\bibfnamefont {I.~R.}\ \bibnamefont {Best}}, \bibinfo {author} {\bibfnamefont {T.~J.}\ \bibnamefont {Sullivan}},\ and\ \bibinfo {author} {\bibfnamefont {J.~R.}\ \bibnamefont {Kermode}},\ }\bibfield  {title} {\bibinfo {title} {{Uncertainty quantification in atomistic simulations of silicon using interatomic potentials}},\ }\href@noop {} {\bibfield  {journal} {\bibinfo  {journal} {The Journal of Chemical Physics}\ }\textbf {\bibinfo {volume} {161}},\ \bibinfo {pages} {064112} (\bibinfo {year} {2024})}\BibitemShut {NoStop}%
\bibitem [{\citenamefont {Gibbs}\ \emph {et~al.}(2025)\citenamefont {Gibbs}, \citenamefont {Cherian},\ and\ \citenamefont {Cand{\`e}s}}]{gibbs2025conformal}%
  \BibitemOpen
  \bibfield  {author} {\bibinfo {author} {\bibfnamefont {I.}~\bibnamefont {Gibbs}}, \bibinfo {author} {\bibfnamefont {J.~J.}\ \bibnamefont {Cherian}},\ and\ \bibinfo {author} {\bibfnamefont {E.~J.}\ \bibnamefont {Cand{\`e}s}},\ }\bibfield  {title} {\bibinfo {title} {Conformal prediction with conditional guarantees},\ }\href@noop {} {\bibfield  {journal} {\bibinfo  {journal} {Journal of the Royal Statistical Society Series B: Statistical Methodology}\ ,\ \bibinfo {pages} {qkaf008}} (\bibinfo {year} {2025})}\BibitemShut {NoStop}%
\bibitem [{\citenamefont {Lakshminarayanan}\ \emph {et~al.}(2017)\citenamefont {Lakshminarayanan}, \citenamefont {Pritzel},\ and\ \citenamefont {Blundell}}]{lakshminarayanan2017simple}%
  \BibitemOpen
  \bibfield  {author} {\bibinfo {author} {\bibfnamefont {B.}~\bibnamefont {Lakshminarayanan}}, \bibinfo {author} {\bibfnamefont {A.}~\bibnamefont {Pritzel}},\ and\ \bibinfo {author} {\bibfnamefont {C.}~\bibnamefont {Blundell}},\ }\bibfield  {title} {\bibinfo {title} {Simple and scalable predictive uncertainty estimation using deep ensembles},\ }\href@noop {} {\bibfield  {journal} {\bibinfo  {journal} {Advances in neural information processing systems}\ }\textbf {\bibinfo {volume} {30}} (\bibinfo {year} {2017})}\BibitemShut {NoStop}%
\bibitem [{\citenamefont {Shaked}(1977)}]{shaked1977concept}%
  \BibitemOpen
  \bibfield  {author} {\bibinfo {author} {\bibfnamefont {M.}~\bibnamefont {Shaked}},\ }\bibfield  {title} {\bibinfo {title} {A concept of positive dependence for exchangeable random variables},\ }\href@noop {} {\bibfield  {journal} {\bibinfo  {journal} {The Annals of Statistics}\ ,\ \bibinfo {pages} {505}} (\bibinfo {year} {1977})}\BibitemShut {NoStop}%
\bibitem [{\citenamefont {Auddy}\ \emph {et~al.}(2024)\citenamefont {Auddy}, \citenamefont {Zou}, \citenamefont {Rahnamarad},\ and\ \citenamefont {Maleki}}]{auddy2024approximate}%
  \BibitemOpen
  \bibfield  {author} {\bibinfo {author} {\bibfnamefont {A.}~\bibnamefont {Auddy}}, \bibinfo {author} {\bibfnamefont {H.}~\bibnamefont {Zou}}, \bibinfo {author} {\bibfnamefont {K.}~\bibnamefont {Rahnamarad}},\ and\ \bibinfo {author} {\bibfnamefont {A.}~\bibnamefont {Maleki}},\ }\bibfield  {title} {\bibinfo {title} {Approximate leave-one-out cross validation for regression with $\ell_1$ regularizers},\ }in\ \href@noop {} {\emph {\bibinfo {booktitle} {International Conference on Artificial Intelligence and Statistics}}}\ (\bibinfo {organization} {PMLR},\ \bibinfo {year} {2024})\ pp.\ \bibinfo {pages} {2377--2385}\BibitemShut {NoStop}%
\bibitem [{\citenamefont {Bai}\ \emph {et~al.}(2022)\citenamefont {Bai}, \citenamefont {Mei}, \citenamefont {Wang}, \citenamefont {Zhou},\ and\ \citenamefont {Xiong}}]{bai2022efficient}%
  \BibitemOpen
  \bibfield  {author} {\bibinfo {author} {\bibfnamefont {Y.}~\bibnamefont {Bai}}, \bibinfo {author} {\bibfnamefont {S.}~\bibnamefont {Mei}}, \bibinfo {author} {\bibfnamefont {H.}~\bibnamefont {Wang}}, \bibinfo {author} {\bibfnamefont {Y.}~\bibnamefont {Zhou}},\ and\ \bibinfo {author} {\bibfnamefont {C.}~\bibnamefont {Xiong}},\ }\bibfield  {title} {\bibinfo {title} {Efficient and differentiable conformal prediction with general function classes},\ }\href@noop {} {\bibfield  {journal} {\bibinfo  {journal} {arXiv preprint arXiv:2202.11091}\ } (\bibinfo {year} {2022})}\BibitemShut {NoStop}%
\bibitem [{\citenamefont {Jung}\ \emph {et~al.}(2022)\citenamefont {Jung}, \citenamefont {Noarov}, \citenamefont {Ramalingam},\ and\ \citenamefont {Roth}}]{jung2022batch}%
  \BibitemOpen
  \bibfield  {author} {\bibinfo {author} {\bibfnamefont {C.}~\bibnamefont {Jung}}, \bibinfo {author} {\bibfnamefont {G.}~\bibnamefont {Noarov}}, \bibinfo {author} {\bibfnamefont {R.}~\bibnamefont {Ramalingam}},\ and\ \bibinfo {author} {\bibfnamefont {A.}~\bibnamefont {Roth}},\ }\bibfield  {title} {\bibinfo {title} {Batch multivalid conformal prediction},\ }\href@noop {} {\bibfield  {journal} {\bibinfo  {journal} {arXiv preprint arXiv:2209.15145}\ } (\bibinfo {year} {2022})}\BibitemShut {NoStop}%
\bibitem [{\citenamefont {Weinan}\ \emph {et~al.}(2019)\citenamefont {Weinan}, \citenamefont {Ma},\ and\ \citenamefont {Wu}}]{weinan2019barron}%
  \BibitemOpen
  \bibfield  {author} {\bibinfo {author} {\bibfnamefont {E.}~\bibnamefont {Weinan}}, \bibinfo {author} {\bibfnamefont {C.}~\bibnamefont {Ma}},\ and\ \bibinfo {author} {\bibfnamefont {L.}~\bibnamefont {Wu}},\ }\bibfield  {title} {\bibinfo {title} {Barron spaces and the compositional function spaces for neural network models},\ }\href@noop {} {\bibfield  {journal} {\bibinfo  {journal} {arXiv preprint arXiv:1906.08039}\ } (\bibinfo {year} {2019})}\BibitemShut {NoStop}%
\bibitem [{\citenamefont {Foygel~Barber}\ \emph {et~al.}(2021)\citenamefont {Foygel~Barber}, \citenamefont {Candes}, \citenamefont {Ramdas},\ and\ \citenamefont {Tibshirani}}]{foygel2021limits}%
  \BibitemOpen
  \bibfield  {author} {\bibinfo {author} {\bibfnamefont {R.}~\bibnamefont {Foygel~Barber}}, \bibinfo {author} {\bibfnamefont {E.~J.}\ \bibnamefont {Candes}}, \bibinfo {author} {\bibfnamefont {A.}~\bibnamefont {Ramdas}},\ and\ \bibinfo {author} {\bibfnamefont {R.~J.}\ \bibnamefont {Tibshirani}},\ }\bibfield  {title} {\bibinfo {title} {The limits of distribution-free conditional predictive inference},\ }\href@noop {} {\bibfield  {journal} {\bibinfo  {journal} {Information and Inference: A Journal of the IMA}\ }\textbf {\bibinfo {volume} {10}},\ \bibinfo {pages} {455} (\bibinfo {year} {2021})}\BibitemShut {NoStop}%
\bibitem [{\citenamefont {Vovk}(2012)}]{vovk2012conditional}%
  \BibitemOpen
  \bibfield  {author} {\bibinfo {author} {\bibfnamefont {V.}~\bibnamefont {Vovk}},\ }\bibfield  {title} {\bibinfo {title} {Conditional validity of inductive conformal predictors},\ }in\ \href@noop {} {\emph {\bibinfo {booktitle} {Asian conference on machine learning}}}\ (\bibinfo {organization} {PMLR},\ \bibinfo {year} {2012})\ pp.\ \bibinfo {pages} {475--490}\BibitemShut {NoStop}%
\bibitem [{\citenamefont {Perez}\ \emph {et~al.}(2025)\citenamefont {Perez}, \citenamefont {Subramanyam}, \citenamefont {Maliyov},\ and\ \citenamefont {Swinburne}}]{perez2025uncertainty}%
  \BibitemOpen
  \bibfield  {author} {\bibinfo {author} {\bibfnamefont {D.}~\bibnamefont {Perez}}, \bibinfo {author} {\bibfnamefont {A.~P.}\ \bibnamefont {Subramanyam}}, \bibinfo {author} {\bibfnamefont {I.}~\bibnamefont {Maliyov}},\ and\ \bibinfo {author} {\bibfnamefont {T.~D.}\ \bibnamefont {Swinburne}},\ }\bibfield  {title} {\bibinfo {title} {Uncertainty quantification for misspecified machine learned interatomic potentials},\ }\href@noop {} {\bibfield  {journal} {\bibinfo  {journal} {npj Computational Materials}\ }\textbf {\bibinfo {volume} {11}},\ \bibinfo {pages} {263} (\bibinfo {year} {2025})}\BibitemShut {NoStop}%
\bibitem [{\citenamefont {Bart{\'o}k}\ \emph {et~al.}(2013)\citenamefont {Bart{\'o}k}, \citenamefont {Kondor},\ and\ \citenamefont {Cs{\'a}nyi}}]{bartok2013representing}%
  \BibitemOpen
  \bibfield  {author} {\bibinfo {author} {\bibfnamefont {A.~P.}\ \bibnamefont {Bart{\'o}k}}, \bibinfo {author} {\bibfnamefont {R.}~\bibnamefont {Kondor}},\ and\ \bibinfo {author} {\bibfnamefont {G.}~\bibnamefont {Cs{\'a}nyi}},\ }\bibfield  {title} {\bibinfo {title} {On representing chemical environments},\ }\href@noop {} {\bibfield  {journal} {\bibinfo  {journal} {Physical Review B—Condensed Matter and Materials Physics}\ }\textbf {\bibinfo {volume} {87}},\ \bibinfo {pages} {184115} (\bibinfo {year} {2013})}\BibitemShut {NoStop}%
\bibitem [{\citenamefont {Drautz}(2020)}]{drautz2020atomic}%
  \BibitemOpen
  \bibfield  {author} {\bibinfo {author} {\bibfnamefont {R.}~\bibnamefont {Drautz}},\ }\bibfield  {title} {\bibinfo {title} {Atomic cluster expansion of scalar, vectorial, and tensorial properties including magnetism and charge transfer},\ }\href@noop {} {\bibfield  {journal} {\bibinfo  {journal} {Physical Review B}\ }\textbf {\bibinfo {volume} {102}},\ \bibinfo {pages} {024104} (\bibinfo {year} {2020})}\BibitemShut {NoStop}%
\bibitem [{\citenamefont {Liu}\ \emph {et~al.}(2022)\citenamefont {Liu}, \citenamefont {Tyagin}, \citenamefont {Ushijima-Mwesigwa}, \citenamefont {Ghosh},\ and\ \citenamefont {Safro}}]{liu2022towards}%
  \BibitemOpen
  \bibfield  {author} {\bibinfo {author} {\bibfnamefont {X.}~\bibnamefont {Liu}}, \bibinfo {author} {\bibfnamefont {I.}~\bibnamefont {Tyagin}}, \bibinfo {author} {\bibfnamefont {H.}~\bibnamefont {Ushijima-Mwesigwa}}, \bibinfo {author} {\bibfnamefont {I.}~\bibnamefont {Ghosh}},\ and\ \bibinfo {author} {\bibfnamefont {I.}~\bibnamefont {Safro}},\ }\bibfield  {title} {\bibinfo {title} {Towards practical explainability with cluster descriptors},\ }in\ \href@noop {} {\emph {\bibinfo {booktitle} {2022 IEEE International Conference on Data Mining Workshops (ICDMW)}}}\ (\bibinfo {organization} {IEEE},\ \bibinfo {year} {2022})\ pp.\ \bibinfo {pages} {1--10}\BibitemShut {NoStop}%
\bibitem [{\citenamefont {Sakai}\ \emph {et~al.}(2024)\citenamefont {Sakai}, \citenamefont {Matsumura}, \citenamefont {Inoue}, \citenamefont {Kawaguchi}, \citenamefont {Thang}, \citenamefont {Ishikawa}, \citenamefont {H{\"o}skuldsson},\ and\ \citenamefont {Sk{\'u}lason}}]{sakai2024active}%
  \BibitemOpen
  \bibfield  {author} {\bibinfo {author} {\bibfnamefont {Y.}~\bibnamefont {Sakai}}, \bibinfo {author} {\bibfnamefont {N.}~\bibnamefont {Matsumura}}, \bibinfo {author} {\bibfnamefont {A.}~\bibnamefont {Inoue}}, \bibinfo {author} {\bibfnamefont {H.}~\bibnamefont {Kawaguchi}}, \bibinfo {author} {\bibfnamefont {D.}~\bibnamefont {Thang}}, \bibinfo {author} {\bibfnamefont {A.}~\bibnamefont {Ishikawa}}, \bibinfo {author} {\bibfnamefont {{\'A}.~B.}\ \bibnamefont {H{\"o}skuldsson}},\ and\ \bibinfo {author} {\bibfnamefont {E.}~\bibnamefont {Sk{\'u}lason}},\ }\bibfield  {title} {\bibinfo {title} {Active learning for graph neural networks training in catalyst energy prediction},\ }in\ \href@noop {} {\emph {\bibinfo {booktitle} {2024 International Joint Conference on Neural Networks (IJCNN)}}}\ (\bibinfo {organization} {IEEE},\ \bibinfo {year} {2024})\ pp.\ \bibinfo {pages} {1--8}\BibitemShut {NoStop}%
\bibitem [{\citenamefont {Wen}\ and\ \citenamefont {Tadmor}(2020)}]{wen2020uncertainty}%
  \BibitemOpen
  \bibfield  {author} {\bibinfo {author} {\bibfnamefont {M.}~\bibnamefont {Wen}}\ and\ \bibinfo {author} {\bibfnamefont {E.~B.}\ \bibnamefont {Tadmor}},\ }\bibfield  {title} {\bibinfo {title} {Uncertainty quantification in molecular simulations with dropout neural network potentials},\ }\href@noop {} {\bibfield  {journal} {\bibinfo  {journal} {npj computational materials}\ }\textbf {\bibinfo {volume} {6}},\ \bibinfo {pages} {124} (\bibinfo {year} {2020})}\BibitemShut {NoStop}%
\bibitem [{\citenamefont {Perego}\ and\ \citenamefont {Bonati}(2024)}]{perego2024data}%
  \BibitemOpen
  \bibfield  {author} {\bibinfo {author} {\bibfnamefont {S.}~\bibnamefont {Perego}}\ and\ \bibinfo {author} {\bibfnamefont {L.}~\bibnamefont {Bonati}},\ }\bibfield  {title} {\bibinfo {title} {Data efficient machine learning potentials for modeling catalytic reactivity via active learning and enhanced sampling},\ }\href@noop {} {\bibfield  {journal} {\bibinfo  {journal} {npj Computational Materials}\ }\textbf {\bibinfo {volume} {10}},\ \bibinfo {pages} {291} (\bibinfo {year} {2024})}\BibitemShut {NoStop}%
\bibitem [{\citenamefont {Bigi}\ \emph {et~al.}(2024)\citenamefont {Bigi}, \citenamefont {Chong}, \citenamefont {Ceriotti},\ and\ \citenamefont {Grasselli}}]{bigi2024prediction}%
  \BibitemOpen
  \bibfield  {author} {\bibinfo {author} {\bibfnamefont {F.}~\bibnamefont {Bigi}}, \bibinfo {author} {\bibfnamefont {S.}~\bibnamefont {Chong}}, \bibinfo {author} {\bibfnamefont {M.}~\bibnamefont {Ceriotti}},\ and\ \bibinfo {author} {\bibfnamefont {F.}~\bibnamefont {Grasselli}},\ }\bibfield  {title} {\bibinfo {title} {A prediction rigidity formalism for low-cost uncertainties in trained neural networks},\ }\href@noop {} {\bibfield  {journal} {\bibinfo  {journal} {Machine Learning: Science and Technology}\ }\textbf {\bibinfo {volume} {5}},\ \bibinfo {pages} {045018} (\bibinfo {year} {2024})}\BibitemShut {NoStop}%
\bibitem [{\citenamefont {Podryabinkin}\ \emph {et~al.}(2023)\citenamefont {Podryabinkin}, \citenamefont {Garifullin}, \citenamefont {Shapeev},\ and\ \citenamefont {Novikov}}]{podryabinkin2023mlip}%
  \BibitemOpen
  \bibfield  {author} {\bibinfo {author} {\bibfnamefont {E.}~\bibnamefont {Podryabinkin}}, \bibinfo {author} {\bibfnamefont {K.}~\bibnamefont {Garifullin}}, \bibinfo {author} {\bibfnamefont {A.}~\bibnamefont {Shapeev}},\ and\ \bibinfo {author} {\bibfnamefont {I.}~\bibnamefont {Novikov}},\ }\bibfield  {title} {\bibinfo {title} {Mlip-3: Active learning on atomic environments with moment tensor potentials},\ }\href@noop {} {\bibfield  {journal} {\bibinfo  {journal} {The Journal of Chemical Physics}\ }\textbf {\bibinfo {volume} {159}} (\bibinfo {year} {2023})}\BibitemShut {NoStop}%
\bibitem [{\citenamefont {Fan}\ \emph {et~al.}(2022)\citenamefont {Fan}, \citenamefont {Wang}, \citenamefont {Ying}, \citenamefont {Song}, \citenamefont {Wang}, \citenamefont {Wang}, \citenamefont {Zeng}, \citenamefont {Xu}, \citenamefont {Lindgren}, \citenamefont {Rahm} \emph {et~al.}}]{fan2022gpumd}%
  \BibitemOpen
  \bibfield  {author} {\bibinfo {author} {\bibfnamefont {Z.}~\bibnamefont {Fan}}, \bibinfo {author} {\bibfnamefont {Y.}~\bibnamefont {Wang}}, \bibinfo {author} {\bibfnamefont {P.}~\bibnamefont {Ying}}, \bibinfo {author} {\bibfnamefont {K.}~\bibnamefont {Song}}, \bibinfo {author} {\bibfnamefont {J.}~\bibnamefont {Wang}}, \bibinfo {author} {\bibfnamefont {Y.}~\bibnamefont {Wang}}, \bibinfo {author} {\bibfnamefont {Z.}~\bibnamefont {Zeng}}, \bibinfo {author} {\bibfnamefont {K.}~\bibnamefont {Xu}}, \bibinfo {author} {\bibfnamefont {E.}~\bibnamefont {Lindgren}}, \bibinfo {author} {\bibfnamefont {J.~M.}\ \bibnamefont {Rahm}}, \emph {et~al.},\ }\bibfield  {title} {\bibinfo {title} {Gpumd: A package for constructing accurate machine-learned potentials and performing highly efficient atomistic simulations},\ }\href@noop {} {\bibfield  {journal} {\bibinfo  {journal} {The Journal of Chemical Physics}\ }\textbf {\bibinfo {volume} {157}} (\bibinfo {year} {2022})}\BibitemShut {NoStop}%
\bibitem [{\citenamefont {Zhu}\ \emph {et~al.}(2023)\citenamefont {Zhu}, \citenamefont {Batzner}, \citenamefont {Musaelian},\ and\ \citenamefont {Kozinsky}}]{zhu2023fast}%
  \BibitemOpen
  \bibfield  {author} {\bibinfo {author} {\bibfnamefont {A.}~\bibnamefont {Zhu}}, \bibinfo {author} {\bibfnamefont {S.}~\bibnamefont {Batzner}}, \bibinfo {author} {\bibfnamefont {A.}~\bibnamefont {Musaelian}},\ and\ \bibinfo {author} {\bibfnamefont {B.}~\bibnamefont {Kozinsky}},\ }\bibfield  {title} {\bibinfo {title} {Fast uncertainty estimates in deep learning interatomic potentials},\ }\href@noop {} {\bibfield  {journal} {\bibinfo  {journal} {The Journal of Chemical Physics}\ }\textbf {\bibinfo {volume} {158}} (\bibinfo {year} {2023})}\BibitemShut {NoStop}%
\bibitem [{\citenamefont {Beck}\ \emph {et~al.}(2025)\citenamefont {Beck}, \citenamefont {Simko}, \citenamefont {Schaaf}, \citenamefont {Marsalek},\ and\ \citenamefont {Schran}}]{beck2025multi}%
  \BibitemOpen
  \bibfield  {author} {\bibinfo {author} {\bibfnamefont {H.}~\bibnamefont {Beck}}, \bibinfo {author} {\bibfnamefont {P.}~\bibnamefont {Simko}}, \bibinfo {author} {\bibfnamefont {L.~L.}\ \bibnamefont {Schaaf}}, \bibinfo {author} {\bibfnamefont {O.}~\bibnamefont {Marsalek}},\ and\ \bibinfo {author} {\bibfnamefont {C.}~\bibnamefont {Schran}},\ }\bibfield  {title} {\bibinfo {title} {Multi-head committees enable direct uncertainty prediction for atomistic foundation models},\ }\href@noop {} {\bibfield  {journal} {\bibinfo  {journal} {arXiv preprint arXiv:2508.09907}\ } (\bibinfo {year} {2025})}\BibitemShut {NoStop}%
\bibitem [{\citenamefont {Swinburne}\ and\ \citenamefont {Perez}(2025)}]{swinburne2025}%
  \BibitemOpen
  \bibfield  {author} {\bibinfo {author} {\bibfnamefont {T.}~\bibnamefont {Swinburne}}\ and\ \bibinfo {author} {\bibfnamefont {D.}~\bibnamefont {Perez}},\ }\bibfield  {title} {\bibinfo {title} {Parameter uncertainties for imperfect surrogate models in the low-noise regime},\ }\bibfield  {journal} {\bibinfo  {journal} {Machine Learning: Science and Technology}\ }\href {https://doi.org/10.1088/2632-2153/ad9fce} {10.1088/2632-2153/ad9fce} (\bibinfo {year} {2025})\BibitemShut {NoStop}%
\bibitem [{\citenamefont {Batatia}\ \emph {et~al.}(2023)\citenamefont {Batatia}, \citenamefont {Benner}, \citenamefont {Chiang}, \citenamefont {Elena}, \citenamefont {Kov{\'a}cs}, \citenamefont {Riebesell}, \citenamefont {Advincula}, \citenamefont {Asta}, \citenamefont {Avaylon}, \citenamefont {Baldwin} \emph {et~al.}}]{batatia2023foundation}%
  \BibitemOpen
  \bibfield  {author} {\bibinfo {author} {\bibfnamefont {I.}~\bibnamefont {Batatia}}, \bibinfo {author} {\bibfnamefont {P.}~\bibnamefont {Benner}}, \bibinfo {author} {\bibfnamefont {Y.}~\bibnamefont {Chiang}}, \bibinfo {author} {\bibfnamefont {A.~M.}\ \bibnamefont {Elena}}, \bibinfo {author} {\bibfnamefont {D.~P.}\ \bibnamefont {Kov{\'a}cs}}, \bibinfo {author} {\bibfnamefont {J.}~\bibnamefont {Riebesell}}, \bibinfo {author} {\bibfnamefont {X.~R.}\ \bibnamefont {Advincula}}, \bibinfo {author} {\bibfnamefont {M.}~\bibnamefont {Asta}}, \bibinfo {author} {\bibfnamefont {M.}~\bibnamefont {Avaylon}}, \bibinfo {author} {\bibfnamefont {W.~J.}\ \bibnamefont {Baldwin}}, \emph {et~al.},\ }\bibfield  {title} {\bibinfo {title} {A foundation model for atomistic materials chemistry},\ }\href@noop {} {\bibfield  {journal} {\bibinfo  {journal} {arXiv preprint arXiv:2401.00096}\ } (\bibinfo {year} {2023})}\BibitemShut {NoStop}%
\bibitem [{\citenamefont {Rasmussen}(1999)}]{rasmussen1999infinite}%
  \BibitemOpen
  \bibfield  {author} {\bibinfo {author} {\bibfnamefont {C.}~\bibnamefont {Rasmussen}},\ }\bibfield  {title} {\bibinfo {title} {The infinite gaussian mixture model},\ }\href@noop {} {\bibfield  {journal} {\bibinfo  {journal} {Advances in neural information processing systems}\ }\textbf {\bibinfo {volume} {12}} (\bibinfo {year} {1999})}\BibitemShut {NoStop}%
\bibitem [{\citenamefont {Sivaraman}\ \emph {et~al.}(2021)\citenamefont {Sivaraman}, \citenamefont {Guo}, \citenamefont {Ward}, \citenamefont {Hoyt}, \citenamefont {Williamson}, \citenamefont {Foster}, \citenamefont {Benmore},\ and\ \citenamefont {Jackson}}]{sivaraman2021automated}%
  \BibitemOpen
  \bibfield  {author} {\bibinfo {author} {\bibfnamefont {G.}~\bibnamefont {Sivaraman}}, \bibinfo {author} {\bibfnamefont {J.}~\bibnamefont {Guo}}, \bibinfo {author} {\bibfnamefont {L.}~\bibnamefont {Ward}}, \bibinfo {author} {\bibfnamefont {N.}~\bibnamefont {Hoyt}}, \bibinfo {author} {\bibfnamefont {M.}~\bibnamefont {Williamson}}, \bibinfo {author} {\bibfnamefont {I.}~\bibnamefont {Foster}}, \bibinfo {author} {\bibfnamefont {C.}~\bibnamefont {Benmore}},\ and\ \bibinfo {author} {\bibfnamefont {N.}~\bibnamefont {Jackson}},\ }\bibfield  {title} {\bibinfo {title} {Automated development of molten salt machine learning potentials: application to licl},\ }\href@noop {} {\bibfield  {journal} {\bibinfo  {journal} {The Journal of Physical Chemistry Letters}\ }\textbf {\bibinfo {volume} {12}},\ \bibinfo {pages} {4278} (\bibinfo {year} {2021})}\BibitemShut {NoStop}%
\bibitem [{\citenamefont {Jain}\ \emph {et~al.}(2013)\citenamefont {Jain}, \citenamefont {Ong}, \citenamefont {Hautier}, \citenamefont {Chen}, \citenamefont {Richards}, \citenamefont {Dacek}, \citenamefont {Cholia}, \citenamefont {Gunter}, \citenamefont {Skinner}, \citenamefont {Ceder} \emph {et~al.}}]{jain2013commentary}%
  \BibitemOpen
  \bibfield  {author} {\bibinfo {author} {\bibfnamefont {A.}~\bibnamefont {Jain}}, \bibinfo {author} {\bibfnamefont {S.~P.}\ \bibnamefont {Ong}}, \bibinfo {author} {\bibfnamefont {G.}~\bibnamefont {Hautier}}, \bibinfo {author} {\bibfnamefont {W.}~\bibnamefont {Chen}}, \bibinfo {author} {\bibfnamefont {W.~D.}\ \bibnamefont {Richards}}, \bibinfo {author} {\bibfnamefont {S.}~\bibnamefont {Dacek}}, \bibinfo {author} {\bibfnamefont {S.}~\bibnamefont {Cholia}}, \bibinfo {author} {\bibfnamefont {D.}~\bibnamefont {Gunter}}, \bibinfo {author} {\bibfnamefont {D.}~\bibnamefont {Skinner}}, \bibinfo {author} {\bibfnamefont {G.}~\bibnamefont {Ceder}}, \emph {et~al.},\ }\bibfield  {title} {\bibinfo {title} {Commentary: The materials project: A materials genome approach to accelerating materials innovation},\ }\href@noop {} {\bibfield  {journal} {\bibinfo  {journal} {APL materials}\ }\textbf {\bibinfo {volume} {1}} (\bibinfo {year} {2013})}\BibitemShut {NoStop}%
\bibitem [{\citenamefont {Kaplan}\ \emph {et~al.}(2025)\citenamefont {Kaplan}, \citenamefont {Liu}, \citenamefont {Qi}, \citenamefont {Ko}, \citenamefont {Deng}, \citenamefont {Riebesell}, \citenamefont {Ceder}, \citenamefont {Persson},\ and\ \citenamefont {Ong}}]{kaplan2025foundational}%
  \BibitemOpen
  \bibfield  {author} {\bibinfo {author} {\bibfnamefont {A.~D.}\ \bibnamefont {Kaplan}}, \bibinfo {author} {\bibfnamefont {R.}~\bibnamefont {Liu}}, \bibinfo {author} {\bibfnamefont {J.}~\bibnamefont {Qi}}, \bibinfo {author} {\bibfnamefont {T.~W.}\ \bibnamefont {Ko}}, \bibinfo {author} {\bibfnamefont {B.}~\bibnamefont {Deng}}, \bibinfo {author} {\bibfnamefont {J.}~\bibnamefont {Riebesell}}, \bibinfo {author} {\bibfnamefont {G.}~\bibnamefont {Ceder}}, \bibinfo {author} {\bibfnamefont {K.~A.}\ \bibnamefont {Persson}},\ and\ \bibinfo {author} {\bibfnamefont {S.~P.}\ \bibnamefont {Ong}},\ }\bibfield  {title} {\bibinfo {title} {A foundational potential energy surface dataset for materials},\ }\href@noop {} {\bibfield  {journal} {\bibinfo  {journal} {arXiv preprint arXiv:2503.04070}\ } (\bibinfo {year} {2025})}\BibitemShut {NoStop}%
\bibitem [{\citenamefont {Benesty}\ \emph {et~al.}(2009)\citenamefont {Benesty}, \citenamefont {Chen}, \citenamefont {Huang},\ and\ \citenamefont {Cohen}}]{benesty2009pearson}%
  \BibitemOpen
  \bibfield  {author} {\bibinfo {author} {\bibfnamefont {J.}~\bibnamefont {Benesty}}, \bibinfo {author} {\bibfnamefont {J.}~\bibnamefont {Chen}}, \bibinfo {author} {\bibfnamefont {Y.}~\bibnamefont {Huang}},\ and\ \bibinfo {author} {\bibfnamefont {I.}~\bibnamefont {Cohen}},\ }\bibfield  {title} {\bibinfo {title} {Pearson correlation coefficient},\ }in\ \href@noop {} {\emph {\bibinfo {booktitle} {Noise reduction in speech processing}}}\ (\bibinfo  {publisher} {Springer},\ \bibinfo {year} {2009})\ pp.\ \bibinfo {pages} {1--4}\BibitemShut {NoStop}%
\bibitem [{\citenamefont {Schaaf}\ \emph {et~al.}(2023)\citenamefont {Schaaf}, \citenamefont {Fako}, \citenamefont {De}, \citenamefont {Sch{\"a}fer},\ and\ \citenamefont {Cs{\'a}nyi}}]{schaaf2023accurate}%
  \BibitemOpen
  \bibfield  {author} {\bibinfo {author} {\bibfnamefont {L.~L.}\ \bibnamefont {Schaaf}}, \bibinfo {author} {\bibfnamefont {E.}~\bibnamefont {Fako}}, \bibinfo {author} {\bibfnamefont {S.}~\bibnamefont {De}}, \bibinfo {author} {\bibfnamefont {A.}~\bibnamefont {Sch{\"a}fer}},\ and\ \bibinfo {author} {\bibfnamefont {G.}~\bibnamefont {Cs{\'a}nyi}},\ }\bibfield  {title} {\bibinfo {title} {Accurate energy barriers for catalytic reaction pathways: an automatic training protocol for machine learning force fields},\ }\href@noop {} {\bibfield  {journal} {\bibinfo  {journal} {npj Computational Materials}\ }\textbf {\bibinfo {volume} {9}},\ \bibinfo {pages} {180} (\bibinfo {year} {2023})}\BibitemShut {NoStop}%
\bibitem [{\citenamefont {Lopanitsyna}\ \emph {et~al.}(2023)\citenamefont {Lopanitsyna}, \citenamefont {Fraux}, \citenamefont {Springer}, \citenamefont {De},\ and\ \citenamefont {Ceriotti}}]{lopanitsyna2023modeling}%
  \BibitemOpen
  \bibfield  {author} {\bibinfo {author} {\bibfnamefont {N.}~\bibnamefont {Lopanitsyna}}, \bibinfo {author} {\bibfnamefont {G.}~\bibnamefont {Fraux}}, \bibinfo {author} {\bibfnamefont {M.~A.}\ \bibnamefont {Springer}}, \bibinfo {author} {\bibfnamefont {S.}~\bibnamefont {De}},\ and\ \bibinfo {author} {\bibfnamefont {M.}~\bibnamefont {Ceriotti}},\ }\bibfield  {title} {\bibinfo {title} {Modeling high-entropy transition metal alloys with alchemical compression},\ }\href@noop {} {\bibfield  {journal} {\bibinfo  {journal} {Physical Review Materials}\ }\textbf {\bibinfo {volume} {7}},\ \bibinfo {pages} {045802} (\bibinfo {year} {2023})}\BibitemShut {NoStop}%
\bibitem [{\citenamefont {Terentjev}\ \emph {et~al.}(2018)\citenamefont {Terentjev}, \citenamefont {Constantin},\ and\ \citenamefont {Pitarke}}]{terentjev2018dispersion}%
  \BibitemOpen
  \bibfield  {author} {\bibinfo {author} {\bibfnamefont {A.~V.}\ \bibnamefont {Terentjev}}, \bibinfo {author} {\bibfnamefont {L.~A.}\ \bibnamefont {Constantin}},\ and\ \bibinfo {author} {\bibfnamefont {J.~M.}\ \bibnamefont {Pitarke}},\ }\bibfield  {title} {\bibinfo {title} {Dispersion-corrected pbesol exchange-correlation functional},\ }\href@noop {} {\bibfield  {journal} {\bibinfo  {journal} {Physical review B}\ }\textbf {\bibinfo {volume} {98}},\ \bibinfo {pages} {214108} (\bibinfo {year} {2018})}\BibitemShut {NoStop}%
\bibitem [{\citenamefont {Levine}\ \emph {et~al.}(2025)\citenamefont {Levine}, \citenamefont {Shuaibi}, \citenamefont {Spotte-Smith}, \citenamefont {Taylor}, \citenamefont {Hasyim}, \citenamefont {Michel}, \citenamefont {Batatia}, \citenamefont {Cs{\'a}nyi}, \citenamefont {Dzamba}, \citenamefont {Eastman} \emph {et~al.}}]{levine2025open}%
  \BibitemOpen
  \bibfield  {author} {\bibinfo {author} {\bibfnamefont {D.~S.}\ \bibnamefont {Levine}}, \bibinfo {author} {\bibfnamefont {M.}~\bibnamefont {Shuaibi}}, \bibinfo {author} {\bibfnamefont {E.~W.~C.}\ \bibnamefont {Spotte-Smith}}, \bibinfo {author} {\bibfnamefont {M.~G.}\ \bibnamefont {Taylor}}, \bibinfo {author} {\bibfnamefont {M.~R.}\ \bibnamefont {Hasyim}}, \bibinfo {author} {\bibfnamefont {K.}~\bibnamefont {Michel}}, \bibinfo {author} {\bibfnamefont {I.}~\bibnamefont {Batatia}}, \bibinfo {author} {\bibfnamefont {G.}~\bibnamefont {Cs{\'a}nyi}}, \bibinfo {author} {\bibfnamefont {M.}~\bibnamefont {Dzamba}}, \bibinfo {author} {\bibfnamefont {P.}~\bibnamefont {Eastman}}, \emph {et~al.},\ }\bibfield  {title} {\bibinfo {title} {The open molecules 2025 (omol25) dataset, evaluations, and models},\ }\href@noop {} {\bibfield  {journal} {\bibinfo  {journal} {arXiv preprint arXiv:2505.08762}\ } (\bibinfo {year} {2025})}\BibitemShut {NoStop}%
\bibitem [{\citenamefont {Owen}\ \emph {et~al.}(2024)\citenamefont {Owen}, \citenamefont {Torrisi}, \citenamefont {Xie}, \citenamefont {Batzner}, \citenamefont {Bystrom}, \citenamefont {Coulter}, \citenamefont {Musaelian}, \citenamefont {Sun},\ and\ \citenamefont {Kozinsky}}]{owen2024complexity}%
  \BibitemOpen
  \bibfield  {author} {\bibinfo {author} {\bibfnamefont {C.~J.}\ \bibnamefont {Owen}}, \bibinfo {author} {\bibfnamefont {S.~B.}\ \bibnamefont {Torrisi}}, \bibinfo {author} {\bibfnamefont {Y.}~\bibnamefont {Xie}}, \bibinfo {author} {\bibfnamefont {S.}~\bibnamefont {Batzner}}, \bibinfo {author} {\bibfnamefont {K.}~\bibnamefont {Bystrom}}, \bibinfo {author} {\bibfnamefont {J.}~\bibnamefont {Coulter}}, \bibinfo {author} {\bibfnamefont {A.}~\bibnamefont {Musaelian}}, \bibinfo {author} {\bibfnamefont {L.}~\bibnamefont {Sun}},\ and\ \bibinfo {author} {\bibfnamefont {B.}~\bibnamefont {Kozinsky}},\ }\bibfield  {title} {\bibinfo {title} {Complexity of many-body interactions in transition metals via machine-learned force fields from the tm23 data set},\ }\href@noop {} {\bibfield  {journal} {\bibinfo  {journal} {npj Computational Materials}\ }\textbf {\bibinfo {volume} {10}},\ \bibinfo {pages} {92} (\bibinfo {year} {2024})}\BibitemShut {NoStop}%
\bibitem [{\citenamefont {Romano}\ \emph {et~al.}(2019)\citenamefont {Romano}, \citenamefont {Patterson},\ and\ \citenamefont {Candes}}]{romano2019conformalized}%
  \BibitemOpen
  \bibfield  {author} {\bibinfo {author} {\bibfnamefont {Y.}~\bibnamefont {Romano}}, \bibinfo {author} {\bibfnamefont {E.}~\bibnamefont {Patterson}},\ and\ \bibinfo {author} {\bibfnamefont {E.}~\bibnamefont {Candes}},\ }\bibfield  {title} {\bibinfo {title} {Conformalized quantile regression},\ }\href@noop {} {\bibfield  {journal} {\bibinfo  {journal} {Advances in neural information processing systems}\ }\textbf {\bibinfo {volume} {32}} (\bibinfo {year} {2019})}\BibitemShut {NoStop}%
\bibitem [{\citenamefont {de~Swart}(2018)}]{de2018octet}%
  \BibitemOpen
  \bibfield  {author} {\bibinfo {author} {\bibfnamefont {J.~J.}\ \bibnamefont {de~Swart}},\ }\bibfield  {title} {\bibinfo {title} {The octet model and its clebsch-gordan coefficients},\ }in\ \href@noop {} {\emph {\bibinfo {booktitle} {The Eightfold Way}}}\ (\bibinfo  {publisher} {CRC Press},\ \bibinfo {year} {2018})\ pp.\ \bibinfo {pages} {120--143}\BibitemShut {NoStop}%
\bibitem [{\citenamefont {He}\ \emph {et~al.}(2016)\citenamefont {He}, \citenamefont {Zhang}, \citenamefont {Ren},\ and\ \citenamefont {Sun}}]{he2016deep}%
  \BibitemOpen
  \bibfield  {author} {\bibinfo {author} {\bibfnamefont {K.}~\bibnamefont {He}}, \bibinfo {author} {\bibfnamefont {X.}~\bibnamefont {Zhang}}, \bibinfo {author} {\bibfnamefont {S.}~\bibnamefont {Ren}},\ and\ \bibinfo {author} {\bibfnamefont {J.}~\bibnamefont {Sun}},\ }\bibfield  {title} {\bibinfo {title} {Deep residual learning for image recognition},\ }in\ \href@noop {} {\emph {\bibinfo {booktitle} {Proceedings of the IEEE conference on computer vision and pattern recognition}}}\ (\bibinfo {year} {2016})\ pp.\ \bibinfo {pages} {770--778}\BibitemShut {NoStop}%
\bibitem [{\citenamefont {Chong}\ \emph {et~al.}(2023)\citenamefont {Chong}, \citenamefont {Grasselli}, \citenamefont {Ben~Mahmoud}, \citenamefont {Morrow}, \citenamefont {Deringer},\ and\ \citenamefont {Ceriotti}}]{chong2023robustness}%
  \BibitemOpen
  \bibfield  {author} {\bibinfo {author} {\bibfnamefont {S.}~\bibnamefont {Chong}}, \bibinfo {author} {\bibfnamefont {F.}~\bibnamefont {Grasselli}}, \bibinfo {author} {\bibfnamefont {C.}~\bibnamefont {Ben~Mahmoud}}, \bibinfo {author} {\bibfnamefont {J.~D.}\ \bibnamefont {Morrow}}, \bibinfo {author} {\bibfnamefont {V.~L.}\ \bibnamefont {Deringer}},\ and\ \bibinfo {author} {\bibfnamefont {M.}~\bibnamefont {Ceriotti}},\ }\bibfield  {title} {\bibinfo {title} {Robustness of local predictions in atomistic machine learning models},\ }\href@noop {} {\bibfield  {journal} {\bibinfo  {journal} {Journal of Chemical Theory and Computation}\ }\textbf {\bibinfo {volume} {19}},\ \bibinfo {pages} {8020} (\bibinfo {year} {2023})}\BibitemShut {NoStop}%
\bibitem [{\citenamefont {Schraudolph}(2002)}]{schraudolph2002fast}%
  \BibitemOpen
  \bibfield  {author} {\bibinfo {author} {\bibfnamefont {N.~N.}\ \bibnamefont {Schraudolph}},\ }\bibfield  {title} {\bibinfo {title} {Fast curvature matrix-vector products for second-order gradient descent},\ }\href@noop {} {\bibfield  {journal} {\bibinfo  {journal} {Neural computation}\ }\textbf {\bibinfo {volume} {14}},\ \bibinfo {pages} {1723} (\bibinfo {year} {2002})}\BibitemShut {NoStop}%
\bibitem [{\citenamefont {Holzm{\"u}ller}\ \emph {et~al.}(2023)\citenamefont {Holzm{\"u}ller}, \citenamefont {Zaverkin}, \citenamefont {K{\"a}stner},\ and\ \citenamefont {Steinwart}}]{holzmuller2023framework}%
  \BibitemOpen
  \bibfield  {author} {\bibinfo {author} {\bibfnamefont {D.}~\bibnamefont {Holzm{\"u}ller}}, \bibinfo {author} {\bibfnamefont {V.}~\bibnamefont {Zaverkin}}, \bibinfo {author} {\bibfnamefont {J.}~\bibnamefont {K{\"a}stner}},\ and\ \bibinfo {author} {\bibfnamefont {I.}~\bibnamefont {Steinwart}},\ }\bibfield  {title} {\bibinfo {title} {A framework and benchmark for deep batch active learning for regression},\ }\href@noop {} {\bibfield  {journal} {\bibinfo  {journal} {Journal of Machine Learning Research}\ }\textbf {\bibinfo {volume} {24}},\ \bibinfo {pages} {1} (\bibinfo {year} {2023})}\BibitemShut {NoStop}%
\bibitem [{\citenamefont {Levenberg}(1944)}]{levenberg1944method}%
  \BibitemOpen
  \bibfield  {author} {\bibinfo {author} {\bibfnamefont {K.}~\bibnamefont {Levenberg}},\ }\bibfield  {title} {\bibinfo {title} {A method for the solution of certain non-linear problems in least squares},\ }\href@noop {} {\bibfield  {journal} {\bibinfo  {journal} {Quarterly of applied mathematics}\ }\textbf {\bibinfo {volume} {2}},\ \bibinfo {pages} {164} (\bibinfo {year} {1944})}\BibitemShut {NoStop}%
\bibitem [{\citenamefont {Marquardt}(1963)}]{marquardt1963algorithm}%
  \BibitemOpen
  \bibfield  {author} {\bibinfo {author} {\bibfnamefont {D.~W.}\ \bibnamefont {Marquardt}},\ }\bibfield  {title} {\bibinfo {title} {An algorithm for least-squares estimation of nonlinear parameters},\ }\href@noop {} {\bibfield  {journal} {\bibinfo  {journal} {Journal of the society for Industrial and Applied Mathematics}\ }\textbf {\bibinfo {volume} {11}},\ \bibinfo {pages} {431} (\bibinfo {year} {1963})}\BibitemShut {NoStop}%
\end{thebibliography}%

\end{document}